\documentclass[preprint, twocolumn]{aastex631}

\pagecolor{white}

\usepackage{graphicx}
\usepackage{bm}

\usepackage{multirow,xcolor}
\usepackage{color}
\usepackage{xfrac}

\begin{document}

\title{AGNBoost: A Machine Learning Approach to AGN Identification with JWST/NIRCam+MIRI Colors and Photometry}

\author[0000-0002-6292-4589]{Kurt Hamblin}
\affiliation{Department of Physics and Astronomy, University of Kansas, Lawrence, KS 66045, USA}

\author[0000-0002-5537-8110]{Allison Kirkpatrick}
\affiliation{Department of Physics and Astronomy, University of Kansas, Lawrence, KS 66045, USA}

\author[0000-0001-8534-7502]{Bren E. Backhaus} 
\affiliation{Department of Physics and Astronomy, University of Kansas, Lawrence, KS 66045, USA} 

\author[0000-0001-5930-0532]{Gregory Troiani}
\affil{Department of Physics and Astronomy, University of Kansas, Lawrence, KS 66045, USA}

\author[0000-0001-9187-3605]{Jeyhan S. Kartaltepe}
\affiliation{Laboratory for Multiwavelength Astrophysics, School of Physics and Astronomy, Rochester Institute of Technology, 84 Lomb Memorial Drive, Rochester, NY 14623, USA}
 
\author[0000-0002-8360-3880]{Dale D. Kocevski}
\affiliation{Department of Physics and Astronomy, Colby College, Waterville, ME 04901, USA}

\author[0000-0002-6610-2048]{Anton M. Koekemoer} 
\affiliation{Space Telescope Science Institute, 3700 San Martin Drive, Baltimore, MD 21218, USA} 

\author[0000-0003-3216-7190]{Erini Lambrides}\altaffiliation{NPP Fellow}
\affiliation{NASA-Goddard Space Flight Center, Code 662, Greenbelt, MD, 20771, USA}

\author[0000-0001-7503-8482]{Casey Papovich} 
\affiliation{Department of Physics and Astronomy, Texas A\&M University, College Station, TX, 77843-4242 USA} 
\affiliation{George P.\ and Cynthia Woods Mitchell Institute for Fundamental Physics and Astronomy, Texas A\&M University, College Station, TX, 77843-4242 USA} 

\author[0000-0001-5749-5452]{Kaila Ronayne} 
\affiliation{Department of Physics and Astronomy, Texas A\&M University, College Station, TX, 77843-4242 USA} 
\affiliation{George P.\ and Cynthia Woods Mitchell Institute for Fundamental Physics and Astronomy, Texas A\&M University, College Station, TX, 77843-4242 USA} 

\author[0000-0001-8835-7722]{Guang Yang}
\affiliation{Nanjing Institute of Astronomical Optics \& Technology, Chinese Academy of Sciences, Nanjing 210042, China}
\affiliation{CAS Key Laboratory of Astronomical Optics \& Technology, Nanjing Institute of Astronomical Optics \& Technology, Nanjing 210042, China}
 
\author[0000-0002-9921-9218]{Micaela B. Bagley}
\affiliation{Astrophysics Science Division, NASA Goddard Space Flight Center, 8800 Greenbelt Rd, Greenbelt, MD 20771, USA}
\affiliation{Department of Astronomy, The University of Texas at Austin, Austin, TX, USA}

\author[0000-0001-5414-5131]{Mark Dickinson}
\affiliation{NSF's National Optical-Infrared Astronomy Research Laboratory, 950 N. Cherry Ave., Tucson, AZ 85719, USA}

\author[0000-0001-8519-1130]{Steven L. Finkelstein}
\affiliation{Department of Astronomy, The University of Texas at Austin, Austin, TX, USA}
\affiliation{Cosmic Frontier Center, The University of Texas at Austin, Austin, TX, USA}

\author[0000-0002-7959-8783]{Pablo Arrabal Haro}
\altaffiliation{NASA Postdoctoral Fellow}
\affiliation{Astrophysics Science Division, NASA Goddard Space Flight Center, 8800 Greenbelt Rd, Greenbelt, MD 20771, USA}

\author[0000-0001-9879-7780]{Fabio Pacucci}
\affiliation{Center for Astrophysics $\vert$ Harvard \& Smithsonian, 60 Garden St, Cambridge, MA 02138, USA}
\affiliation{Black Hole Initiative, Harvard University, 20 Garden St, Cambridge, MA 02138, USA}

\author[0000-0002-1410-0470]{Jonathan R. Trump}
\affiliation{Department of Physics, 196 Auditorium Road, Unit 3046, University of Connecticut, Storrs, CT 06269, USA}

\author[0000-0003-3382-5941]{Nor Pirzkal}
\affiliation{ESA/AURA Space Telescope Science Institute}

\author[0000-0002-6219-5558]{Alexander de la Vega}
\affiliation{Department of Physics and Astronomy, University of California, 900 University Ave, Riverside, CA 92521, USA}

\author[0009-0002-2209-4813]{Edgar Perez Vidal}
\affiliation{Department of Physics and Astronomy, Tufts University, Medford, MA 02155, USA}

\author[0000-0003-3466-035X]{{L. Y. Aaron} {Yung}}
\affiliation{Space Telescope Science Institute, 3700 San Martin Drive, Baltimore, MD 21218, USA}

\begin{abstract}

We present {\tt AGNBoost}, a machine learning framework utilizing {\tt XGBoostLSS} to identify AGN and estimate redshifts from JWST NIRCam and MIRI photometry. {\tt AGNBoost} constructs 66 input features from 7 NIRCam and 4 MIRI bands to predict the fraction of mid-IR $3$--$30\,\mu$m emission attributable to an AGN power law ($\text{frac}_{\text{AGN}}$) and photometric redshift. Each model is trained on $10^6$ simulated galaxies from {\tt CIGALE}. Models are tested on mock {\tt CIGALE} galaxies, an independent set of empirically-derived templates, and 748 observations from the JWST MIRI EGS Galaxy and AGN (MEGA) survey. On idealized noise-free mock {\tt CIGALE} galaxies, {\tt AGNBoost} achieves $15\%$ outlier fractions of $1.63\%$ ($\text{frac}_{\text{AGN}}$) and $0.15\%$ (redshift), with $\sigma_{\text{RMSE}} = 0.045$ for $\text{frac}_{\text{AGN}}$ and $\sigma_{\text{NMAD}} = 0.004$ for redshift. When realistic photometric uncertainties are introduced, performance remains robust with median predictions on the 1:1 relation, though outlier fractions increase to $4.38\%$ and $3.35\%$, respectively. On the independent template set, {\tt AGNBoost} identifies $92.6\%$ of AGN candidates with $\text{frac}_{\text{AGN}} > 0.3$ and $100\%$ with $\text{frac}_{\text{AGN}} > 0.5$, demonstrating generalization beyond the training distribution. On MEGA galaxies with spectroscopic redshifts, {\tt AGNBoost} achieves $\sigma_{\text{NMAD}} = 0.056$ and $19.79\%$ outliers. {\tt AGNBoost} $\text{frac}_{\text{AGN}}$ estimates broadly agree with {\tt CIGALE} fitting ($\sigma_{\text{RMSE}} = 0.178$, $11.96\%$ outliers). The flexible framework allows straightforward incorporation of additional photometric bands and re-training for other variables. {\tt AGNBoost}'s computational efficiency makes it well-suited for wide-sky surveys requiring rapid AGN identification and redshift estimation.
\end{abstract}

\section{Introduction} \label{sec:intro}
The James Webb Space Telescope (JWST; \citealp{Gardner_2006}) has revolutionized our ability to study galaxies in the mid-infrared, achieving sensitivity levels an order of magnitude deeper than Spitzer/MIPS in significantly shorter integration times \citep{rigby_science_2023, Gardner_2006, jwst_2023}. The Mid-Infrared Instrument (MIRI; \citealp{Rieke_miri}) provides unprecedented wavelength coverage from  $5-28.5\,\mu$m across nine photometric filters, enabling detailed characterization of dusty galaxies out to the so-called ``cosmic noon'' epoch ($z\sim 1-3$), during which the bulk of cosmic star formation and black hole growth occurred  \citep{heckman_present-day_2004, hopkins_evolution_2004, fontana_galaxy_2006, perez-gonzalez_stellar_2008, shankar_self-consistent_2009, aird_evolution_2010, kormendy_coevolution_2013, madau_cosmic_2014, forster_schreiber_star-forming_2020}. This enhanced sensitivity and spectral coverage is particularly essential for studying the growth of galaxies and active galactic nuclei (AGN) over cosmic time and, notably, at cosmic noon. JWST/MIRI is able to identify heavily obscured AGNs at cosmic noon that are entirely missed in traditional optical surveys, allowing for the study of supermassive black hole-host coevolution \citep[e.g.][]{yang_ceers_2023,kirkpatrick_ceers_2023}. This requires disentangling the emission from AGN and highly star-forming galaxies (SFGs), a process that requires accurate models and often significant computational resources, especially in the case of large surveys. 

The mid-IR emission from AGNs and SFGs originates from fundamentally different physical processes. In AGN, the central supermassive black hole is surrounded by a dusty torus that absorbs and reprocesses radiation from the accretion disk. This dust, heated to temperatures over $1000\,$K, produces a characteristic power-law continuum ($f_{\nu} \propto \nu^{-\alpha}$) in the mid-IR \citep{stern_mid-infrared_2005}. The slope and intensity of this continuum reflect both the intrinsic AGN luminosity and the geometry of the obscuring torus \citep[e.g.]{laurent_mid-infrared_2000, toba_luminosity_2014,stalevski_dust_2016}. In contrast, SFGs exhibit strong emission features from polycyclic aromatic hydrocarbons (PAHs), with the most prominent features at 6.2, 7.7, 11.2, and 12.7 $\mu$m, excited by UV photons from young stellar populations \citep{Draine_2001, peeters_polycyclic_2004, magdis_evolving_2012}. The relative strengths of these PAH features to the underlying continuum provide key diagnostics to distinguish active from non-active galaxies.  

\begin{figure*}[t!]
\begin{interactive}{animation}{sed_color_animation.mp4}
\plotone{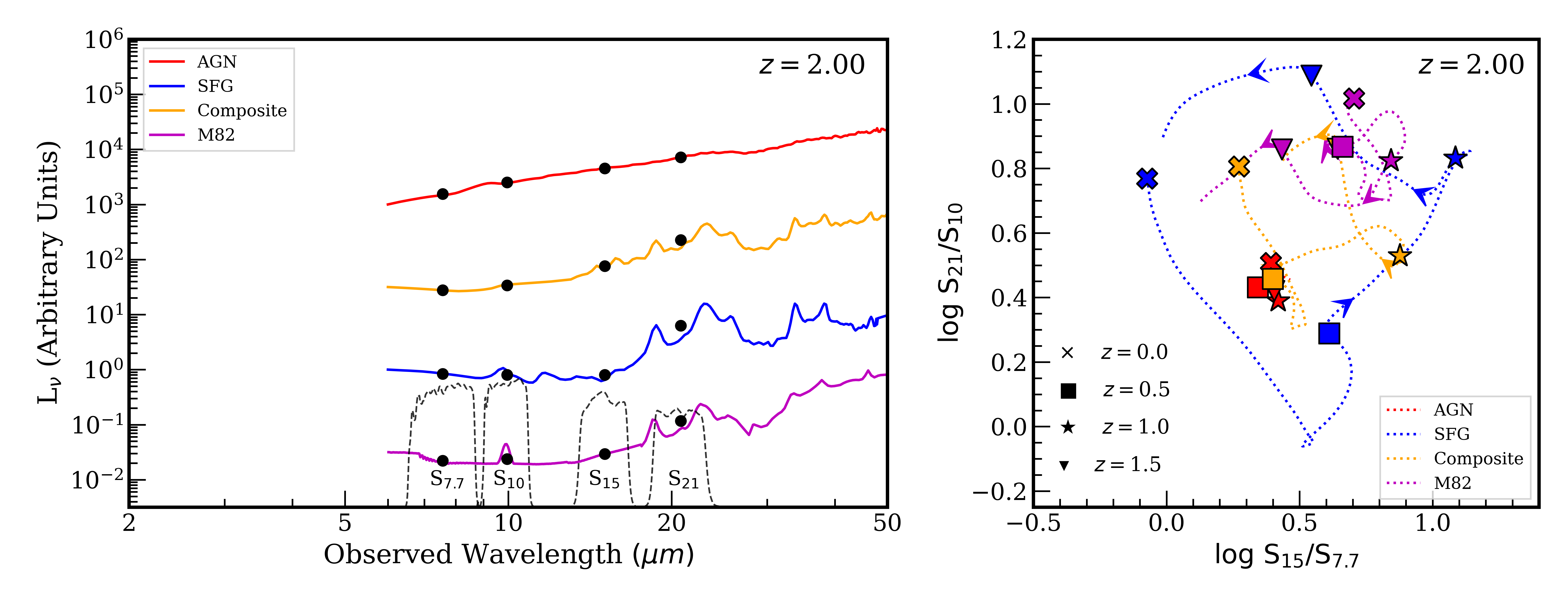}
\end{interactive}
\caption{A single frame of an animation illustrating the mid-IR color properties of SF+AGN composites, SFGs, and mid-IR weak galaxies. The AGN, Composite, and SFG templates are from \citet{kirkpatrick_2015}. Here, M82 \citep{m82_paper} serves as a proxy for a SFG with weak PAH features. Without prior redshift information, it is not possible to draw boundaries in this color space that robustly separate these classes of sources. The full animation can be viewed here: \url{https://hamblin-ku.github.io/ColorAnimation/sed_color_animation.mp4}.
\label{fig:sed_animation}}
\end{figure*}

With the improved mid-IR coverage and sensitivity of JWST, MIRI surveys are also finding populations of mid-IR weak galaxies at cosmic noon. These populations are most likely mid-IR weak due to intrinsically low luminosities, and these sources dominate the MIRI population at $L_{IR} < 10^{10}\,L_{\odot}$. The mid-IR emission of these mid-IR weak galaxies can look very similar to the mid-IR power law of an AGN torus, making it difficult to distinguish between the two. AGN and mid-IR weak galaxies can only be distinguished from comparison of mid-IR emission and rest-frame near-IR emission. In particular, the ability to identify the stellar bump at $\lambda \sim 1.6\,\mu$m and the stellar emission minimum at $\lambda \sim 5\,\mu$m has proven crucial to discriminate between AGN and mid-IR weak galaxies \citep{kirkpatrick_ceers_2023}.

The interplay between spectral features and redshift makes the identification of AGN with mid-IR photometry difficult unless the redshift is known a priori. Observing the hot dust grains of the AGNs torus at restframe wavelengths of $\lambda \approx 3-10\,\mu$m is crucial for identifying AGN and constraining the IR AGN luminosity \citep{Durodola_2024}, since the resulting power-law emission is distinct from mid-IR SFG emission \citep{Franceschini_1991, padovani_2017}. Importantly, the optical depth is low at mid-IR wavelengths so the AGN emission is not strongly diminished by obscuring dust \citep{hickox_araa}. AGN identification cannot reliably be done in the near-IR due to the higher optical depth caused by obscuring dust and contamination from the peak of galaxy stellar emission at near-IR wavelengths. Additionally, systematic redshifting of PAH features in and out of JWST/MIRI bands can cause  SFGs to mimic the rising power-law signature characteristic of AGN \citep{kirkpatrick_ceers_2023}. Figure \ref{fig:sed_animation}, which is derived from a viewable animation,  illustrates this degeneracy in the mid-IR emission of SFGs, AGNs, and mid-IR weak galaxies.

Historically, AGNs have been photometrically identified in the mid-IR with various color selections \citep{lacy_obscured_2004, sajina_simulating_2005, stern_mid-infrared_2005, lacy_optical_2007, donley_identifying_2012, wu_sdss_2012, messias_new_2012, assef_mid-infrared_2013, kirkpatrick_2013, alonso-herrero_mid-infrared_2016, assef_2018}. These color selections make use of the intrinsic spectral differences between AGNs and other sources to divide them into distinct regions of color-space. While mid-IR color selections are capable, their effectiveness is notably limited to galaxies with AGNs that dominate the spectral energy distribution (SED), where $\gtrsim50\%$ of the mid-IR emission is attributable to an AGN \citep{assef_mid-infrared_2013}. AGNs that do not dominate the SED can mix with prominent PAH features to produce a similar spectral shape as a SFG with intrinsically weak PAH features \citep{pope_mid-infrared_2008, sajina_spitzer-_2012, kirkpatrick_goods-herschel_2012}. This degeneracy in spectral shape means that weak PAH features alone are not a robust indicator of the presence of an AGN.

SED-fitting methods (see \citealp{conroy_sed_review, Iyer_2025} for a review) have been utilized as an alternative method for AGN identification \citep{Bovy_2011}. These methods simultaneously model multiple physical components from the UV to the far-IR by generating libraries of theoretical templates and statistically comparing them to observed photometry. While  more comprehensive than simple color cuts, traditional parametric SED fitting suffers from two notable limitations that impede its application to large galaxy surveys. First, the computational demands can be prohibitive. For CIGALE \citep{boquien_cigale}, which has been widely adopted for recent JWST AGN studies \citep{yang_ceers_2023, Perez_2024_dark, Kocevski_2025}, fitting a catalog of order 1000 galaxies can take hours to days on standard hardware, depending on the chosen parameter grid. Second, the results are highly sensitive to the chosen parameter grid or parameter priors, meaning that the SED fitting process is rarely a one-time activity and can require multiple iterations, further increasing the needed computation time \citep{Pacifici_2023}. Non-parametric approaches, such as the dense basis method of \citet{Iyer_2017}, have emerged to address these limitations, offering substantial improvements in computational efficiency ($\sim30$ minutes for 1000 galaxies) and reduced sensitivity to parameter priors. While these advances represent significant progress, there remains value in developing alternative efficient methods, namely for AGN-related tasks, that can provide immediate feedback to identify small sub-populations of interest in large catalogs for follow-up study.

Machine learning based techniques are well-suited for such tasks when speed is a priority. Unlike traditional color-selection methods, machine learning algorithms can simultaneously analyze multiple colors across all available bands, enabling the identification of subtle spectral features that distinguish AGNs from SFGs. This multi-dimensional approach is potentially powerful for breaking degeneracies that arise when PAH features are redshifted between bands, as the models can learn to recognize patterns in how multiple colors change simultaneously. Recently, machine learning algorithms have been successfully applied to color selection tasks and redshift estimation in a variety of wide-field optical and infrared  surveys \citep{bai_machine_2018, Fotopoulou_2018,Duncan_2018, Brescia_2019, holwerda_2021, Saxena_2024,Roster_2024,mechbal_machine_2024, Merz_2025}. The abundance of accurately labeled data and spectroscopic coverage makes wide field surveys natural candidates for the application of machine learning models, but there have been few attempts to apply these models to surveys with sparse spectroscopic coverage and smaller sample sizes (N${\sim}100\text{--}1000$). This lack of spectroscopic coverage and overall limited amounts of data poses a challenge to the application of machine learning algorithms, which are generally prone to over-fitting and poor generalization on small datasets \citep{hastie01statisticallearning}. 

In this paper, we explore the potential of machine learning models to supplant traditional color selection when applied to small datasets (N${\sim}100\text{--}1000$) with our machine learning model {\tt AGNBoost}, based on the {\tt XGBoostLSS} algorithm \citep{marz_xgboostlss}. {\tt AGNBoost} is trained on the  JWST/  NIRCam+MIRI photometric observations and derived quantities of a sample of mock galaxies from {\tt CIGALE} \citep{boquien_cigale}, and we test it against both the mock galaxies and a set of real JWST/NIRCam+MIRI observations from the MIRI EGS Galaxy and AGN (MEGA) survey \citep{backhaus_2025}. In Section \ref{data} we describe our training sample construction and the MEGA observations. Section \ref{methodology} details the {\tt AGNBoost} methodology and implementation. Section \ref{results} presents our model performance and assessment. Section \ref{summary} summarizes our conclusions and outlines future applications.

\section{Data and Sample Selection} \label{data}

\subsection{{\tt CIGALE} Training Sample} \label{subsec:CIGALE}
The foundation of {\tt AGNBoost}'s training relies on a comprehensive set of mock galaxies generated using the Code Investigating GALaxy Emission ({\tt CIGALE}; \citealp{boquien_cigale}). We use {\tt CIGALE} v2022.1 in the `save\_fluxes' mode to generate $\approx 10^9$ unique galaxy templates using {\tt CIGALE}'s physical models, which include both stellar and AGN emission components. The total adopted parameters are summarized in Table \ref{table:cigale_params}.

\begin{deluxetable*}{llll}[t!]
\tablecaption{{\tt CIGALE} Simulation Parameters\label{table:cigale_params}}
\tablehead{
\colhead{Module} & \colhead{Parameter} & \colhead{Symbol} & \colhead{Values}
}
\startdata
Star formation history & Stellar e-folding time & $\tau_\mathrm{star}$ & 0.1, 0.5, 1, 5 Gyr \\
\texttt{sfhdelayed} & Stellar age & $t_\mathrm{star}$ & 0.5, 1, 3, 5, 7 Gyr \\
& Burst mass fraction & $f_\mathrm{burst}$ & 0.0, 0.02, 0.1 \\
\hline
Simple stellar population & Initial mass function & -- & Chabrier (2003) \\
\texttt{bc03} & Metallicity & $Z$ & 0.02 \\
\hline
Nebular emission & Ionization parameter & $\log U$ & $-2.0$ \\
\texttt{nebular} & Gas metallicity & $Z_\mathrm{gas}$ & 0.02 \\
\hline
Dust attenuation & Color excess of nebular lines & $E(B-V)_\mathrm{line}$ & 0.0, 0.1, 0.2, 0.3, 0.4, 0.5 \\
\texttt{dustatt\_modified\_starburst} & & & 0.6, 0.7, 0.8, 0.9, 1.0 \\
& ratio between line and continuum $E(B-V)$ & $\frac{E(B-V)_\mathrm{line}}{E(B-V)_\mathrm{cont}}$ & 1 \\
\hline
& PAH mass fraction & $q_\mathrm{PAH}$ & 1.77, 3.19, 5.26 \\
Galactic dust emission & Minimum radiation field & $U_\mathrm{min}$ & 0.1, 1.0, 10 \\
\texttt{dl2014}& Fraction of PDR emission & $\gamma$ & 0.01, 0.1, 0.2 \\
\hline
AGN (UV-to-IR) emission & Viewing angle & $\theta_\mathrm{AGN}$ & 70$^\circ$ \\
\texttt{skirtor2016} & AGN contribution to IR luminosity & frac$_\mathrm{AGN}$ & 0.1--0.9 (step 0.1), 0.99 \\
& Wavelength range where frac$_\mathrm{AGN}$ is defined & $\lambda_\mathrm{AGN}$ & 3--30 $\mu$m \\
& Extinction in polar direction & $E(B-V)_\mathrm{polar}$ & 0.03, 0.1 \\
& Polar dust temperature & T$_\mathrm{dust}$ & 100, 500, 900, 1300 K \\
\hline
Redshift+IGM & Source redshift & $z$ & 0.01--8.0 (100 steps) \\
\texttt{redshifting} & & &  \\
\enddata
\tablecomments{For parameters not listed here, we adopt default values.}
\end{deluxetable*}

For star formation, we adopt a standard delayed star formation history (SFH; {\tt sfhdelayed} in {\tt CIGALE}), with $e$-folding times ranging from $0.1\text{--}5$\,Gyr and stellar ages ranging from $0.5\text{--}5$\,Gyr. We also allow for starbursts with an age of $20$\,Myr and a mass fraction of $f_{\text{burst}} = \text{0.0, 0.02, 0.1}$. For the stellar component, we adopt the \citet{bruzual_stellar_2003} stellar population models (module {\tt bc03} in {\tt CIGALE}) with a Chabrier initial mass function \citep{chabrier_2003}. Dust attenuation is modeled using the {\tt dustatt\_modified\_starburst} module which is based on the attenuation curve of \citet{calzetti_dust_2000} but extended to short wavelengths ($91.2\text{--}150$\,nm) with \citet{leitherer_global_2002}. The allowed range of color excess is $E(B\text{--}V)=0\text{--}1$. Galactic dust emission is modeled with the {\tt dl2014} module \citep{draine_andromedas_2014}, which models dust emission with a diffused emission component and a photodissociating component associated with star formation. We use a PAH mass fraction ($q_\text{PAH}$) varying from $q_\text{PAH} = 1.77\text{--}5.26$, a minimum radiation parameter of diffuse dust ($U_\text{min}$) ranging from $U_\text{min}=1\text{--}10$, and relative strengths of diffuse and star forming components ($\gamma$) of $\gamma=0.01\text{--}0.2$. These parameter ranges were chosen to envelop reasonable values expected for MEGA galaxies, and the gridding was chosen to permit {\tt CIGALE} fitting of the MEGA catalog in a reasonable timeframe (${\sim}$1 day on a standard laptop). 

AGN emission is modeled using the {\tt SKIRTOR} clumpy torus model, based on the clumpy torus models from \citet{stalevski_3d_2012,stalevski_dust_2016}. In the {\tt SKIRTOR} module, the relative strength of AGN in the IR is set by the frac$_{\text{AGN}}$ parameter, which corresponds to the fraction of $3\text{--}30\,\mu$m mid-IR light attributable to an AGN power law. We let frac$_{\text{AGN}}$ range from frac$_\text{AGN}=0.0\text{--}1$ in steps of $0.1$. We set the viewing angle to $70^\circ$, a typical value for type 2 AGN, since we expect very few type 1 AGNs in small-area JWST surveys \citep{yang_ceers_2023, kirkpatrick_ceers_2023}. We also vary the AGN polar dust temperature, allowing T$_{\text{dust}}=\text{100, 500, 900, 1300}$\,K. The remaining {\tt SKIRTOR} parameters generally have minor effects on SED shapes \citep{yang_2020}, so we adopt default values to reduce the necessary computation time. Finally, we choose a redshift grid evenly spaced in $\log\left(1+z\right)$ with 100 steps from $z=0.01\text{--}8.0$. We adopt this redshift range to fully cover the spread of MEGA redshifts. In total, this {\tt CIGALE} parameter space created $\approx 10^9$ galaxies, which we uniformly sample the mock set down to N$=10^6$ due to the computational limitations of training machine learning models on large datasets\footnote{We have made this mock {\tt CIGALE} catalog publicly available on the Harvard Dataverse \citep{dataverse_mockcatalog} }.

\begin{figure*}[t!]
\plotone{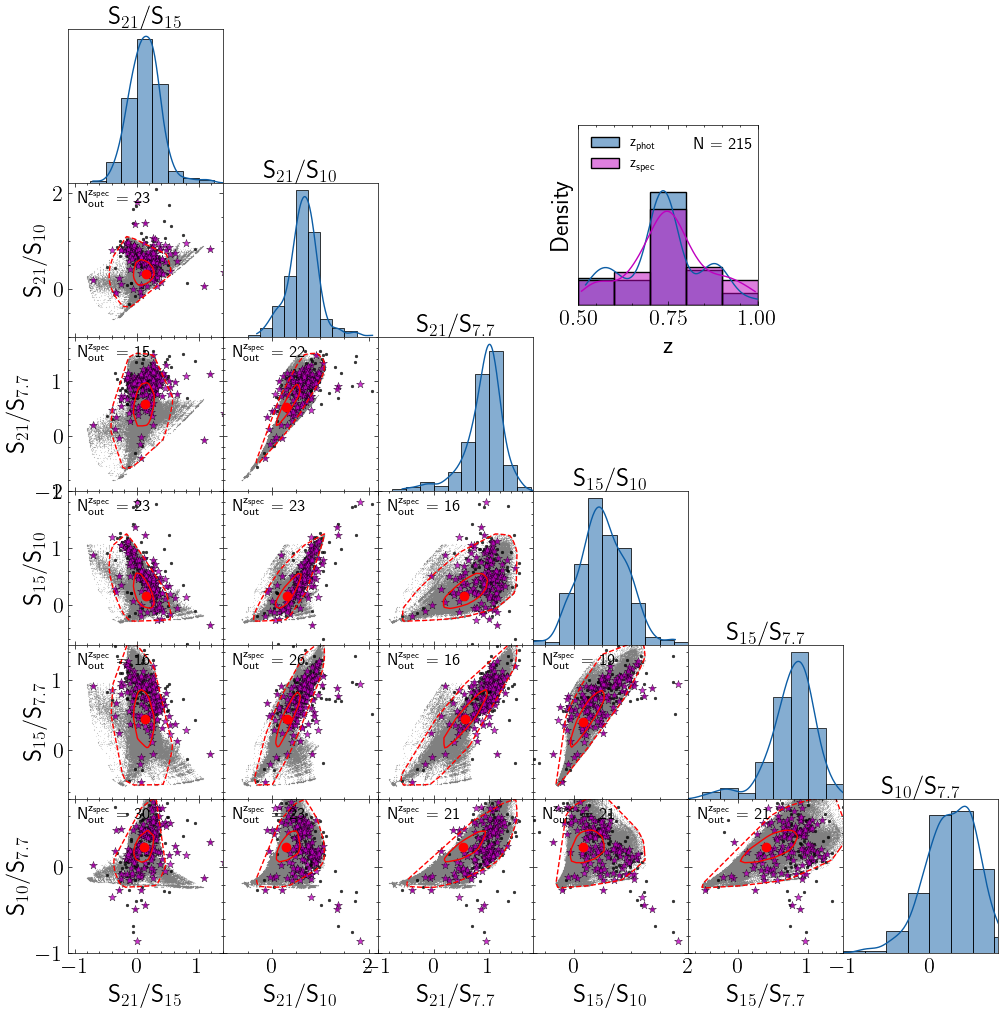}
\caption{JWST/MIRI color comparison between mock {\tt CIGALE} galaxies and MEGA observations in the $0.5 \leq z < 1.0$ redshift bin. MEGA sources with spectroscopic redshifts are shown as purple stars, those with photometric redshifts as black points, and individual {\tt CIGALE} galaxies as gray points. Red regions show bagplots (the bivariate analog of boxplots; \citealp{Roussee_bagplot}), where the filled area contains $50\%$ of the {\tt CIGALE} data, the central point marks the depth median, and the dashed line indicates the outlier boundary. The MEGA galaxy colors show good agreement with the {\tt CIGALE} mock galaxy color distribution, validating the representativeness of our training data. AGNBoost performance for MEGA galaxies lying outside the {\tt CIGALE} color space is examined in Figure \ref{fig:outside_of_colorspace} in Appendix \ref{appendix:colorspace_effects}.
\label{fig:color_compare}}
\end{figure*}

\begin{figure*}[t!]
\plotone{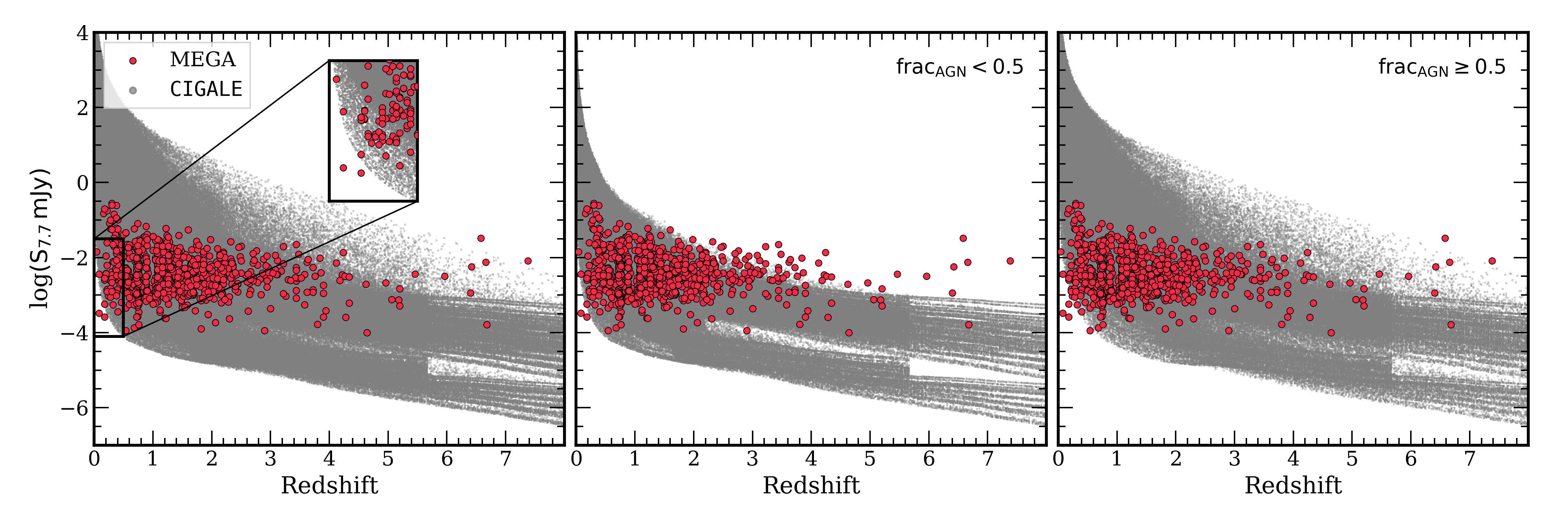}
\caption{JWST/MIRI F770W photometry distributions comparing mock {\tt CIGALE} galaxies (gray) and MEGA observations (red). The left panel shows the full {\tt CIGALE} mock set, the middle panel shows {\tt CIGALE} non-AGN (frac$_\text{AGN} < 0.5$), and the right panel shows {\tt CIGALE} AGN (frac$_\text{AGN} \geq 0.5$). The same MEGA sample is plotted in all three panels for comparison. The {\tt CIGALE} simulations provide excellent coverage of the MEGA F770W photometry, with only 6 ($\sim0.9\%$) very low-redshift sources ($z\sim0.01$) falling outside the simulated range.
\label{fig:phot_compare}}
\end{figure*}

To validate the mock set of {\tt CIGALE} galaxies generated from the above {\tt CIGALE} parameter space, we compare the colors and photometric observations of the mock galaxies to those of galaxies observed in MEGA. Figure \ref{fig:color_compare} demonstrates that the mock galaxies reproduce the observed color distributions in the $0.5 \leq z < 1.0$ bin. We verified that this good color agreement holds across all redshift bins out to $z=8$. In Appendix \ref{appendix:colorspace_effects}, we investigate the {\tt AGNBoost} performance of MEGA galaxies that lie outside the {\tt CIGALE} color space. We also directly compare the photometric observations of the {\tt CIGALE} simulations to MEGA in Figure \ref{fig:phot_compare}. The {\tt CIGALE} simulations fully cover the range of MEGA photometric observations, but with different distribution shapes. In our testing, we found that sampling the distributions of mock photometry to match that of the real MEGA galaxies had no major effect on {\tt AGNBoost} predictive performance.

The complete mock sample is divided following standard machine learning practices \citep{hastie01statisticallearning}:  60$\%$ for training,  20$\%$ for model validation during model tuning, 20$\%$ for final testing. This division ensures sufficient statistics in each subset while maintaining independence between training and evaluation. Each subset is carefully constructed to maintain the same distribution of physical parameters (frac$_{\text{AGN}}$, redshift) as the full sample.

\subsection{MEGA Observations}\label{subsec:mega_data}
We test the performance of our models on data from the MIRI EGS Galaxy and AGN (MEGA) survey (PI A. Kirkpatrick; \citealp{backhaus_2025}), a four-band MIRI survey covering 25 pointings in the Extended Groth Strip (EGS) field. Three pointings used only the three reddest filters (F1000W, F1500W, F2100W), while the remainder also included the blue filter (F770W). MEGA builds upon existing coverage of the EGS field \citep{EGS_2007}, providing MIRI observations for $68.9\%$ of the NIRCam area from the Cosmic Evolution Early Release Science Survey (CEERS; \citealp{finkelstein_cosmic_2025}). All MEGA data used in this paper can be found in MAST:\dataset[10.17909/9nab-af49]{https://doi.org/10.17909/9nab-af49}.

We use the most recent MEGA reductions from \citet{backhaus_2025}, which adopt a detection threshold of SNR~$\geq$~2.2 over 4 pixels. Following \citet{backhaus_2025}, we remove sources with false positive detections from extended features of bright galaxies, stars, or noise at map edges, and apply an SNR~$>$~1 cut in the bluest available MIRI band (F770W or F1000W). This results in a final MEGA catalog of 6450 sources. 

To obtain NIRCam photometry, we use photometric catalogs from the UNICORN project (Finkelstein et al. in preparation). These catalogs follow the photometric procedures of \citet{Finkelstein_2024}, to which we refer the reader for more information. In brief, the catalogs are based on Source Extractor \citep{SourceExtractor} using F277W+F356W as the detection image, and provide robust estimates of colors, total fluxes, and photometric uncertainties. Photometric redshifts are estimated using {\tt Lazy}\footnote{\url{https://github.com/hollisakins/Lazy.jl}} \citep{lazy_jl}, a Julia-based version of the template-fitting code {\tt EAZY} \citep{brammer_eazy}. We remove all UNICORN sources flagged for poor photometric redshift fits and those z~$\geq$~14 to ensure reliable photometric redshifts, resulting in 105,135 UNICORN sources available for matching.

We cross-match the MEGA and UNICORN catalogs by matching to the nearest object within $0.5\arcsec$, yielding 3792 matches. We then apply an SNR~$>$~3 cut in F770W+F1500W+F2100W, resulting in 853 sources. Of these, 105 are missing at least one photometric observation; we exclude these from our analysis, though in Sections \ref{methodology:imputation} and \ref{results:imputaton} we test the ability to reliably characterize sources with missing photometry using statistical imputation. We also cross-match within $0.5\arcsec$ to the CANDELS \citep{candels_2011_anton, candels_2011} EGS redshift catalog of \citet{Kodra_2023} to obtain spectroscopic redshifts where available. Our final sample comprises 748 MEGA galaxies with complete photometry and photometric redshifts, 288 of which have spectroscopic redshifts. Since the MEGA sources lack existing $\text{frac}_{\text{AGN}}$ estimates, in Section \ref{subsec:mega_results} we use {\tt CIGALE} to fit these sources with the parameters from Table \ref{table:cigale_params}.

\section{Methodology} \label{methodology}

\subsection{{\tt XGBoostLSS} Framework} \label{sec:xgboostlss}

The framework of {\tt AGNBoost} is based on the machine learning algorithm  {\tt XGBoostLSS} \citep{marz_xgboostlss}, which extends {\tt XGBoost} (eXtreme Gradient Boosting; \citealp{chen_xgboost}), an ensemble boosted decision tree algorithm, to probabilistic forecasting. 

Decision trees are a non-parametric method with a hierarchical tree-like structure that learn simple decision rules from data. Broadly, ensemble models are machine learning methods that combine the predictions of multiple base learners, such as decision trees, in some way (e.g., averaging, voting, stacking, etc.) to achieve a more effective overall model \citep{Zhou_2012, Ganaie_2021}. The generalization ability (i.e., the model's ability to perform on unseen data) of the overall ensemble is generally better than that of any of the base learners, and ensemble methods are less prone to falling into local optima \citep{Dietterrich_2000}. Boosting methods, such as {\tt XGBoost}, are a class of ensemble techniques that aim to reduce model bias by combining sequentially trained base learners into one powerful learner. 

While {\tt XGBoost} is a powerful algorithm, it is statistically limited to modeling only the conditional mean $\mathbb{E}(Y|X=x)$ and treats higher moments of the conditional distribution as fixed nuisance parameters \citep{marz_xgboostlss}. These assumptions are only valid when data is symmetrically Gaussian with a constant variance, but real observations are not generally so well behaved and exhibit characteristics of variable higher moments (e.g., heteroskedasticity, skewness, or kurtosis). {\tt XGBoostLSS} bridges this statistical modeling gap by connecting {\tt XGBoost} to Generalized Additive Models for Location Scale and Shape (GAMLSS; \citealp{stasinopoulos_generalized_2007}) to predict the entire conditional distribution $F_Y(y\vert x)$.

{\tt XGBoostLSS} has a few built-in features that make it appealing for application to astronomical data. There is built-in support for both L1 (Lasso regression) and L2 (Ridge regression) regularization to avoid model over-fitting to spurious features or noise. It also has the ability to handle missing data through a technique known as sparsity-aware split finding, in which a default direction is assigned at every branch in a tree that results in the lowest corresponding loss. 

\begin{deluxetable*}{ccccc}[t!]
\tablecaption{{\tt AGNBoost} Hyperparameters\label{table:xgboost_params}}
\tablehead{
\colhead{Hyperparameter} & \colhead{Description} & \colhead{Range Explored} & \colhead{Model} & \colhead{Optimal Value}
}
\startdata
\multirow{2}{*}{max\_depth} & \multirow{2}{*}{Max depth of trees} & \multirow{2}{*}{[3,10]} & frac$_\mathrm{AGN}$ & 8 \\
& & & $z$ & 10 \\
\hline
\multirow{2}{*}{gamma} & Min loss reduction to further & \multirow{2}{*}{[10$^{-8}$, 40]} & frac$_\mathrm{AGN}$ & 1.87e-02 \\
& partition leaf node & & $z$ & 1.28e-1 \\
\hline
\multirow{2}{*}{min\_child\_weight} & Min sum of hessian needed & \multirow{2}{*}{[1,250]} & frac$_\mathrm{AGN}$ & 80 \\
& to partition & & $z$ & 21 \\
\hline
\multirow{2}{*}{lambda} & \multirow{2}{*}{L2 regularization on weights} & \multirow{2}{*}{[1,150]} & frac$_\mathrm{AGN}$ & 3.98 \\
& & & $z$ & 76.9 \\
\hline
\multirow{2}{*}{alpha} & \multirow{2}{*}{L1 regularization on weights} & \multirow{2}{*}{[10$^{-3}$,100]} & frac$_\mathrm{AGN}$ & 1.47e-4 \\
& & & $z$ & 1.85e-2 \\
\hline
\multirow{3}{*}{subsample} & Subsample ratio of training & \multirow{3}{*}{[0.7, 1.0]} & frac$_\mathrm{AGN}$ & 0.784 \\
& instances per boosting & & $z$ & 0.948 \\
& iteration & & & \\
\hline
\multirow{2}{*}{tree\_method} & Tree construction algorithm & \multirow{2}{*}{[`hist',`approx']} & frac$_\mathrm{AGN}$ & `hist' \\
& in XGBoost & & $z$ & `hist' \\
\hline
\multirow{3}{*}{stabilization} & Stabilization of gradients and & \multirow{3}{*}{[`none',`L2', `MAD']} & frac$_\mathrm{AGN}$ & `MAD' \\
& Hessians to improve model & & $z$ & `L2'\\
& convergence & & & \\
\hline
\multirow{2}{*}{Response\_fn} & Transformation function of & \multirow{2}{*}{[`softplus', `exp']} & frac$_\mathrm{AGN}$ & `softplus' \\
& distribution parameters & & $z$ & `softplus'\\
\enddata
\tablecomments{For {\tt XGBoostLSS}/{\tt XGBoost} hyperparameters not listed here, we use default values.}
\end{deluxetable*}
\vspace{-2\baselineskip}

We train separate {\tt XGBoostLSS} models for estimation of frac$_{\text{AGN}}$ and photometric redshift. Joint training of both quantities in a single model is not possible because {\tt XGBoostLSS} only supports multivariate regression for normally distributed variables, which neither frac$_{\text{AGN}}$ nor photometric redshift follow. Since frac$_{\text{AGN}}$ is intrinsically bounded $[0,1)$, we choose to estimate a zero-inflated beta distribution for frac$_{\text{AGN}}$. The beta distribution, commonly used in Bayesian inference, is particularly flexible at modeling parameters bounded on the interval (0,1), but is not defined at zero. Per the {\tt CIGALE} modeling performed in this work, galaxies are allowed frac$_{\text{AGN}} = 0 $ (i.e., entirely SFG), which the  zero-inflated beta distribution accounts for by adding a Bernoulli distribution at frac$_{\text{AGN}} = 0$. Conversely, redshift is theoretically bounded $[0,\inf)$, but since this work is focused on a large redshift range ($0.01< z < 8$), we choose to transform redshift to the range (0,1) with a modified sigmoid transformation ($\Phi(z) = 2/(1 + e^{-az}) - 1$; $a = 0.4$) so that we can model the transformed redshift with a beta distribution. This transformation provides variable compression of redshift onto the (0,1) interval, controlled by the parameter $a$ which was chosen to create better resolution in the intermediate redshift range ($2<z<5$), where PAH features redshift out of the NIRCam+MIRI band coverage, making redshift estimation more challenging due to the loss of key spectral diagnostics. Compared to the scale factor transformation ($a(z)=1/(1 + z)$), this modified sigmoid transformation better spreads redshifts over the unit interval.

Each model shares the same 66 inputs: 7 NIRCam bands (F115W, F150W, F200W, F277W, F356W, F410M, F444W) + 4 MIRI bands (F770W, F1000W, F1500W, F2100W), and 55 derived colors from all unique filter combinations to avoid using unnecessary reciprocal colors. We package our trained{\tt XGBoostLSS} models into a repository of publicly available Python code, {\tt AGNBoost}\footnote{\url{https://github.com/hamblin-ku/AGNBoost}}.

\begin{figure*}[t!]
\centering
\plotone{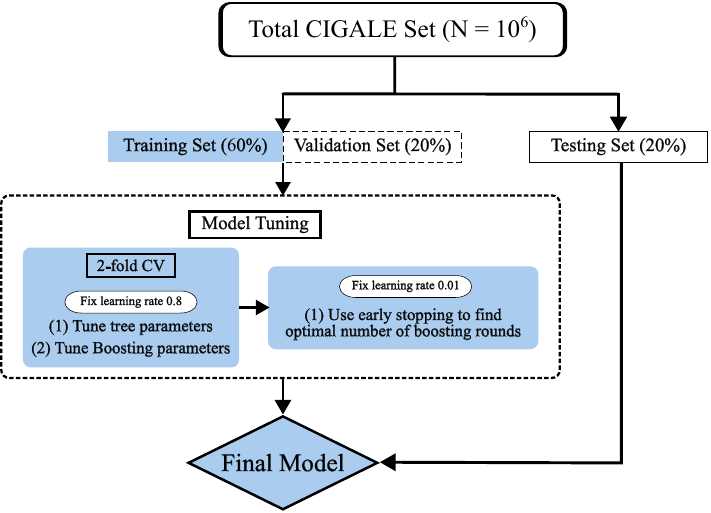}
\caption{Flowchart illustrating the {\tt AGNBoost} training procedure. The multi-stage approach optimizes computational efficiency by first performing coarse hyperparameter tuning, then fine-tuning the number of boosting rounds. In the first stage of tuning, the training and validation sets are combined for 2-fold cross-validation (CV) to identify optimal hyperparameters. In the second stage of tuning, these hyperparameters are used to train models on the training set alone, with the validation set used for early stopping to determine the optimal number of boosting rounds (i.e., when validation loss begins to increase). The final  model is trained using these optimized parameters on the combined training and validation datasets to maximize the use of available training data.}
\label{fig:AGNboost_flowchart}
\end{figure*}

\subsection{ Model Training and Optimization}\label{methodology:model_tuning}
Due to the large size of our training set, a methodical approach was necessary in order to tune {\tt AGNBoost}'s models in a reasonable amount of time. Broadly, to tune the configurable variables of our models, known as hyperparameters, we employ a multi-stage search procedure, where in the first stage we separately tune tree parameters (the last three rows of Table \ref{table:xgboost_params})  and boosting parameters (the remaining rows of Table \ref{table:xgboost_params}), and in the final stage we find the optimal number of boosting iterations. The first stage of tuning is performed with 2-fold cross-validation and a high learning rate ($0.8$) on the combination of the training and validation sets, and the second tuning stage is performed with a low training rate ($0.01$) on just the training set with the validation set used to stop the training procedure when the loss on the validation set increases (known as early stopping). The primary benefit of this approach is that the high learning rate of the first two stages allows fast model tuning, while the final stage can use these now confidently identified parameters to find the optimal boosting duration. This process is outlined  Figure \ref{fig:AGNboost_flowchart}.

For each stage of hyperparameter tuning, we adopt a Bayesian approach with {\tt Optuna} \citep{optuna}, an open source hyperparameter optimization framework. Unlike traditional grid search methods, which try all possible combinations within a given hyperparameter space, Bayesian hyperparameter tuning efficiently explores the hyperparameter space by continuously updating the parameter probability density functions based on the results of all trials.

\subsection{Uncertainty Quantification}

{\tt XGBoostLSS} provides two primary means of quantifying uncertainties of estimates. The conditional distributions modeled by {\tt XGBoostLSS} capture the uncertainty that arises due to the inherent and unpredictable randomness of the data, known as the aleatoric uncertainty \citep{hullermeier_aleatoric_2021}. The epistemic uncertainty, or the uncertainty due to a lack of model knowledge, is estimated with the virtual ensemble method of \citet{malinin_uncertainty}. In brief, a virtual ensemble is created from $M$ number of sub-models within each trained {\tt XGBoostLSS} model and each sub-model is used to make predictions. The variance of these $M$ observations is used as a robust estimate of the epistemic uncertainty.  Notably, virtual ensembles are obtained from \textit{one} already trained model, while creating a true ensemble of $N$ {\tt XGBoostLSS} models would require $N$ times the computational cost, so this approach can be considered computationally ``free.''

\begin{figure*}[t!]
\plottwo{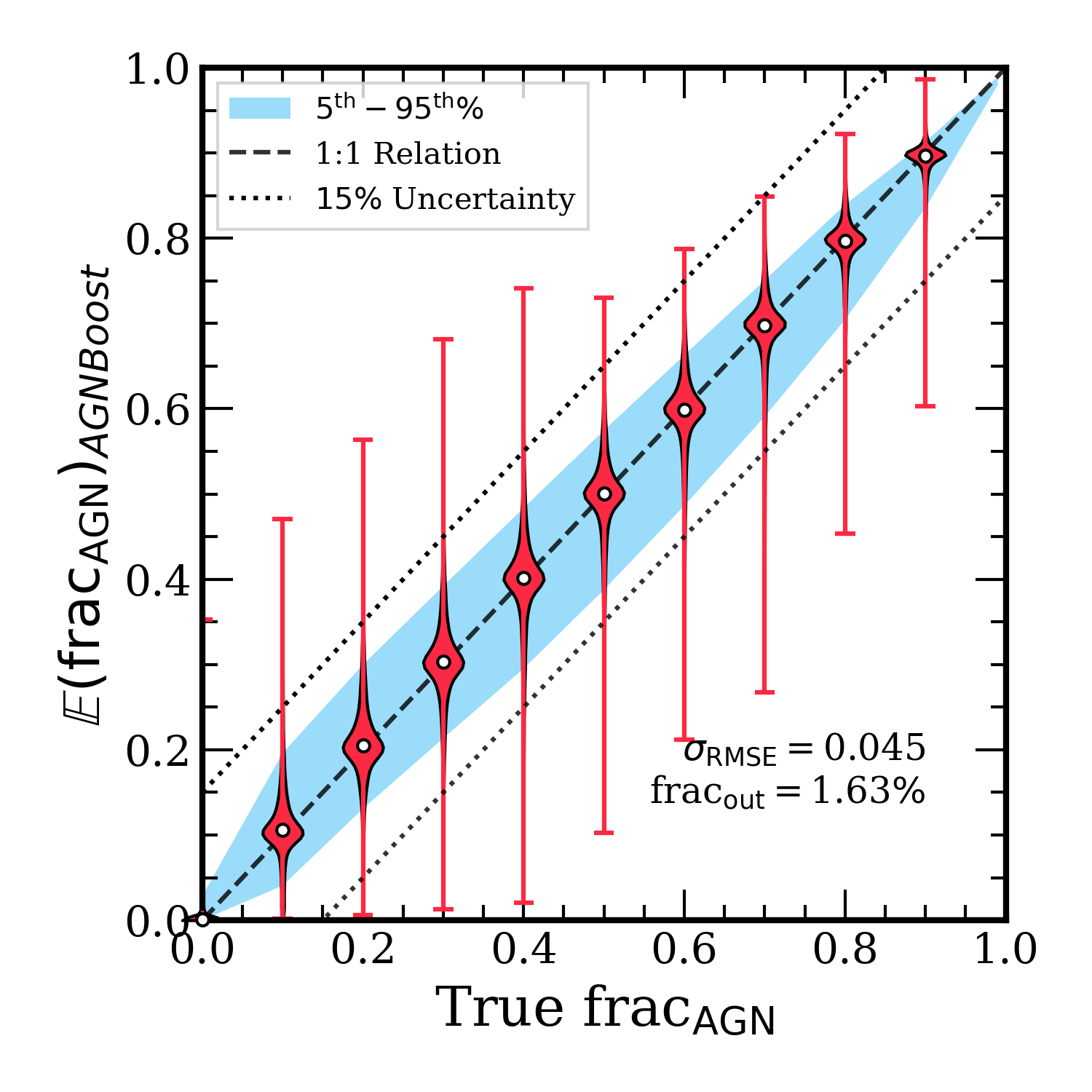}{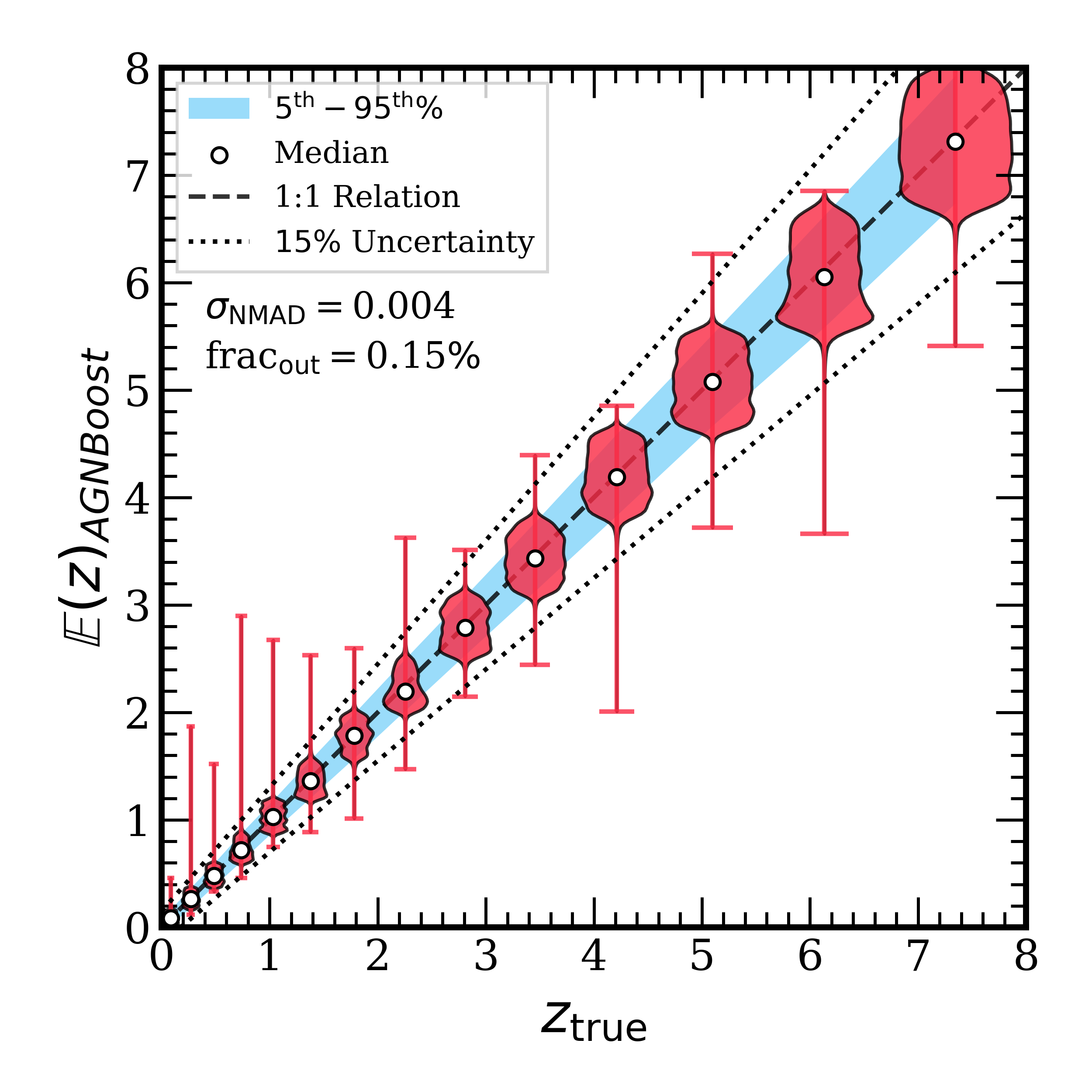}
\caption{Expectation values of {\tt AGNBoost} predictions on the test set of mock {\tt CIGALE} data, for frac$_\text{AGN}$ (left) and redshift (right) models. The violin plots show the distribution of predicted expectation values at each bin of true value, with white dots marking medians, and whiskers extending to the extrema. The light blue shading represents the 5th-95th percentile interval, and the black dashed and dotted lines correspond to the 1:1 relation and $15\%$ uncertainty regions, respectively. Both models demonstrate excellent performance with predictions tightly centered on the 1:1 relation and sub-$1\%$ outlier fractions.
\label{fig:sim_perf}}
\end{figure*}

We also derive prediction uncertainties due to source photometric uncertainty. For each source, we create 100 mock sources with Monte Carlo, by considering the measured fluxes and associated flux errors as the mean and standard deviation of a normal flux distribution. With these 100 mock sources we create 100 point estimates from each model, corresponding to the peak of their beta distributions, and use the standard deviation of these point estimates to quantify an uncertainty due to photometric uncertainty for every source in each {\tt AGNBoost} model. The total uncertainty for each source is derived by summing these three  measures in quadrature.

\subsection{Missing Photometry Imputation}\label{methodology:imputation}

It is common for observed MIRI sources to be missing detections in one or more bands, due to the way that bright PAH and absorption features shift through the MIRI bands. While {\tt AGNBoost} is able to make predictions for sources with missing photometric observations, we also test whether replacing these missing values with statistical imputation can improve the model performance. We implement and test a Generative Adversarial Network (GAN) for photometry imputation, using the {\tt SGAIN} algorithm of \citet{SGAIN_paper}\footnote{\url{https://github.com/dtneves/ICCS_2021}}. {\tt SGAIN} benefits from increased stability and performance over comparative GAN-based methods. {\tt SGAIN} consists of a generator and a discriminator, where the generator's goal is to accurately impute missing data and the discriminator's goal is to distinguish between observed and imputed data. This adversarial training process ensures that the generator's imputation performance increases over training. 

We integrate {\tt SGAIN} into a separate module of {\tt AGNBoost}, allowing users to impute missing photometry from their catalogs before performing data analysis. We use cross-validation to optimize the mini-batch size, the $\alpha$ parameter, and the optimal number of iterations of {\tt SGAIN}. Following standard practices, photometric observations are standardized to the range (0,1) to avoid problems due to scale. The {\tt SGAIN} imputation process is performed 100 times. To account for imputation uncertainty due to photometric uncertainty, at each iteration the existing flux measurements are randomly sampled  from a normal distribution of mean equal to the flux value and standard deviation equal to the flux error. The final imputed value is calculated as the average of these imputation trials, and its error is given by the standard deviation of these trials.
\section{ Results and Discussion } \label{results}

Before presenting results, we clarify two important aspects of {\tt AGNBoost}'s methodology. First, we train separate {\tt XGBoostLSS} models for $\text{frac}_{\text{AGN}}$ and redshift estimation and do not jointly predict both parameters simultaneously. Second, redshift is not provided as an input feature when estimating $\text{frac}_{\text{AGN}}$. Both models use only the 66 photometric features derived from NIRCam and MIRI observations described in Section \ref{sec:xgboostlss}.

\begin{figure*}[t!]
\plottwo{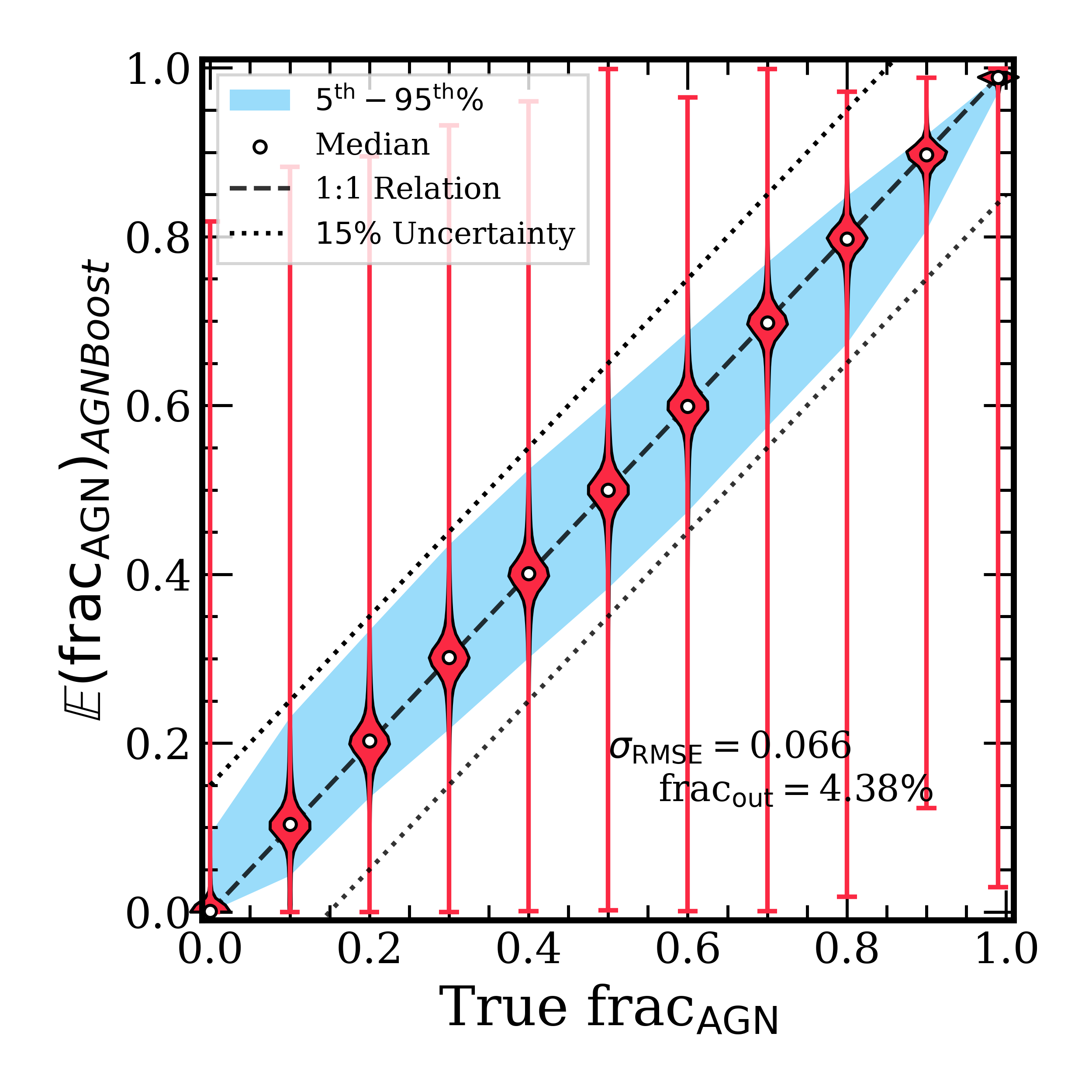}{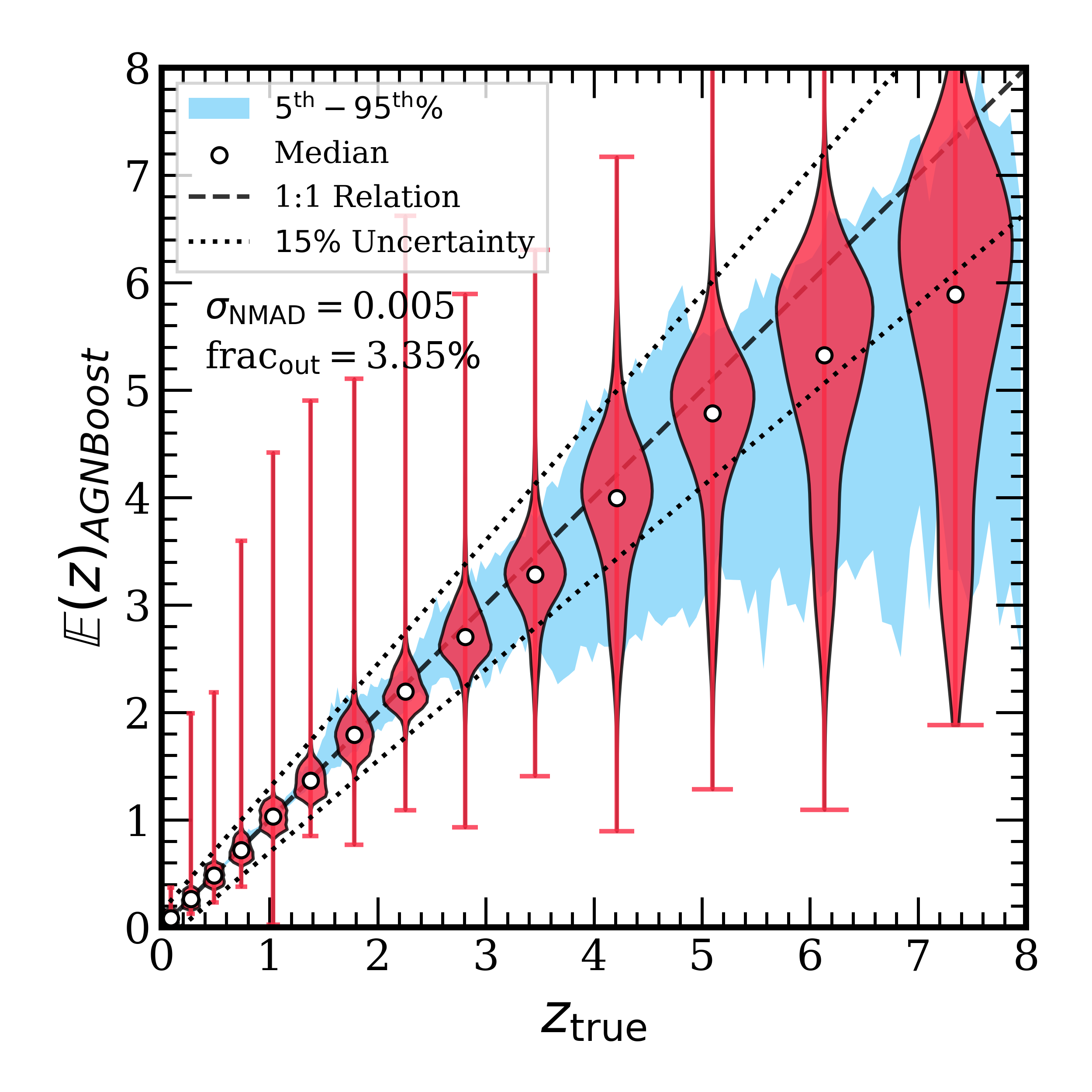}
\caption{Expectation values of {\tt AGNBoost} predictions on the test set of mock CIGALE data with realistic photometric uncertainty added, for frac$_\text{AGN}$ (left) and redshift (right) models. The Violin plot elements are the same as in Figure \ref{fig:sim_perf}. For frac$_\text{AGN}$ estimation, introducing photometric uncertainty opens up the outlier ranges considerably, but the median 1:1 relation performance is maintained.  Photometric redshift estimation degrades significantly at $z > 4$, where critical spectral features shift outside JWST band coverage; with fewer diagnostic features available to anchor the redshift estimate, photometric uncertainties have a more pronounced impact on model performance.
\label{fig:sim_perf_witherr}}
\end{figure*}

\subsection{Performance on Mock {\tt CIGALE} Data}

The results in Figure \ref{fig:sim_perf} show the model performance on the {\tt CIGALE} test set ($20\%$ of the total sample). To evaluate the performance of each model, we calculate the root mean square error (RMSE; $\sigma_{\text{RMSE}} = \sqrt{\text{mean}\{ (x_{\text{pred}} - x_{\text{true}})^2 \}}$), and the outlier fraction (frac$_{\text{out}}$). For the frac$_{\text{AGN}}$ model, we define frac$_{\text{out}}$ as $\left|\Delta\text{frac}_{\text{AGN}}\right| > 0.15$, and for the redshift model we define the frac$_{\text{out}}$ as $\left|\Delta z\right| / \left(1 + z_{\text{true}} \right) > 0.15$. Across all values of frac$_{\text{AGN}}$, only $\sim1.63\%$ of the test sample lie outside the $15\%$ uncertainty region, and  we find the RMSE to be $\sigma_{\text{RMSE}} = 0.045$. The redshift model shows similar performance, with a tight spread around the 1:1 relation out to $z=8$ and a normalized median absolute deviation (NMAD; $\sigma_{\text{NMAD}} = 1.48 \times \text{median}\{ \vert \Delta z - \text{median}(\Delta z)\vert / (1 + z_{\text{spec}} ) \}$)  of $\sigma_{\text{NMAD}} = 0.004$. Only $\sim0.15\%$ of mock sources lie outside the drawn $15\%$ uncertainty region.

We determine how sensitive {\tt AGNBoost} is to photometric uncertainty using MEGA observations. We first determine the correlation between SNR and flux from the MEGA photometric catalog. For each mock photometric point, we then assign an appropriate uncertainty based on flux. Finally, we use Monte Carlo to perturb the mock flux given the uncertainty following a normal distribution. When mock fluxes fall below or above the distribution of MEGA fluxes, we extrapolate the fitted linear relation in log-log space. We then apply the same selection criteria used for the creation of our MEGA set: S/N $>3$ in F770W, F1500W, and F2100W. Figure \ref{fig:sim_perf_witherr} shows the resulting {\tt AGNBoost} predictive performance with the added photometric uncertainties. 

For frac$_{\text{AGN}}$ estimation, {\tt AGNBoost} maintains median predictions on the 1:1 relation with $\sigma_{\text{RMSE}} = 0.066$, and a $5^{th}-95^{th}\%$ spread within the $15\%$ uncertainty region. However, the $15\%$ outlier fraction has increased to $\text{frac}_\text{out} = 4.38\%$, and these outliers show larger deviations from the 1:1 relation than in the noise-free case, a reflection of the degeneracy inherent to photometric-based AGN selection methods.

For photometric redshift estimation, the median predictions follow the 1:1 relation for $z\leq5$ and fall below the $15\%$ uncertainty region at $z>7$. The $15\%$ outlier fraction has increased to $3.35\%$, but the NMAD has only increased to $\sigma_{\text{NMAD}} = 0.004$. This is because the majority of mock sources have $z<2$, where the predicted photometric redshift spread is low. Photometric redshift estimation degrades significantly at $z>4$ when photometric uncertainty is introduced, which we attribute to the shifting of critical dust spectral features outside of the coverage of JWST/NIRCam+MIRI. With fewer diagnostic features available to anchor the redshift estimate, photometric uncertainties have a more pronounced impact on model performance.

\subsection{Performance on Independent Mock Data}

\begin{figure*}[t!]
\plottwo{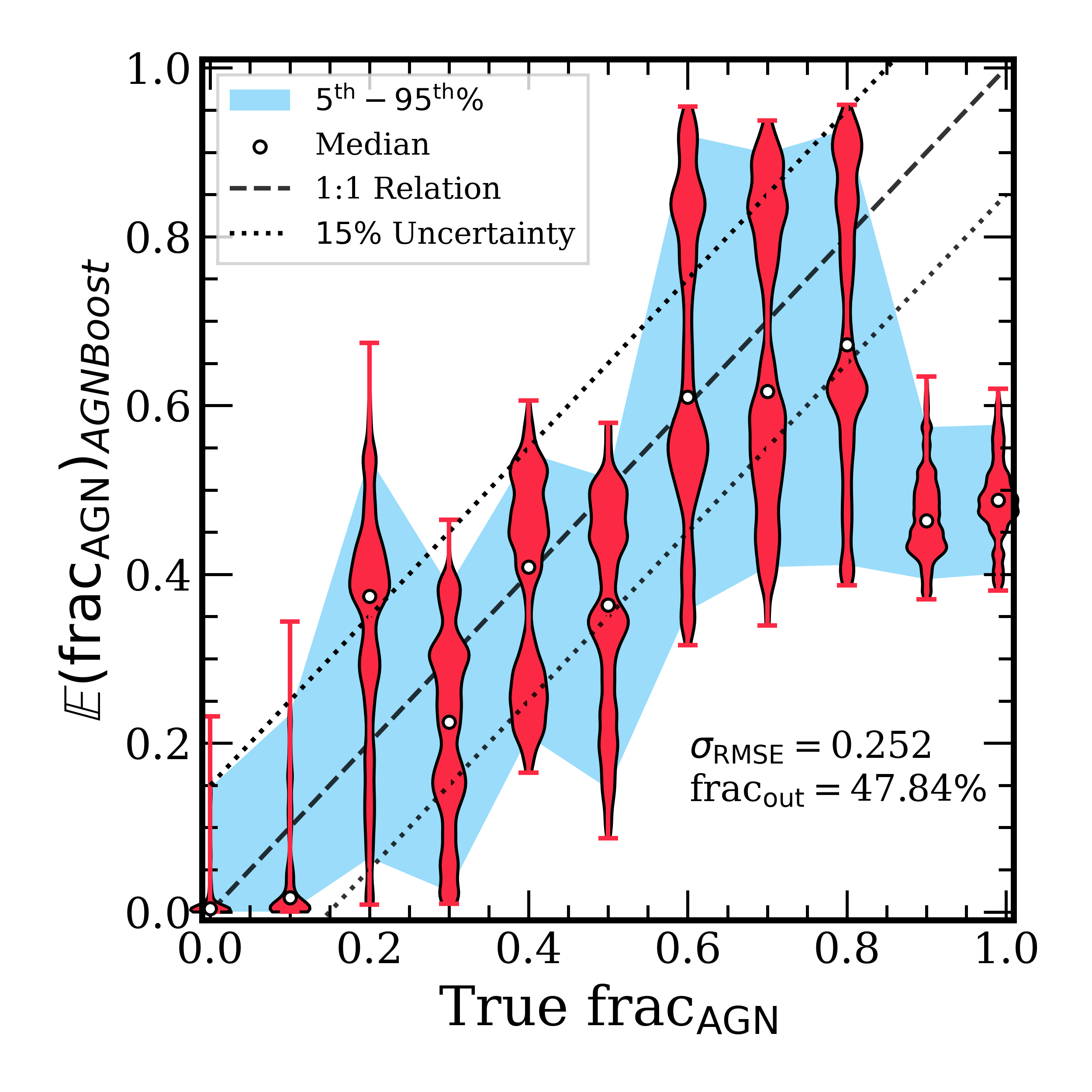}{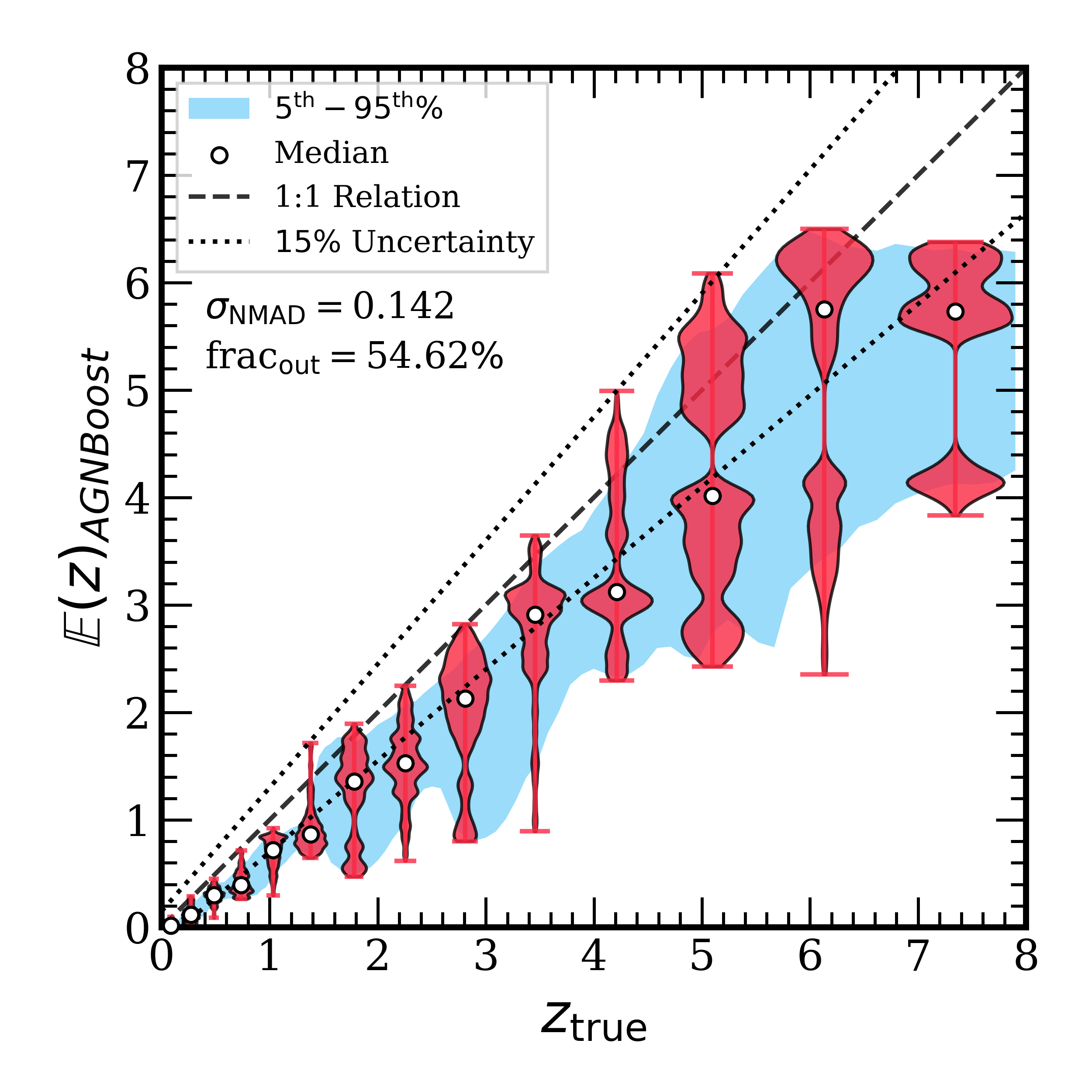}
\caption{Same as Figure \ref{fig:sim_perf}, but for the set of mock galaxies generated from \citet{vidal_2025} templates. While outlier fractions are high for both models, {\tt AGNBoost} successfully identifies AGN candidates, correctly selecting $92.6\%$ of sources with $\text{frac}_{\text{AGN}} > 0.3$ and $100\%$ of sources with $\text{frac}_{\text{AGN}} > 0.5$. Redshifts are systematically underestimated with a median offset of $-16.5\%$, and $\text{frac}_{\text{AGN}}$ is notably underestimated for the two highest bins ($\text{frac}_{\text{AGN}} = 0.9, 0.99$), which more closely resemble lower $\text{frac}_{\text{AGN}}$ {\tt CIGALE} templates (see Figure \ref{fig:sed_compare}).
\label{fig:vidal_perf}}
\end{figure*}

To test the generalization ability of {\tt AGNBoost}, we evaluate its performance on an independent set of mock galaxies generated from the AGN SED templates of \citet{vidal_2025}. These templates extend the IR ($2$--$1000\,\mu$m) AGN library of \citet{kirkpatrick_2015} to optical wavelengths ($0.1$--$1000\,\mu$m). The \citet{kirkpatrick_2015} library is based on 300+ sources ($z\sim1-3$, $L_{\rm IR}\sim10^{11}-10^{12}\,L_\odot$) with Spitzer-IRS mid-IR spectra and Herschel far-IR photometry, binned by $\text{frac}_{\text{AGN,MIR}}$ (the fraction of $5$--$15\,\mu$m emission attributable to AGN) in steps of 0.1 to produce 11 empirical SED templates. \citet{vidal_2025} extended these templates to optical wavelengths by stitching them to SWIRE templates \citep{SWIRE} at wavelengths that vary depending on $\text{frac}_{\text{AGN,MIR}}$: \textit{Sd} at $5.6\,\mu$m for $\text{frac}_{\text{AGN,MIR}} < 0.2$, \textit{NGC 6240} at $2.1\,\mu$m for $\text{frac}_{\text{AGN,MIR}} \in [0.2, 0.6)$, \textit{I19254-S} at $3\,\mu$m for $\text{frac}_{\text{AGN,MIR}} \in [0.6, 0.8]$, and \textit{QSO2} at $3\,\mu$m for $\text{frac}_{\text{AGN,MIR}} > 0.8$. While \citet{vidal_2025} interpolated these templates to a finer grid of $\text{frac}_{\text{AGN,MIR}}$ with $\Delta\text{frac}_{\text{AGN,MIR}} = 0.0005$, we adopt only those 11 templates on the original frac$_\text{AGN,MIR}$ grid of \citet{kirkpatrick_2015} for ease of comparison.

We generated $10^6$ mock galaxies from these 11 templates by randomly redshifting each template according to a uniform redshift distribution over $z \in [0.01, 8]$. Since the wavelength definition of $\text{frac}_{\text{AGN}}$ differs between the \citet{vidal_2025} templates ($5$--$15\,\mu$m) and the {\tt CIGALE} training set ($3$--$30\,\mu$m), we created an entirely new set of mock {\tt CIGALE} galaxies with $\text{frac}_{\text{AGN}}$ defined over $5$--$15\,\mu$m, using the same parameter grid as Table \ref{table:cigale_params} except for the wavelength definition. We then retrained new {\tt AGNBoost} models on these mock {\tt CIGALE} galaxies, following the same training procedures (Section \ref{methodology:model_tuning}) as the original models.

Figure \ref{fig:vidal_perf} shows {\tt AGNBoost} performance on this independent test set. Photometric redshift estimation is notably poor, with $\sigma_{\text{NMAD}} = 0.142$ and $\text{frac}_{\text{out}} = 54.62\%$, and predicted redshifts are systematically underestimated with a median offset of $-16.5\%$. However, at $z < 4$, photometric redshift estimates are more reasonable, with predicted distributions remaining mostly unimodal and median predictions tracking just below the $15\%$ uncertainty region.  For $\text{frac}_{\text{AGN}}$ estimation, the scatter is high ($\sigma_{\text{RMSE}} = 0.252$, $\text{frac}_{\text{out}} = 47.84\%$), but {\tt AGNBoost} successfully identifies AGN candidates: using $\text{frac}_{\text{AGN}} > 0.3$ as the threshold, {\tt AGNBoost} correctly selects $92.6\%$ of all sources with true $\text{frac}_{\text{AGN}} > 0.3$ and $100\%$ of sources with $\text{frac}_{\text{AGN}} > 0.5$. Only $20.2\%$ of sources with $\text{frac}_{\text{AGN}} \leq 0.3$ are incorrectly classified as candidate AGNs.

\begin{figure*}[t!]
\plotone{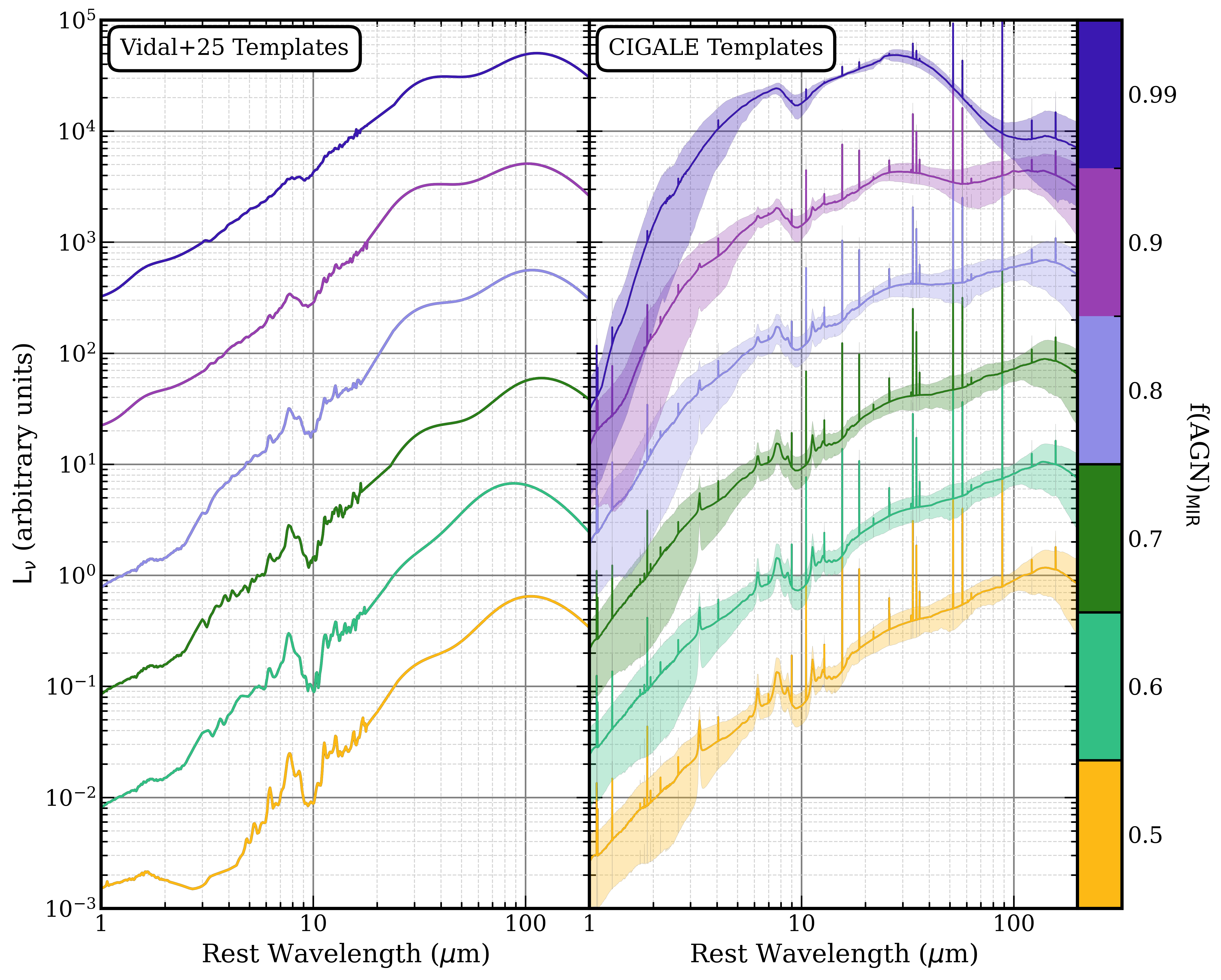}
\caption{Comparison of \citet{vidal_2025} AGN templates (left) and {\tt CIGALE} AGN SEDs (right) at $z = 2$. The {\tt CIGALE} SEDs were generated using the parameter grid from Table \ref{table:cigale_params}, producing 77,760 SEDs for each $\text{frac}_{\text{AGN}}$ value. Solid lines show the median {\tt CIGALE} SED at each wavelength, with shaded regions indicating the median absolute deviation. The two highest $\text{frac}_{\text{AGN}}$ bins show notable differences: {\tt CIGALE} SEDs exhibit much redder slopes below $10\,\mu$m compared to the \citet{vidal_2025} templates, with the $\text{frac}_{\text{AGN}} = 0.9$ and $0.99$ \citet{vidal_2025} templates more closely resembling the {\tt CIGALE} $\text{frac}_{\text{AGN}} = 0.5$ template.
\label{fig:sed_compare}
}
\end{figure*}

Notably, {\tt AGNBoost} systematically underestimates $\text{frac}_{\text{AGN}}$ for the two highest $\text{frac}_{\text{AGN}}$ bins ($\text{frac}_{\text{AGN}} = 0.9, 0.99$). To investigate this behavior, Figure \ref{fig:sed_compare} compares the \citet{vidal_2025} templates to {\tt CIGALE} AGN SEDs. The two highest $\text{frac}_{\text{AGN}}$ bins show substantial differences: the {\tt CIGALE} SEDs exhibit much redder slopes below $10\,\mu$m compared to the \citet{vidal_2025} templates. The $\text{frac}_{\text{AGN}} = 0.9$ and $0.99$ \citet{vidal_2025} templates more closely resemble the {\tt CIGALE} $\text{frac}_{\text{AGN}} = 0.5$ template, explaining why {\tt AGNBoost} produces median estimates of $\text{frac}_{\text{AGN}} \sim 0.5$ for these bins. This can be attributed to the original set of galaxies that the \citet{kirkpatrick_2015} templates were created from. By requiring Spitzer/IRS spectra and Herschel detections, \citet{kirkpatrick_2015} was naturally biased towards dustier star-forming galaxies. The authors found that even galaxies classified as $\text{frac}_{\text{AGN}} > 0.9$ had star formation rates around 100\,$M_\odot$/yr. The {\tt CIGALE} AGN show a noticeable lack of far-IR emission.

\subsection{Performance on MEGA Sources with Complete Observations} \label{subsec:mega_results}

\begin{figure}[t!]
\plotone{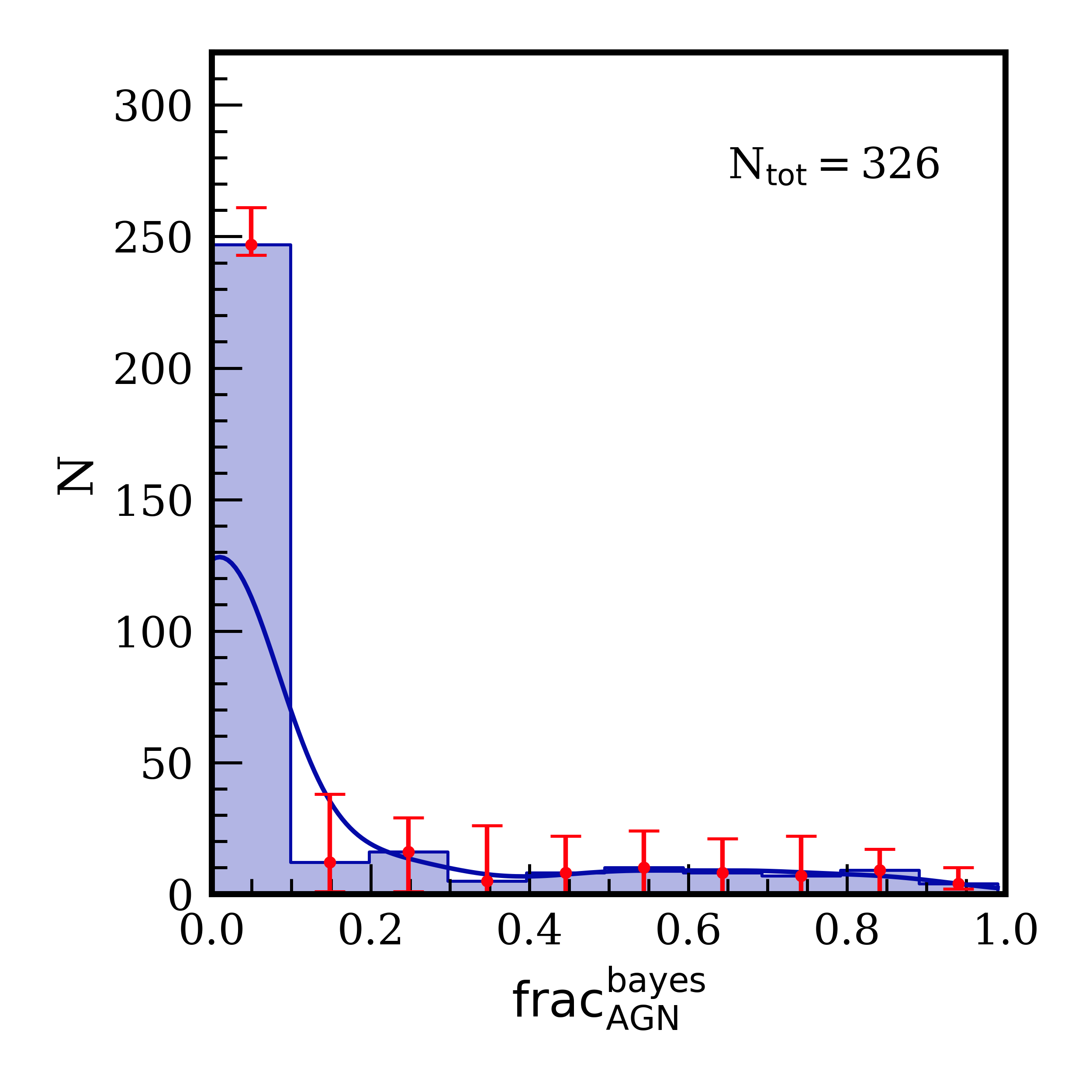}
\caption{Histogram of {\tt CIGALE}-derived frac$_\text{AGN}$ values for the 326 MEGA sources with well-constrained {\tt CIGALE} fits ($0.5 < \chi^2_{\text{red}} \leq 2$). Error bars represent uncertainties propagated from the CIGALE Bayesian frac$_\text{AGN}$ uncertainties. The vertical dashed line indicates the median frac$_\text{AGN}$ value, and the blue curve shows a kernel density estimate (KDE) fit to the distribution.
\label{fig:mega_cigale_fagn}}
\end{figure}

Here, we evaluate how {\tt AGNBoost} performs on the final set of 748 MEGA galaxies with no missing photometric observations. Since the MEGA catalog does not have existing estimates of frac$_{\text{AGN}}$, we use {\tt CIGALE} to fit the MEGA catalog. We adopt the same {\tt CIGALE} parameter space presented in Table \ref{table:cigale_params}, but use the existing redshifts for each source so that {\tt CIGALE} does not fit for redshift. In total, $326$ ($\sim44\%$) sources have fits with an acceptable reduced chi-squared ($\chi^2_{\text{red}}$) of $0.5< \chi^2_{\text{red}} < 2$. We adopt the Bayesian output values of frac$_{\text{AGN}}$ instead of the best-fit output. These Bayesian outputs are the probability-weighted values across all the {\tt CIGALE} models, and are generally more robust than the best-fit values from the minimum-$\chi^2$ model alone \citep{yang_ceers_2023}. Figure \ref{fig:mega_cigale_fagn} shows the distribution of Bayesian frac$_{\text{AGN}}$ estimates. The results of this SED-fitting should not be interpreted as ground-truth, but instead serve as a point of comparison for {\tt CIGALE} SED-fitting when no parameters have been fine-tuned.

\begin{figure*}[t!]
\plottwo{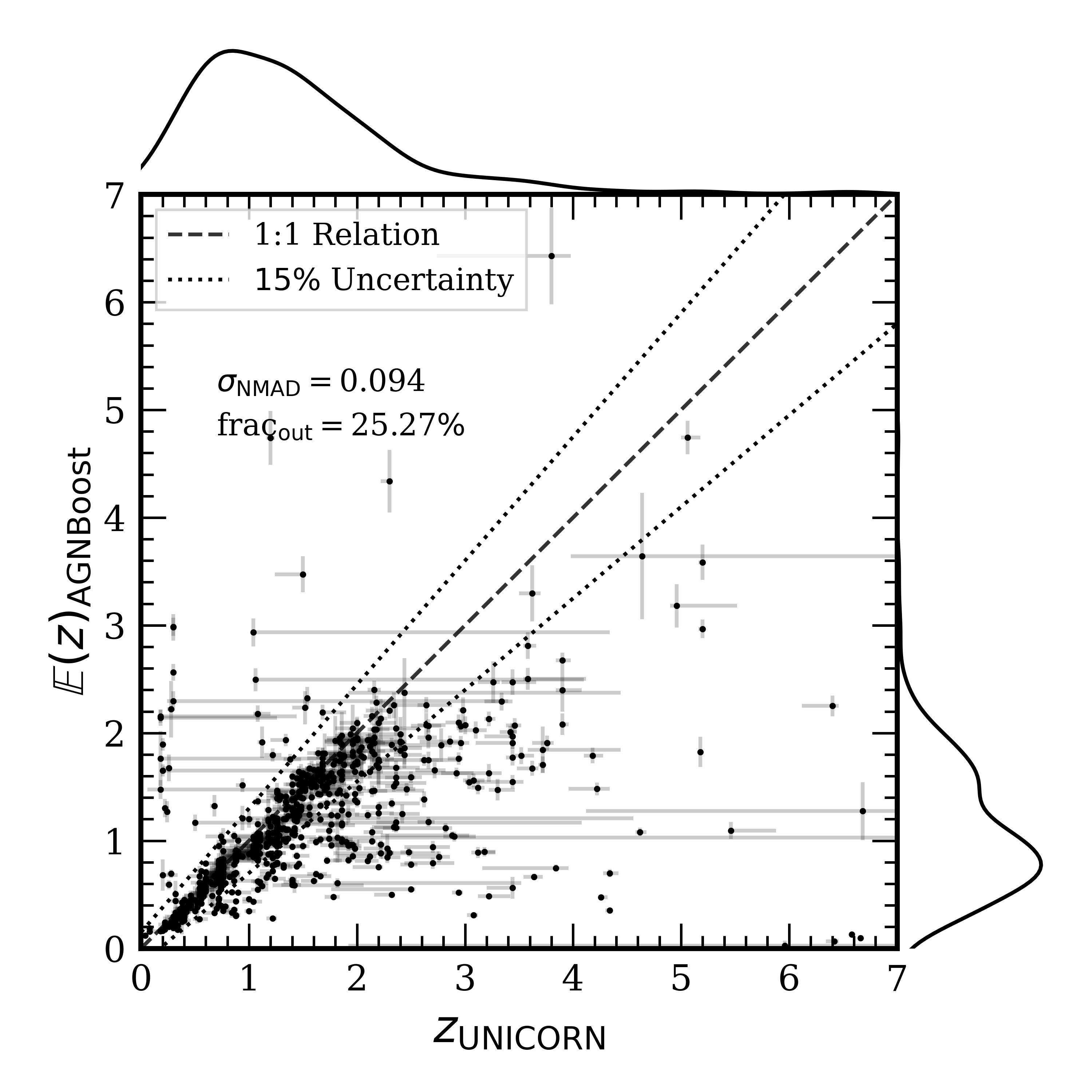}{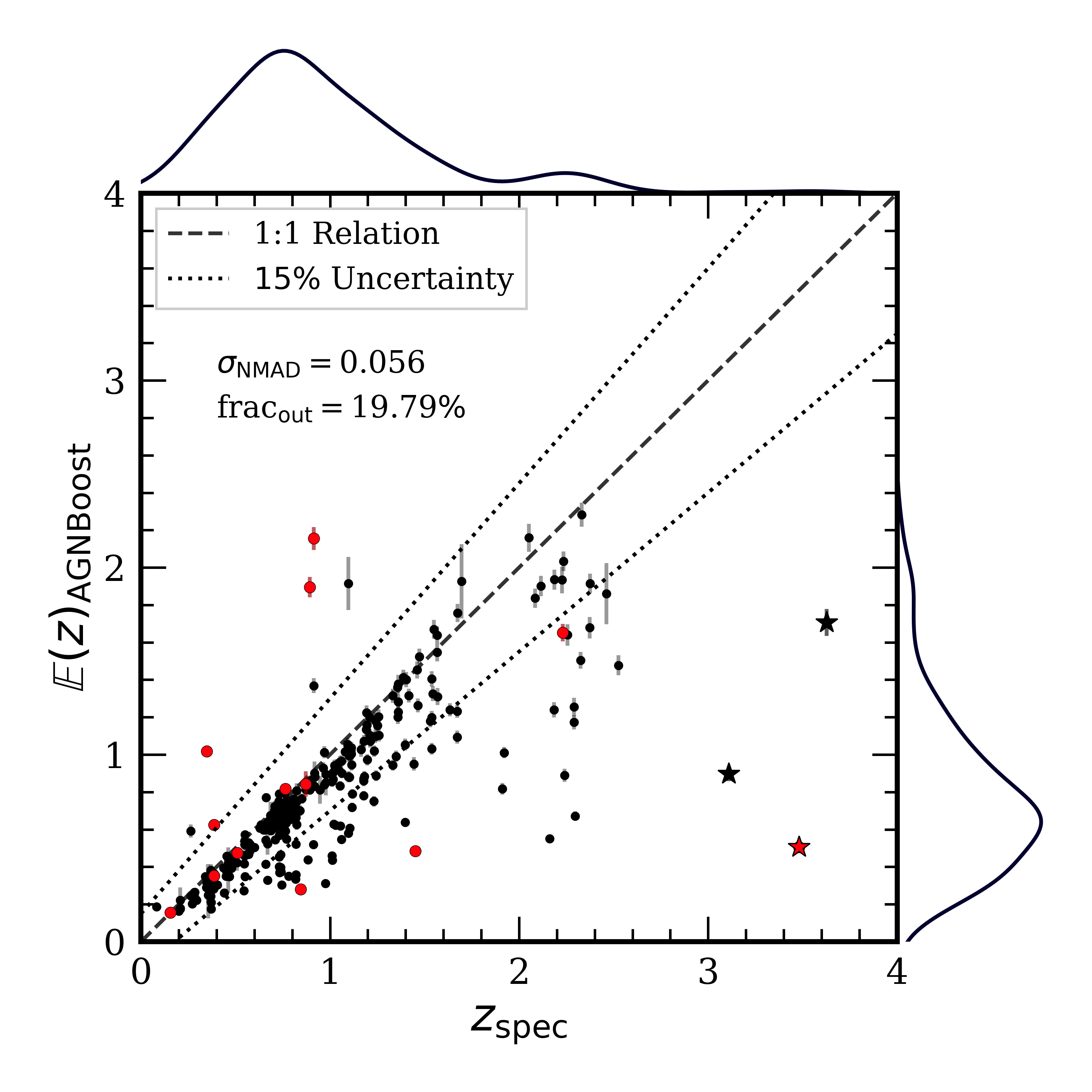}
\caption{Expectation values of {\tt AGNBoost} redshift predictions compared to existing photometric redshifts (left) and spectroscopic redshifts (right). The left panel shows {\tt AGNBoost} predictions against photometric redshifts from the UNICORN catalog (Finkelstein et al. in prep) for the full MEGA sample, with photo-z uncertainties indicating $68\%$ confidence intervals. The right panel compares {\tt AGNBoost} predictions to spectroscopic redshifts for 288 MEGA sources, where red points indicate sources for which UNICORN photometric redshifts disagree with the corresponding spectroscopic redshifts (outside the $15\%$ uncertainty region). The stars mark the 3 catastrophic outliers at $z>3$. The black dashed lines show the 1:1 relation and dotted lines indicate $15\%$ uncertainty regions. Marginal distributions along each axis show the underlying redshift distributions, with redshift quality metrics ($\sigma_{\text{NMAD}}$ and $\text{frac}_{\text{out}}$) displayed. {\tt AGNBoost} redshifts show good agreement with spectroscopic values, particularly for $z < 2$, but tend to underestimate redshifts at $z > 3.5$ compared to UNICORN photometric redshift estimates.
\label{fig:mega_redshift_performance}}
\end{figure*}

The left side of Figure \ref{fig:mega_redshift_performance} compares the redshift estimates of {\tt AGNBoost} to the photometric redshifts from the UNICORN catalog (Finkelstein et al. in prep). The UNICORN photometric redshifts were estimated using HST+NIRCam observations ($\sim0.4\text{--}4.7\,\mu$m observed frame), enabling them to trace the Balmer break from $z\sim 0.2$ out to $z\sim 11.7$. In contrast, {\tt AGNBoost} uses NIRCam+MIRI coverage ($1.2\text{--}21\,\mu$m observed frame), which allows it to trace mid-IR PAH features at $z\leq 2.4$ and the Balmer break for $z\gtrsim 2.1$. We find good agreement between {\tt AGNBoost} and UNICORN redshift estimates at $z < 2$, which we attribute to {\tt AGNBoost} successfully identifying PAH features in this regime. However, there is significant disagreement at $z>3$, where {\tt AGNBoost} should in principle also trace the Balmer break and maintain good agreement with UNICORN. We suspect this discrepancy arises from the CIGALE training SEDs exhibiting weak stellar bumps and PAH features, leading to systematic errors in {\tt AGNBoost}'s redshift estimates at high-$z$.

The right side of Figure \ref{fig:mega_redshift_performance} shows {\tt AGNBoost} photometric redshift performance compared to spectroscopic redshifts. For the 288 MEGA galaxies with spectroscopic redshifts, we obtain $\sigma_{\text{NMAD}} = 0.056$ and an outlier fraction of $\text{frac}_{\text{out}} = 19.79\%$, with a median outlier spectroscopic redshift of $z \approx 1.1$. The red points highlight 13 sources where UNICORN photometric redshifts are outliers relative to the spectroscopic redshifts (outside the $15\%$ uncertainty region). Of these, {\tt AGNBoost} successfully recovers redshift estimates that agree with the spectroscopic redshifts for 5 sources, while the remaining 8 sources are outliers for both methods. The 3 catastrophic outliers at $z_{\text{spec}} > 3$ (marked as stars in Figure \ref{fig:mega_redshift_performance}) exhibit weak Balmer breaks in their SEDs, likely causing {\tt AGNBoost} to systematically underestimate the spectroscopic redshift.

\begin{figure}[t!]
\plotone{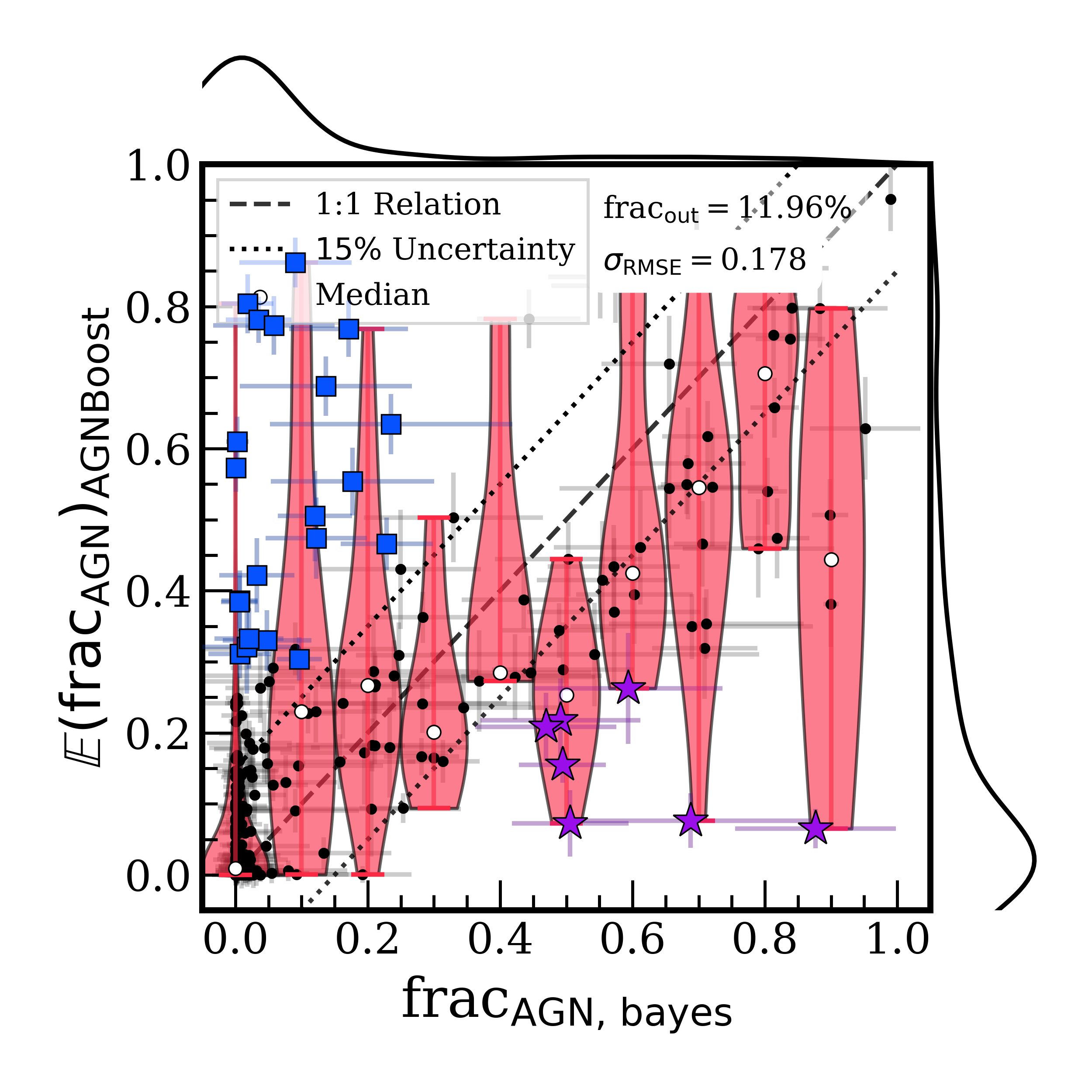}
\caption{Expectation values of {\tt AGNBoost} frac$_\text{AGN}$ predictions compared to {\tt CIGALE} Bayesian frac$_\text{AGN}$ estimates for the 326 MEGA sources with complete photometry and well-constrained {\tt CIGALE} fits.The Violin plot elements are the same as in Figure \ref{fig:sim_perf}. Blue squares highlight sources classified as AGNs by {\tt AGNBoost} but as SFGs by {\tt CIGALE}, while purple stars show the reverse (i.e., AGN by {\tt CIGALE}, SFG by {\tt AGNBoost}). {\tt AGNBoost} frac$_\text{AGN}$ estimates show broad agreement with {\tt CIGALE}.
\label{fig:mega_agn_performance}}
\end{figure}

Figure \ref{fig:mega_agn_performance} shows the $\text{frac}_{\text{AGN}}$ performance of {\tt AGNBoost} on the subset of 326 MEGA galaxies with $0.5 < \chi^2_{\text{red}} < 2$ and complete photometry. Following \citet{kirkpatrick_2017}, we classify sources with $\text{frac}_{\text{AGN}} > 0.3$ as candidate AGNs. The two methods show broad agreement: 276 ($\sim 85\%$) galaxies are classified as SFGs ($\text{frac}_{\text{AGN}} \leq 0.3$) by both {\tt AGNBoost} and {\tt CIGALE}. {\tt AGNBoost} identifies 62 ($\sim 19\%$) AGN candidates, while {\tt CIGALE} identifies 50 ($\sim 15\%$) AGN candidates, yielding similar AGN fractions overall.

We find 28 sources with significant disagreements in estimated $\text{frac}_{\text{AGN}}$ values: 7 sources (purple stars in Figure \ref{fig:mega_agn_performance}) where {\tt AGNBoost} identifies SFGs but {\tt CIGALE} identifies AGN candidates, and 21 sources (blue squares) where {\tt AGNBoost} identifies AGN candidates but {\tt CIGALE} identifies SFGs. To assess these discrepancies, we refit the 7 AGN candidates without an AGN component and the 21 SFGs with a forced AGN component.

For the 7 sources originally identified as AGN candidates by {\tt CIGALE}, refitting without an AGN component yields a reasonable fit ($\chi^2_{\text{red}} = 2.2$) for 1 source, and $\chi^2_{\text{red}} \leq 5.5$ for the remaining 6 sources. Only one source shows a visually poor fit, with significant missing mid-IR emission. For the 21 sources originally identified as SFGs by {\tt CIGALE}, forcing an AGN component yields $0.5 < \chi^2_{\text{red}} < 2$ for 12 sources and $\chi^2_{\text{red}} < 5$ for 19 sources, with only 2 showing visually poor fits with significant mid-IR residuals. Example SEDs for several of these sources are shown in Appendix \ref{app:ambiguous_seds}.

The similar quality of {\tt CIGALE} fits both with and without AGN components for these 28 sources indicates that the available photometry does not provide sufficient information to uniquely determine the presence or strength of AGN emission. Given the fundamental differences in how the two methods operate ({\tt CIGALE} performs discrete template matching while {\tt AGNBoost} identifies learned patterns across the training population) disagreement in these ambiguous cases is not unexpected. Without spectroscopic or X-ray confirmation, we cannot determine which method provides more accurate estimates for these sources. The disagreement highlights cases where photometric data alone may be insufficient to robustly constrain $\text{frac}_{\text{AGN}}$.

\begin{figure*}[t!]
\plottwo{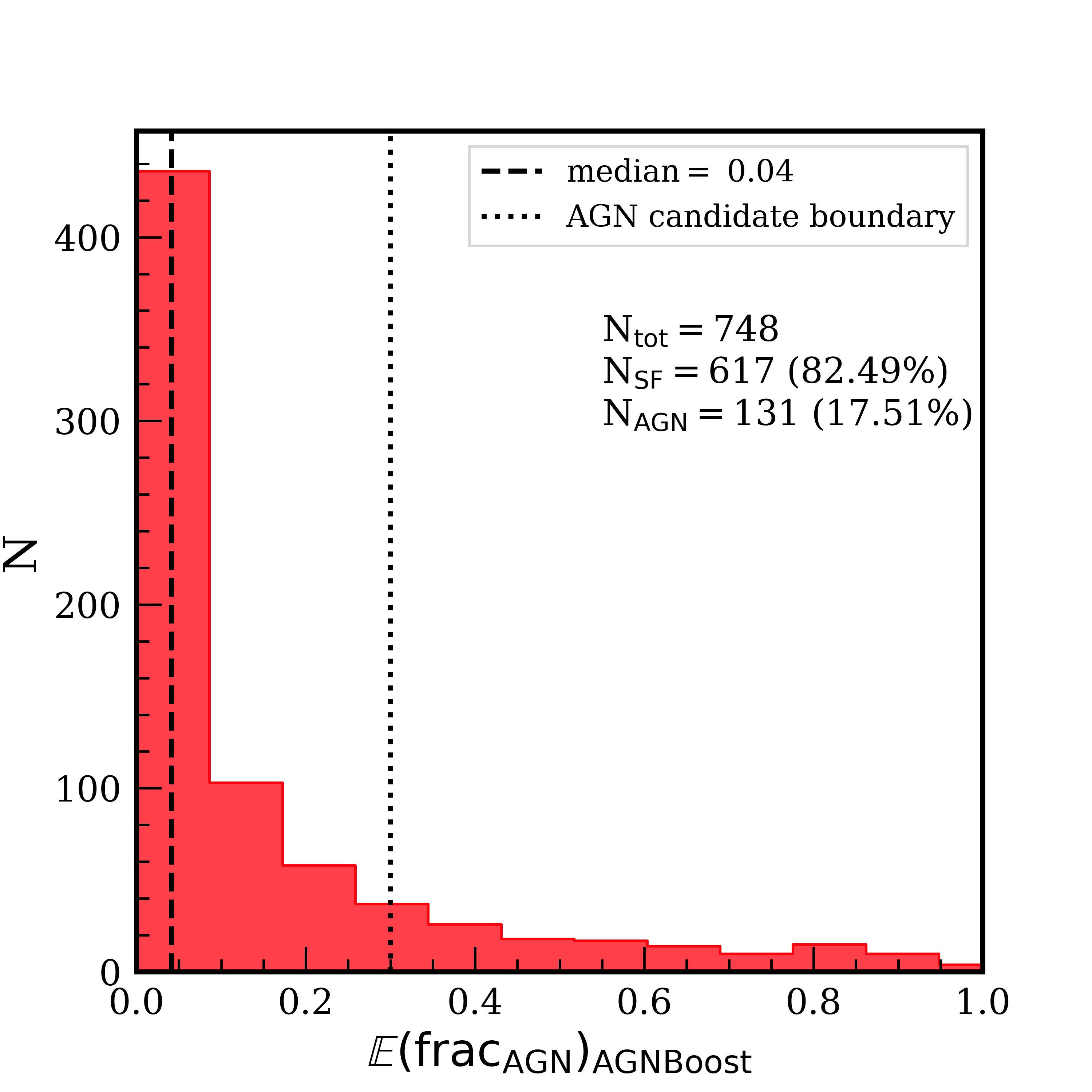}{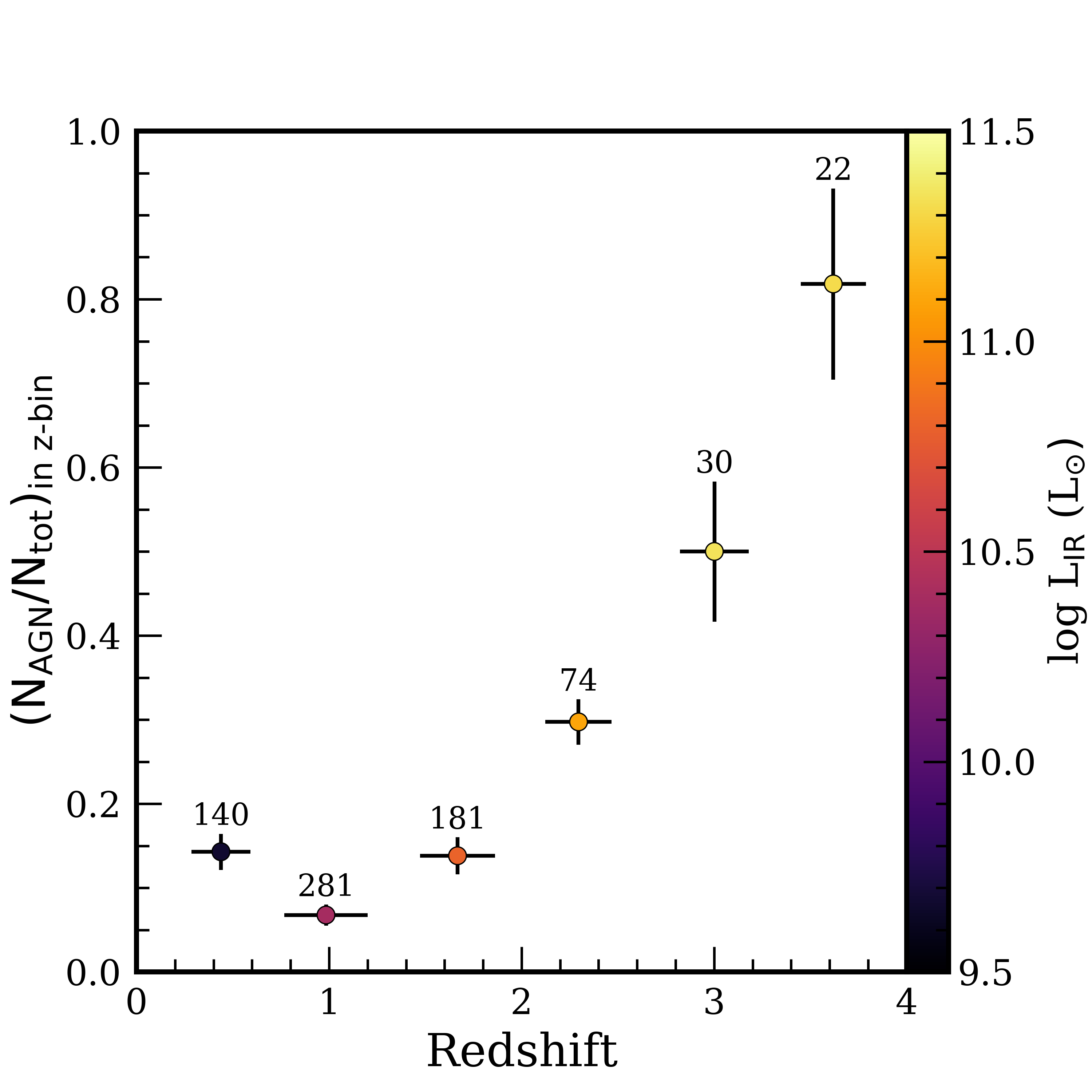}
\caption{Left: Histogram of frac$_\text{AGN}$ values for all 748 MEGA galaxies. Using frac$_\text{AGN} > 0.3$ as the AGN threshold, {\tt AGNBoost} identifies 131 ($\sim 17.5\%$) AGN candidates and 617 ($\sim 82.5\%$) SFGs. Right: AGN candidate population fraction (frac$_\text{AGN} > 0.3$) as a fraction of redshift. Points are plotted at the average redshift of each bin (with horizontal error bars showing the redshift standard deviation within each bin), and are colored by the median $L_{\text{IR}}$ calculated by integrating the best-fit {\tt CIGALE} SEDs from $8\text{--}1000\,\mu$m. Vertical error bars represent the average shift in AGN candidate population fraction when sources move across the $\text{frac}_{\text{AGN}} = 0.3$ threshold due to their {\tt AGNBoost} uncertainty. Redshift bins are determined using the Bayesian blocks algorithm \citep{Bayesian_blocks} with a minimum of 10 galaxies per bin; the number of galaxies in each bin is labeled above each point.
\label{fig:mega_whole_fagn}
}
\end{figure*}

Despite these ambiguities in individual cases, the overall AGN identification performance of {\tt AGNBoost} is robust. Figure \ref{fig:mega_whole_fagn} shows AGN candidate identification across the full MEGA sample of 748 sources. The left panel shows that {\tt AGNBoost} identified 131 ($\sim$$17.5\%$) AGN candidates overall. The right panel shows the AGN candidate population fraction as a function of redshift, revealing AGN fractions exceeding $50\%$ at $z > 3$. This is significantly higher than some previous literature results. For example, \citet{Maiolino_2024} finds a 10\% AGN fraction at $4 < z < 6$. However, \citet{Maiolino_2024} identify AGN spectroscopically through detection of the broad-line region in X-ray luminous sources ($L_{\text{X}} > 10^{43}$ erg/s), selecting only unobscured Type 1 AGN, while {\tt AGNBoost} uses near-IR to mid-IR emission that is less affected by obscuration and sensitive to both Type 1 and Type 2 AGN across a broader luminosity range, including composite systems with modest AGN contributions.

\citet{Lyu_2024} present another instructive comparison, identifying AGN through comprehensive SED fitting of JWST/MIRI observations in the Systematic Mid-infrared Instrument Legacy Extragalactic Survey (SMILES; \citealp{SMILES}). They find 217 AGN candidates from 3273 MIRI-detected sources ($\sim$$6.6\%$ AGN candidate fraction), significantly lower than our $17.5\%$. However, at high redshift ($z > 4$), we find comparable numbers of AGN candidates, suggesting the discrepancy arises primarily from low-redshift sources. 

The discrepancy between AGN fractions can be explained by our different methodological approaches. First, \citet{Lyu_2024} utilize up to 27 photometric bands spanning $0.4$--$25.5\,\mu$m, compared to {\tt AGNBoost}'s 11 bands spanning $1.15$--$21\,\mu$m. The extended wavelength coverage, particularly in the rest-frame UV/optical, provides critical constraints for distinguishing AGN from star-forming galaxies. Second, \citet{Lyu_2024} performed careful fitting of low-metallicity dwarf galaxies, which are known to be particularly challenging to distinguish from AGN photometrically \citep{kirkpatrick_ceers_2023}. To address this degeneracy, they modeled the dust emission in low-mass systems (M$_{*} < 10^{9.5}$ M$_{\odot}$) using the Haro 11 template, a low-metallicity starburst template appropriate for dwarf galaxies \citep{Lyu_2016}, and then rejected low-mass AGN candidates with ambiguous classifications due to low signal-to-noise ratios or photometric inconsistencies. This process eliminated 101 ($\sim54\%$) of 187 initial low-mass AGN candidates, the majority of which are at low redshift. Thus, the elevated AGN candidate fraction from {\tt AGNBoost} likely reflects significant contamination from these low-metallicity dwarf galaxies.

The right panel of Figure \ref{fig:mega_whole_fagn} reveals that the high fraction of AGN at $z > 3$ is most likely a luminosity-dependent selection effect rather than intrinsic evolution. Points are colored by the median infrared luminosity ($8$--$1000\,\mu$m, determined from individual {\tt CIGALE} fits), which increases systematically with redshift. MEGA preferentially detects only the most luminous sources at higher redshifts, and since luminous AGN are easier to detect, the observed AGN candidate population fraction is artificially inflated. Definitive validation of these high-redshift AGN candidates will require X-ray or spectroscopic confirmation through broad lines and high-ionization lines, made easier by the reduced size of this candidate sample.

\subsection{Photometry Imputation Performance}\label{results:imputaton}
\begin{figure*}[t!]
\plotone{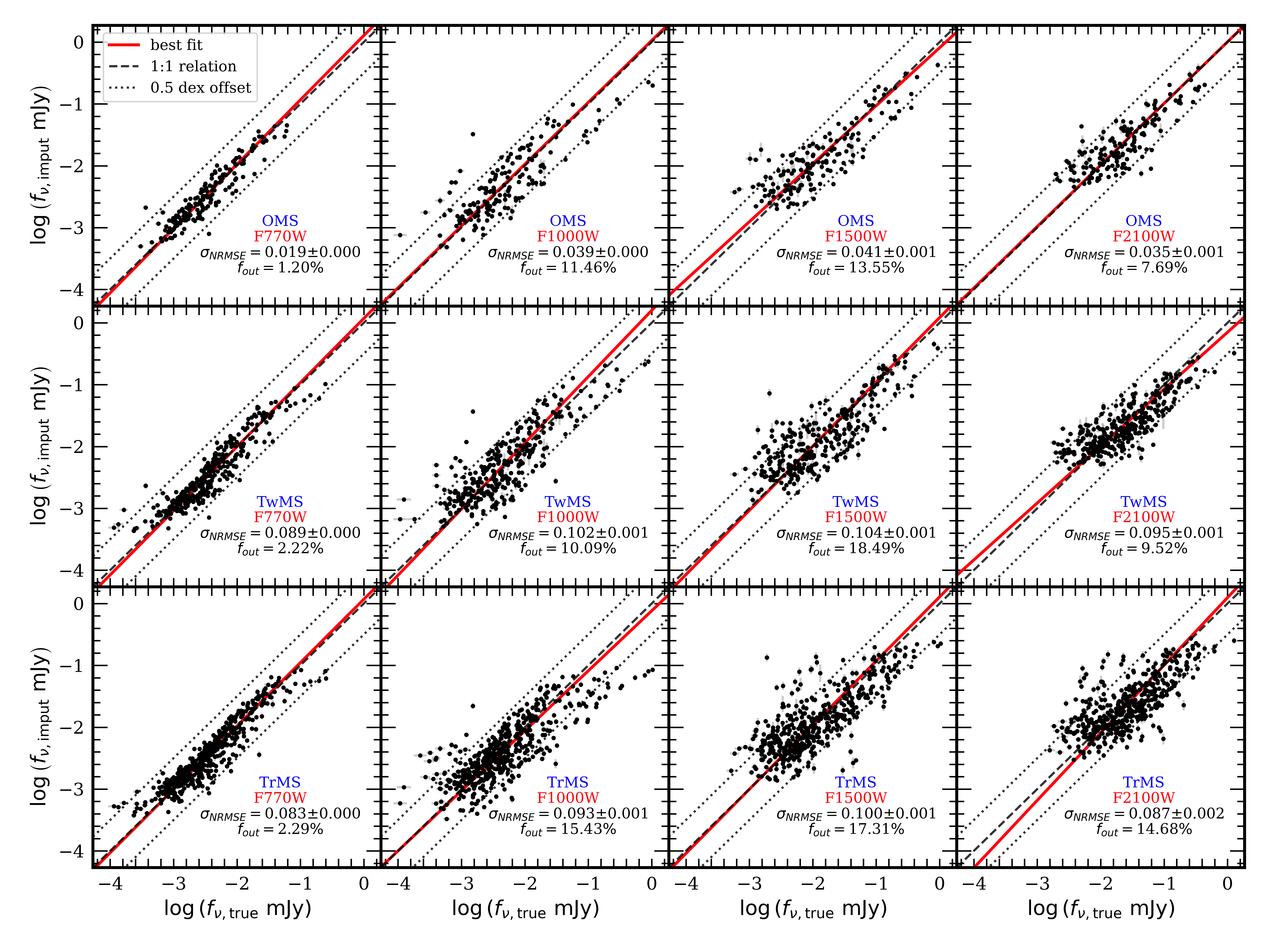}
\caption{{\tt SGAIN} photometry imputation performance for samples with randomly removed MIRI bands. For each source, MIRI bands are randomly removed to create samples missing one band (OMS), two bands (TwMS), and three bands (TrMS). The black dashed line shows the 1:1 relation, the dotted line indicates the 0.5 dex uncertainty region, and the red line shows the line of best fit. The normalized root mean square error ($\sigma_{\text{NRMSE}}$) and outlier fractions are displayed for each band. {\tt SGAIN} demonstrates notably strong performance when only one MIRI band is missing. While imputation performance degrades with the number of missing MIRI bands, {\tt SGAIN} still performs relatively well even when three of the four MIRI bands are missing.
\label{fig:imputation_all_sets}}
\end{figure*}

As discussed in Section \ref{subsec:mega_data}, we test the ability to reliably characterize sources with missing photometry using statistical imputation. Following the procedure of \citet{luo_imputation}, we assess photometry imputation performance by creating new datasets with different missing rates (defined as the number of total missing photometric observations per source) from the final set of 748 MEGA sources with complete photometry. These subsamples are created by randomly deleting photometric observations for each source according to the missing rate of the sub-sample. With this method, we create three new datasets: a one-band missing sample (OMS), a two-band missing sample (TwMS), and a three-band missing sample (TrMS). To assess the accuracy of imputation, we calculate the normalized root mean square error (NRMSE; $\sigma_\text{NRMSE}=\sigma_{\text{RMSE}}/\text{mean}\{x_{\text{pred}}\}$) of each band for each new dataset. Figure \ref{fig:imputation_all_sets} shows the imputation performance for the OMS, TwMS, and TrMS samples. The imputation process becomes more uncertain as the number of missing bands increases, but the outlier fraction (within $0.5$ dex) is still below $20\%$ in all bands when there are three missing MIRI bands in Figure \ref{fig:imputation_all_sets}.

\subsection{Effects of Imputed Photometric Observations on {\tt AGNBoost} Predictions}
\begin{figure*}[t!]
\plotone{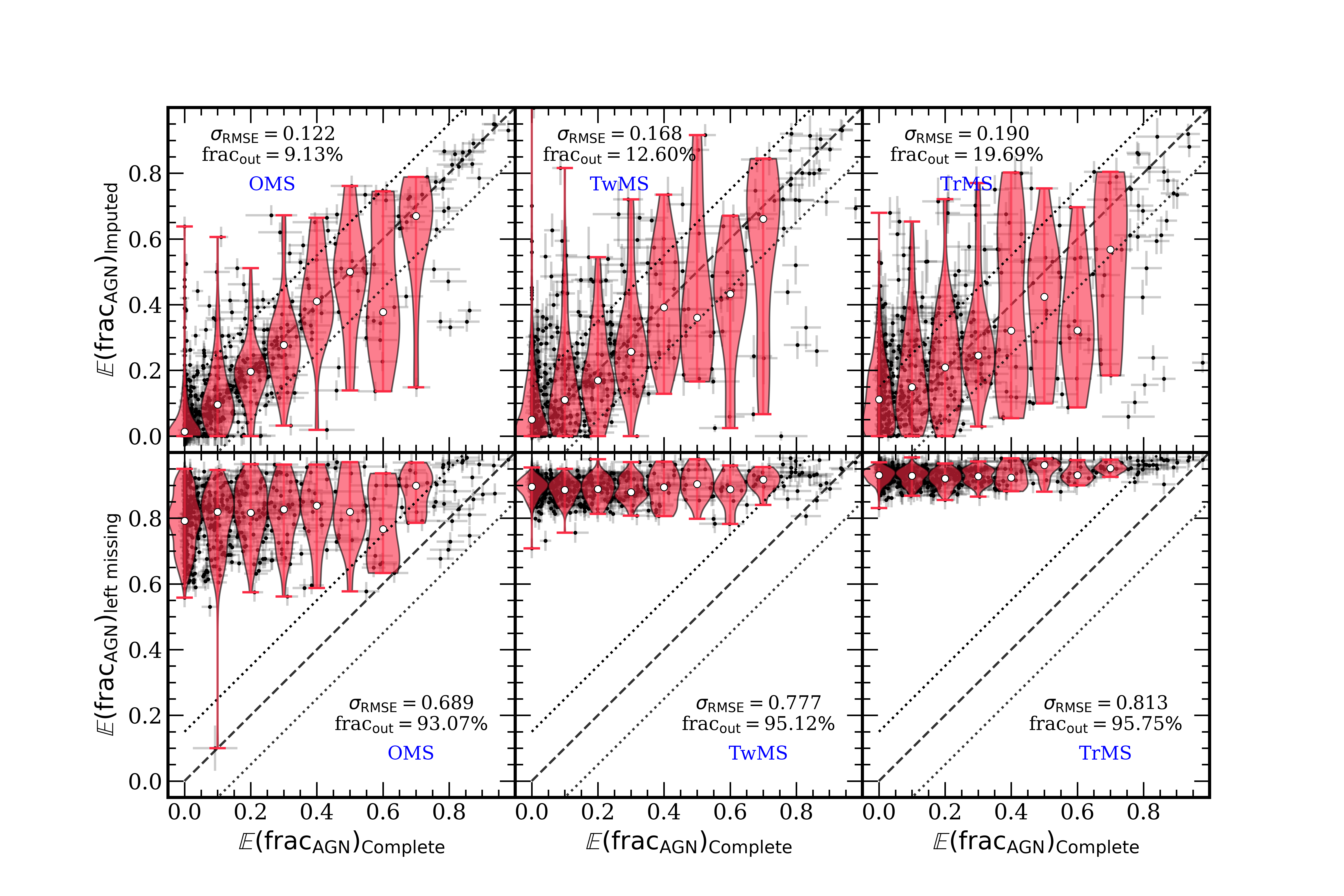}
\caption{ Comparison of {\tt AGNBoost} frac$_\text{AGN}$ estimation performance with and without {\tt SGAIN} imputation for missing MIRI photometry. For each source, MIRI bands are randomly removed to create samples with one missing band (OMS), two missing bands (TwMS), and three missing bands (TrMS). The top row shows {\tt AGNBoost} results when the randomly removed bands are replaced with {\tt SGAIN}-imputed values. The bottom row shows {\tt AGNBoost} performance when missing bands are left blank (i.e., no imputation). {\tt SGAIN} imputation significantly improves frac$_\text{AGN}$ estimation compared to running {\tt AGNBoost} with missing photometric data.
\label{fig:fagn_imput} }
\end{figure*}

\begin{figure*}[t!]
\plotone{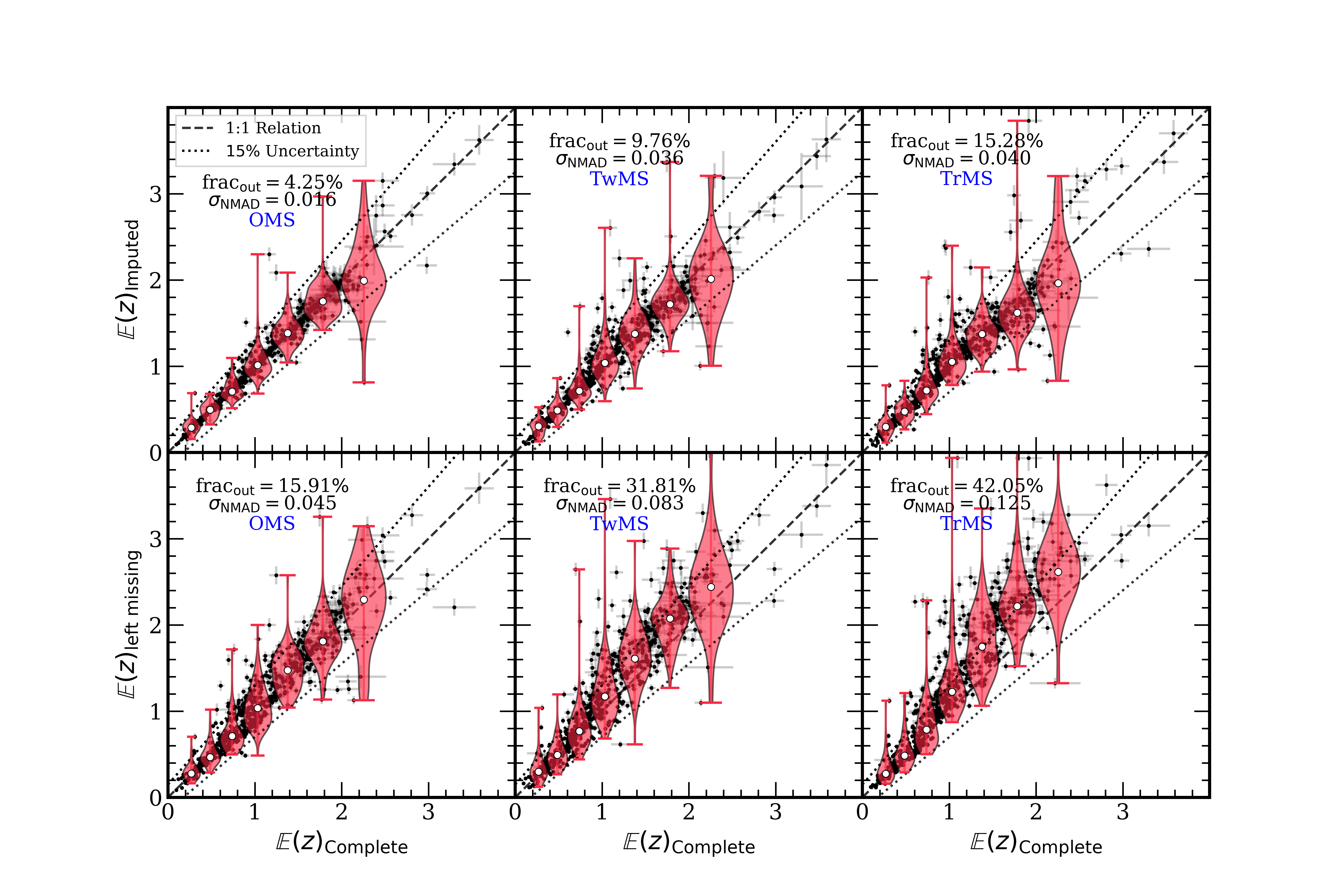}
\caption{Comparison of {\tt AGNBoost} photometric redshift estimation performance with and without {\tt SGAIN} imputation for missing MIRI photometry. {For each source, MIRI bands are randomly removed to create samples with one missing band (OMS), two missing bands (TwMS), and three missing bands (TrMS).} The top row shows {\tt AGNBoost} results using {\tt SGAIN}-imputed values for the randomly removed bands. The bottom row shows {\tt AGNBoost} performance when missing bands are left blank (i.e., no imputation). {\tt SGAIN} imputation reduces the outlier fraction by a factor of 3 in all cases.
\label{fig:z_imput}}
\end{figure*}

Figures \ref{fig:fagn_imput} and \ref{fig:z_imput} show how replacing missing MIRI photometric observations with {\tt SGAIN} imputation improves {\tt AGNBoost}'s ability to estimate frac$_\text{AGN}$ and photometric redshift, respectively. When photometry is missing, frac$_\text{AGN}$ estimation fails completely, defaulting to AGN outputs regardless of the true value (bottom panel of Figure \ref{fig:fagn_imput}). However, when we impute the missing photometry, frac$_\text{AGN}$ estimation recovers the 1:1 relation, though with increasing scatter as more MIRI bands are missing (top panel of Figure \ref{fig:fagn_imput}). In comparison, {\tt AGNBoost}'s redshift estimation does not seem to be significantly affected by missing photometry, though a positive z-bias appears at $z>1$ when two or three MIRI bands are missing (bottom panel of Figure \ref{fig:z_imput}). Photometric imputation removes this bias and reduces outlier fractions by a factor of 3, keeping median estimates on the 1:1 relation.

\subsection{Feature Importance and Model Interpretation} 

\begin{figure*}[t!]
\epsscale{1.1}
\plottwo{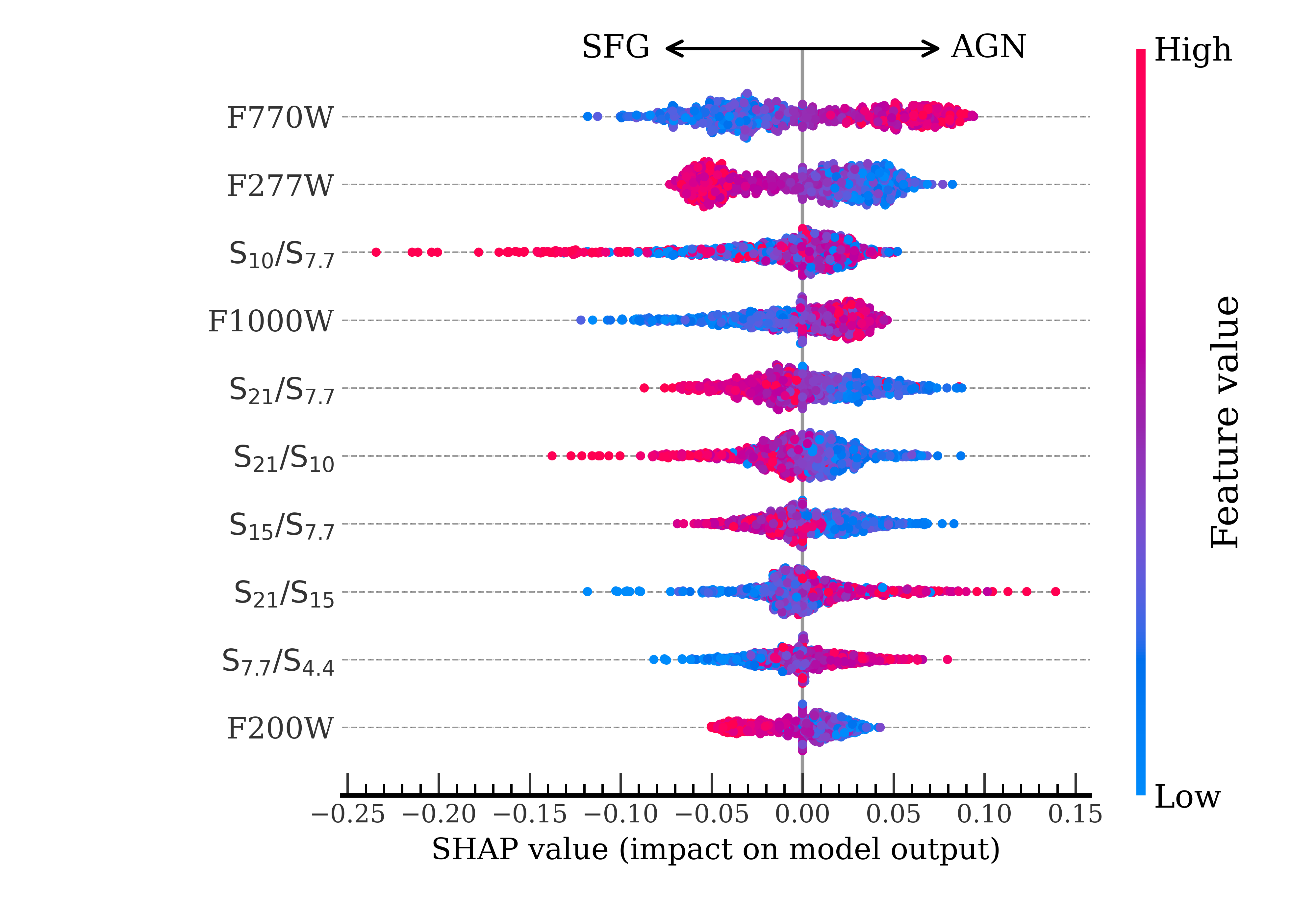}{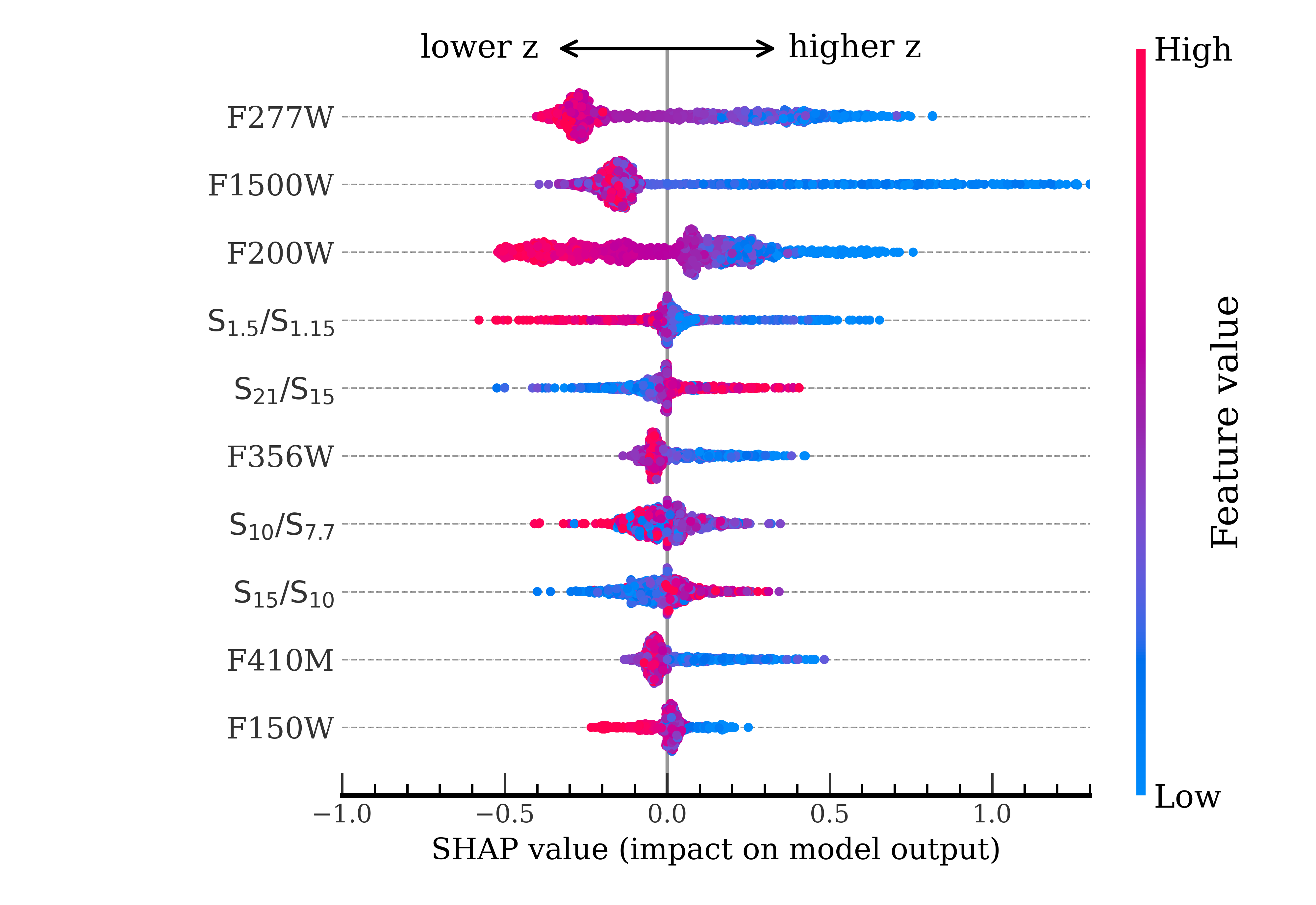}
\caption{SHAP beeswarm plots showing feature importance for the frac$_\text{AGN}$ model (left) and redshift model (right). Features are ranked by importance (top to bottom), with each dot representing an individual source. Colors indicate feature values from low (blue) to high (red), which for color-based features corresponds to bluer and redder colors, respectively. Negative SHAP values push predictions toward low-frac$_\text{AGN}$ (i.e., SFGs) or low redshifts, while positive values push predictions toward high-frac$_\text{AGN}$ (i.e., AGNs) or high redshifts. MIRI-based features rank among the most important predictors for both models, highlighting the critical role of mid-infrared photometry in AGN identification and photometric redshift estimation.
\label{fig:SHAP}}
\end{figure*}

{\tt XGBoostLSS} provides native integration with SHapeley Additive exPlanations (SHAP), a framework for interpreting black box models \citep{lundberg_shap}. SHAP assigns each feature in the model an importance value for individual predictions, allowing interpretation of the model predictions. Of particular use are SHAP's `beeswarm' plots, which rank features by importance and depict how high/low individual feature values affect predictions. In Figure \ref{fig:SHAP} we show beeswarm plots for both the frac$_{\text{AGN}}$ and redshift models of the top 10 most important features, where each point is an individual {\tt CIGALE} galaxy colored by the value of the listed features. 

For the frac$_{\text{AGN}}$ model, we see that F770W, F277W, $S_{10}/S_{7.7}$, and F1000W are the four most important features. We note that {\tt AGNBoost} uses observed fluxes rather than luminosities, and that redshift does not play a role in determining frac$_{\text{AGN}}$. Bright F770W and F1000W photometry pushes the model towards larger values of frac$_{\text{AGN}}$, whereas bright F277W photometry pushes the model towards smaller values of frac$_{\text{AGN}}$. There does not appear to be any trends in how the $S_{10}/S_{7.7}$ color affects {\tt AGNBoost} outputs, as both red and blue colors push the model towards both extremes. Blue $S_{21}/S_{7.7}$, $S_{21}/S_{10}$, and $S_{15}/S_{7.7}$ colors push the model towards larger values of frac$_{\text{AGN}}$, suggesting that {\tt AGNBoost} is successfully identifying PAH features. Red $S_{7.7}/S_{4.4}$ colors push {\tt AGNBoost} towards AGN outputs, suggesting that {\tt AGNBoost} is identifying the characteristic red slope of AGN at near-IR rest wavelengths.

For the redshift model, the top four features are F277W, F1500W, F200W, and $S_{1.5}/S_{1.15}$. As expected, low photometry values tend to push {\tt AGNBoost} towards high-redshift outputs. Blue $S_{15}/S_{10}$ and $S_{21}/S_{15}$ colors push {\tt AGNBoost} towards low-redshift outputs, suggesting that {\tt AGNBoost} is successfully identifying PAH features. Notably, all 4 MIRI bands in MEGA appear in the top 10 features at least once for both the frac$_\text{AGN}$ and redshift models, confirming that mid-IR information is important for finding AGN and estimating their redshifts \citep{yang_ceers_2023, kirkpatrick_ceers_2023, perez_2024_highz, Durodola_2024, Leung_2025, Kocevski_2025}.

\section{Summary} \label{summary}

In this work we presented {\tt AGNBoost}, a publicly-available machine learning model utilizing the {\tt XGBoostLSS} algorithm to estimate $\text{frac}_{\text{AGN}}$ and photometric redshift for NIRCam+MIRI galaxies. {\tt AGNBoost} was trained on a large set of mock galaxies from {\tt CIGALE}, which we verified to be representative of real JWST/NIRCam+MIRI photometric observations and colors. We tested {\tt AGNBoost} on mock {\tt CIGALE} galaxies, an independent set of mock galaxies from the \citet{vidal_2025} templates, and 748 MEGA galaxies with complete photometry. {\tt AGNBoost} provides a computationally fast method of identifying candidate AGNs without needing to devote substantial time to SED fitting. On catalogs of $N{\sim}1000$ sources, {\tt AGNBoost} provided parameter estimates and uncertainties in minutes on a standard laptop. Importantly, {\tt AGNBoost} is able to identify AGN candidates without requiring prior redshift information, which is often unavailable.

{\tt AGNBoost} provides robust estimates of prediction uncertainty, capturing the aleatoric uncertainty due to intrinsic randomness in observations, the epistemic uncertainty due to model knowledge limitations, and the prediction uncertainty due to photometric uncertainty. {\tt AGNBoost} can also provide feature importance estimates with SHAP plots to understand model predictions. From these SHAP plots we found evidence that {\tt AGNBoost} learned to identify PAH features: blue $S_{21}/S_{7.7}$, $S_{21}/S_{10}$, and $S_{15}/S_{7.7}$ colors pushed {\tt AGNBoost} predictions toward lower values of $\text{frac}_{\text{AGN}}$. Additionally, {\tt AGNBoost} provides simple and efficient handling of missing photometric observations through {\tt SGAIN} imputation, enabling $\text{frac}_{\text{AGN}}$ and redshift estimates for all sources.

The adaptable framework of {\tt AGNBoost} allows straightforward incorporation of additional photometric bands and derived quantities as desired. Training new models is simple with the built-in hyperparameter optimization functions of {\tt AGNBoost}, and can be performed on personal computers. It is also possible to extend {\tt AGNBoost} to estimate other parameters of interest such as star formation rates or stellar masses. The computational efficiency and scalability of {\tt AGNBoost} make it ideally suited for the new era of wide-sky surveys, where rapid identification of AGN candidates will be essential for effective target selection and follow-up observations.

The code for {\tt AGNBoost} is publicly available on GitHub at \url{https://github.com/hamblin-ku/AGNBoost}.

\newpage

\begin{acknowledgments}
We thank the anonymous referee for thoughtful comments that have helped to improve the clarity and impact of this work.  K.H. gratefully acknowledges support from a NASA/FINESST award, 80NSSC22K1594. We acknowledge support from program JWST-GO-03794.001. We are grateful to Médéric Boquien for assistance in diagnosing CIGALE configuration issues. The MEGA JWST data used in this paper were obtained from the Mikulski Archive for Space Telescopes (MAST) at the Space Telescope Science Institute.
\end{acknowledgments}

%


\software{AstroPy \citep{astropy:2013, astropy:2018, astropy:2022}, NumPy \citep{numpy}, Matplotlib \citep{matplotlib}, scikit-learn \citep{scikit-learn}, Scipy \citep{2020SciPy-NMeth}, TOPCAT \citep{topcat}, STILTS \citep{STILTS}, pandas \citep{reback2020pandas}, XGBoostLSS \citep{marz_xgboostlss}, Optuna \citep{optuna}, SGAIN \citep{SGAIN_paper}, SHAP \citep{shap_tree}
          }
\newpage
\clearpage



\restartappendixnumbering
\renewcommand{\theHfigure}{\thesection.\arabic{figure}}
\renewcommand{\theHtable}{\thesection.\arabic{table}}
\appendix

\section{Performance on galaxies outside of the {\tt CIGALE} color space} \label{appendix:colorspace_effects}

\begin{figure*}[htb!]
\plottwo{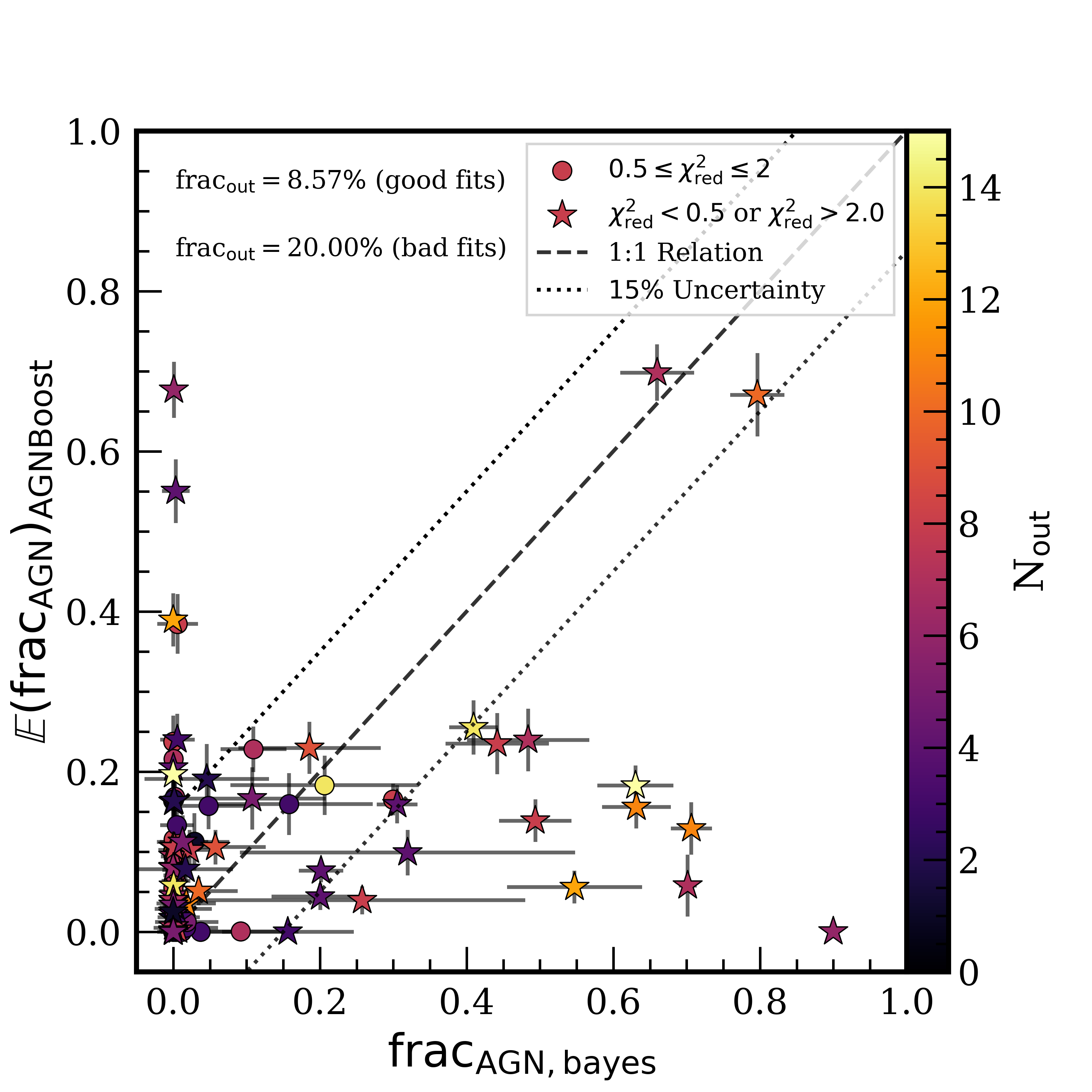}{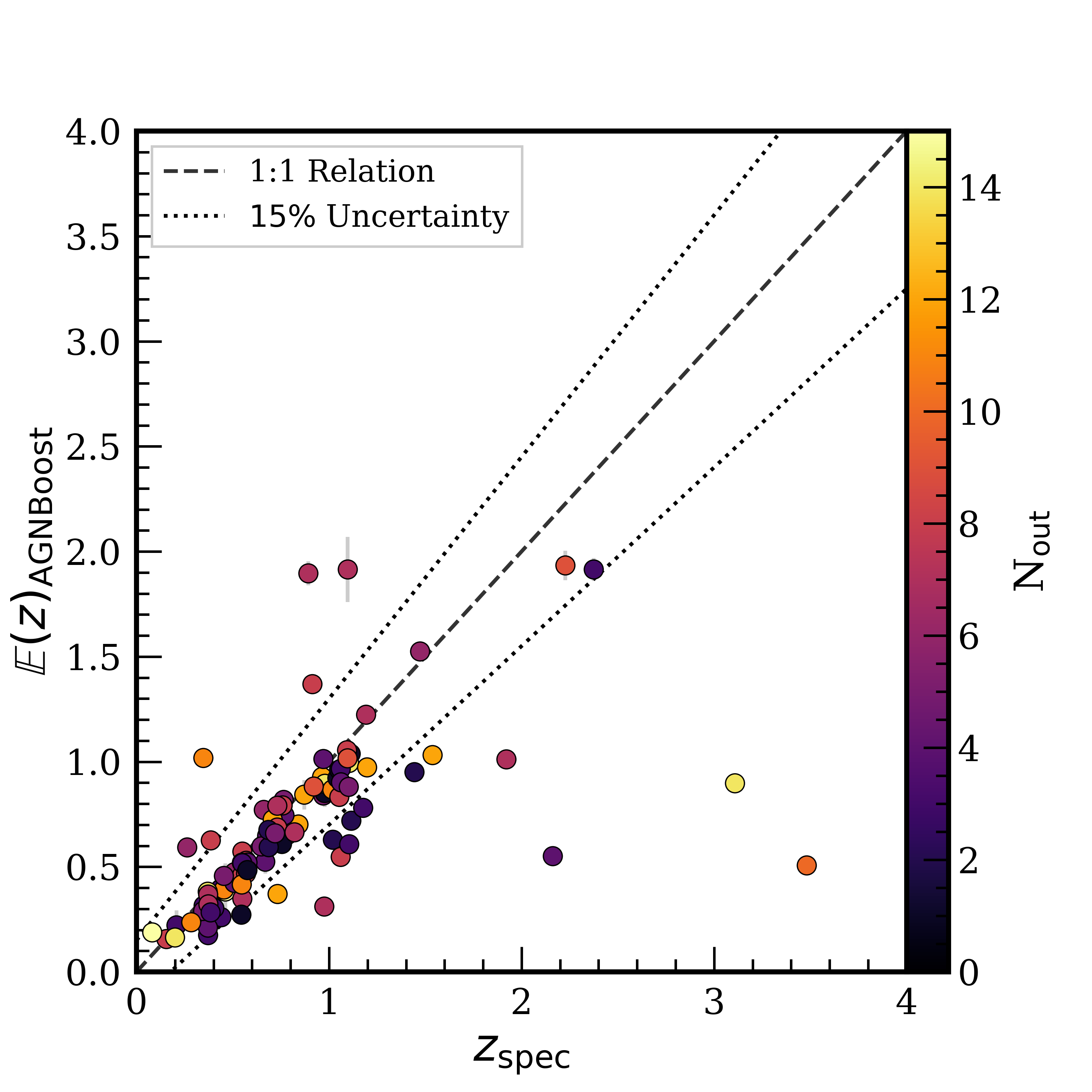}
\caption{Expectation values of {\tt AGNBoost} predictions for MEGA sources, colored by the number of MIRI color combinations for which they fall outside the {\tt CIGALE} color space (as shown in Figure \ref{fig:color_compare}). Left: {\tt AGNBoost} $\text{frac}_{\text{AGN}}$ estimates compared to {\tt CIGALE} Bayesian $\text{frac}_{\text{AGN}}$ estimates. Circles indicate sources with good {\tt CIGALE} fits ($0.5 < \chi^2_{\text{red}} < 2$) and stars indicate poor fits ($\chi^2_{\text{red}} < 0.5$ or $\chi^2_{\text{red}} > 2$). The $15\%$ outlier fractions are $\sim9\%$ and $20\%$ for he good and poor {\tt CIGALE} fits, respectively. Right: {\tt AGNBoost} photometric redshifts compared to spectroscopic redshifts for sources with available spectroscopy. Neither model shows systematic performance degradation as a function of the number of times sources lie outside the {\tt CIGALE} color space.
\label{fig:outside_of_colorspace}}
\end{figure*}

We test whether {\tt AGNBoost} performance is affected for sources that fall outside the {\tt CIGALE} MIRI color space. Figure \ref{fig:outside_of_colorspace} examines {\tt AGNBoost} performance on MEGA sources, colored by the number of MIRI color combinations for which they fall outside the {\tt CIGALE} training distribution. The left panel compares {\tt AGNBoost} $\text{frac}_{\text{AGN}}$ estimates to {\tt CIGALE} Bayesian $\text{frac}_{\text{AGN}}$ estimates for sources with both good ($0.5 < \chi^2_{\text{red}} < 2$) and poor ($\chi^2_{\text{red}} < 0.5$ or $\chi^2_{\text{red}} > 2$) {\tt CIGALE} fits, while the right panel compares {\tt AGNBoost} photometric redshifts to spectroscopic redshifts for sources with available spectroscopy. Neither model shows strong performance trends as a function of how often sources lie outside the {\tt CIGALE} color space, suggesting that {\tt AGNBoost} generalizes reasonably well even to sources not fully represented in the training distribution.

\section{Predicted Uncertainties for mock data}
\begin{figure*}[!htb]
\plottwo{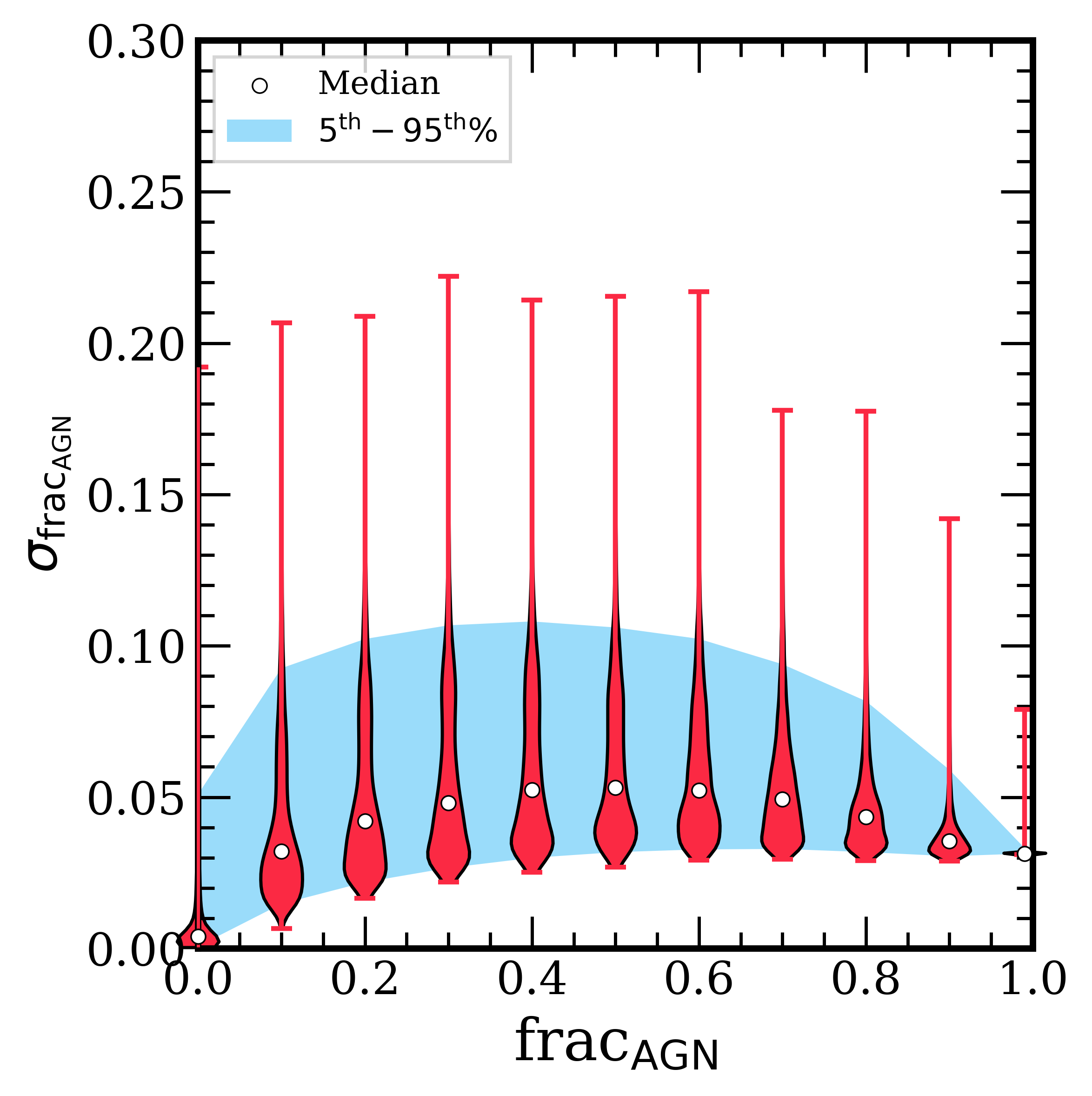}{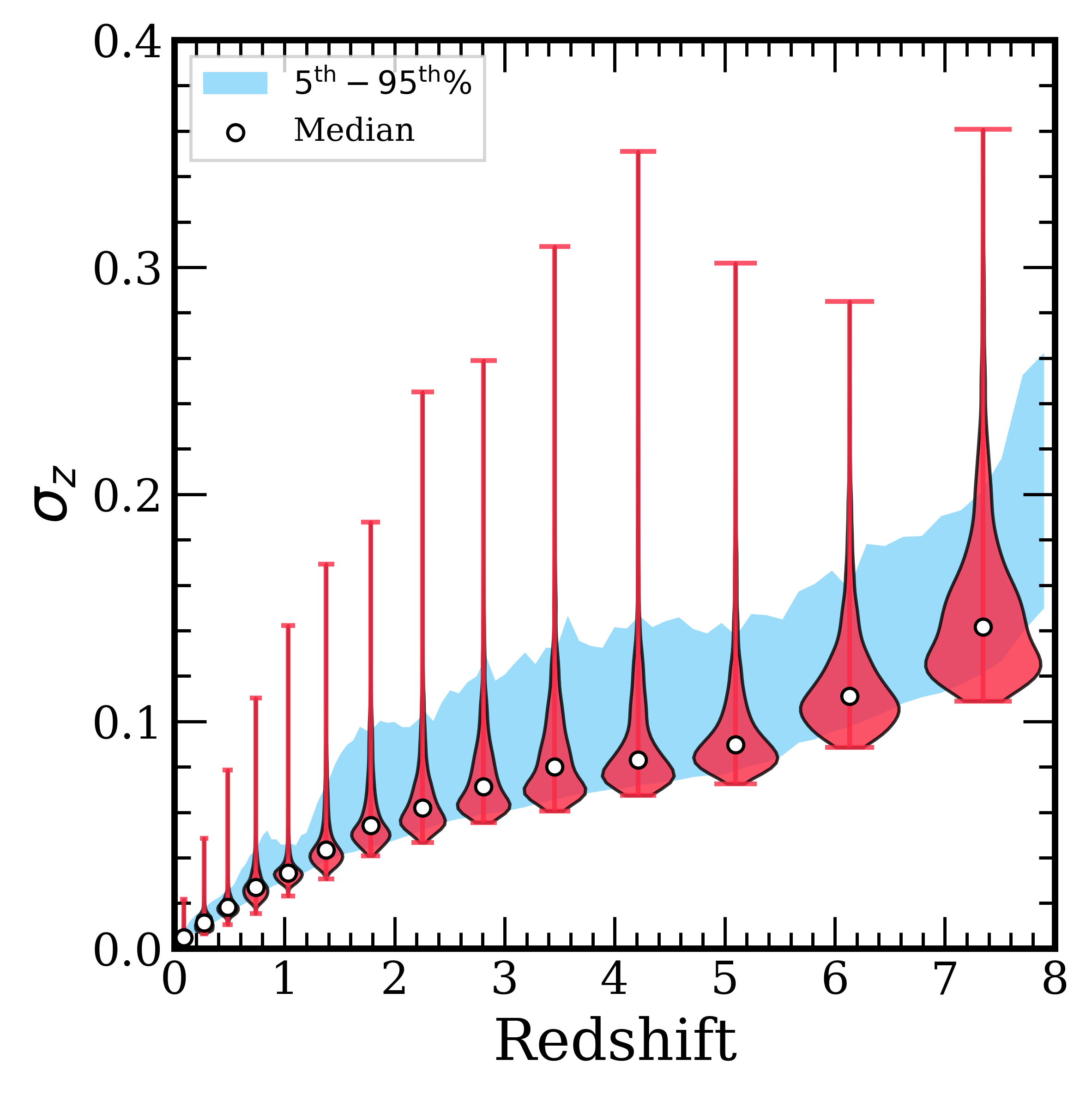}
\caption{Predicted $\text{frac}_{\text{AGN}}$ uncertainty from {\tt AGNBoost}'s conditional zero-inflated beta distribution as a function of true $\text{frac}_{\text{AGN}}$ (left) and predicted photometric redshift uncertainty from  {\tt AGNBoost}'s beta distirbution as a function of true redshift, both for the mock {\tt CIGALE} test set. Violin plot elements are the same as in Figure \ref{fig:sim_perf}. Predicted $\text{frac}_{\text{AGN}}$ uncertainties are lowest when AGN dominate the SED (high $\text{frac}_{\text{AGN}}$) and photometric redshift uncertainties increase with redshift as diagnostic features shift outside the JWST wavelength coverage.\label{fig:sim_perf_unc}}

\plottwo{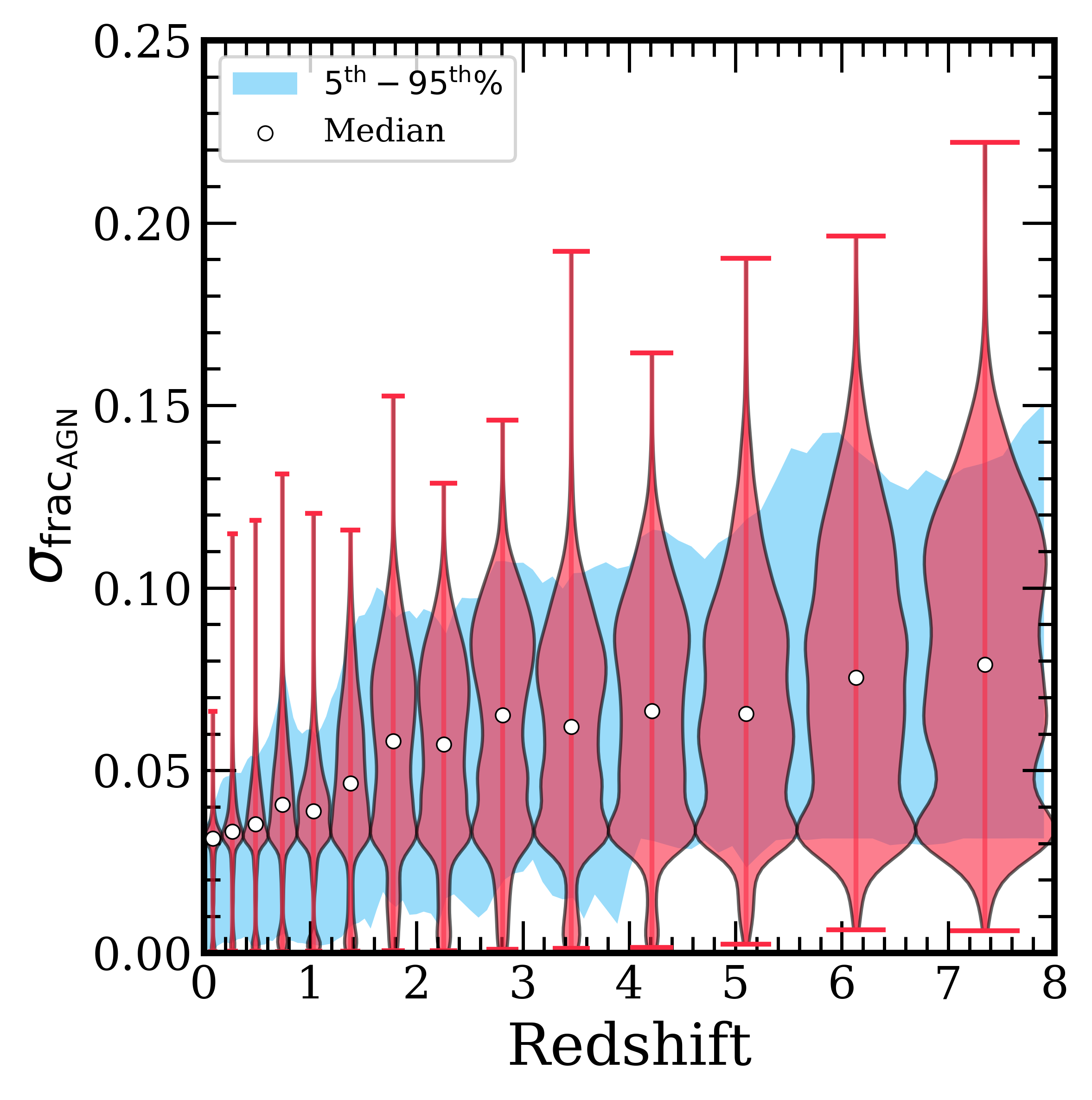}{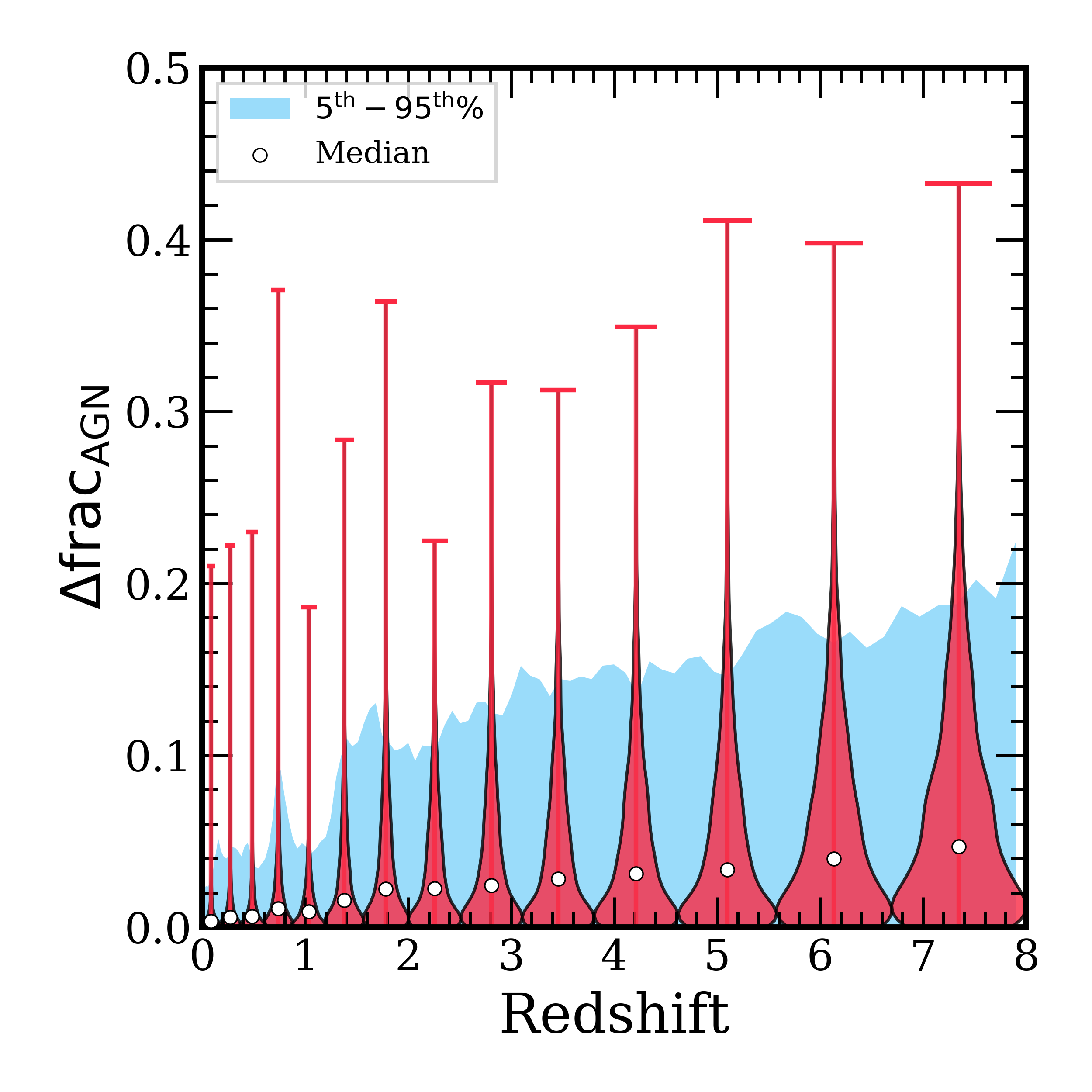}
\caption{Predicted $\text{frac}_{\text{AGN}}$ uncertainty from {\tt AGNBoost}'s conditional zero-inflated beta distribution as a function of redshift (left) and then actual predictive error for $\text{frac}_{\text{AGN}}$ (defined as $|\text{frac}_{\text{AGN}}^{\text{true}} - \text{frac}_{\text{AGN}}^{\text{pred}}|$) as a function of redshift for the mock {\tt CIGALE} test set (left). Violin plot elements are the same as in Figure \ref{fig:sim_perf}. While the predictive error follows the same trend as the predicted uncertainties, the outliers extend roughly twice as far, indicating that {\tt AGNBoost} provides reasonable uncertainty estimates on average but underestimates the tail of the error distribution.
\label{fig:dfagn_vs_redshift}}
\end{figure*}

The violin plots presented in the main text (e.g., Figure \ref{fig:sim_perf}) show distributions of {\tt AGNBoost} expectation values as functions of true parameter values. Here we examine the distributions of predicted uncertainties to assess whether {\tt AGNBoost} properly quantifies its estimation uncertainty. These uncertainties are derived from the conditional distributions predicted by {\tt XGBoostLSS}: a zero-inflated beta distribution for $\text{frac}_{\text{AGN}}$ and a beta distribution for the transformed redshift.

Figure \ref{fig:sim_perf_unc} shows the predicted $\text{frac}_{\text{AGN}}$ uncertainty as a function of true $\text{frac}_{\text{AGN}}$ (left) and the predicted photometric redshift uncertainty as a function of redshift (right) for the mock {\tt CIGALE} test set. The left panel shows that predicted uncertainties are lowest when AGN dominate the SED (high $\text{frac}_{\text{AGN}}$), which is expected since strong AGN are easier to identify from their distinct mid-IR signatures. The right panel shows that predicted photometric redshift uncertainty increases with redshift as diagnostic features shift outside the JWST wavelength coverage.

Figure \ref{fig:dfagn_vs_redshift} shows the predicted $\text{frac}_{\text{AGN}}$ uncertainty as a function (left) and the actual predictive error (i.e., $|\text{frac}_{\text{AGN}}^{\text{true}} - \text{frac}_{\text{AGN}}^{\text{pred}}|$ from Figure \ref{fig:sim_perf}) as a function of redshift. The {\tt AGNBoost} predicted uncertainty increases with redshift, reflecting both increased degeneracies between AGN and star-forming galaxy colors and the redshifting of diagnostic spectral features outside JWST/NIRCam+MIRI wavelength coverage. The actual predictive error follows the same trend as the predicted uncertainty, increasing with redshift. However, the outliers of the actual predictive error extend roughly twice as far as the predicted uncertainties shown in Figure \ref{fig:sim_perf_unc}. This indicates that while {\tt AGNBoost} provides reasonable uncertainty estimates on average, it does not fully capture cases where predictions are notably poor, underestimating the tail of the error distribution.

\section{Wavelength dependent effects of missing photometric bands} \label{appendix:lambda_missing_effects}

\begin{figure*}[!htb]
\epsscale{1.1}
\plotone{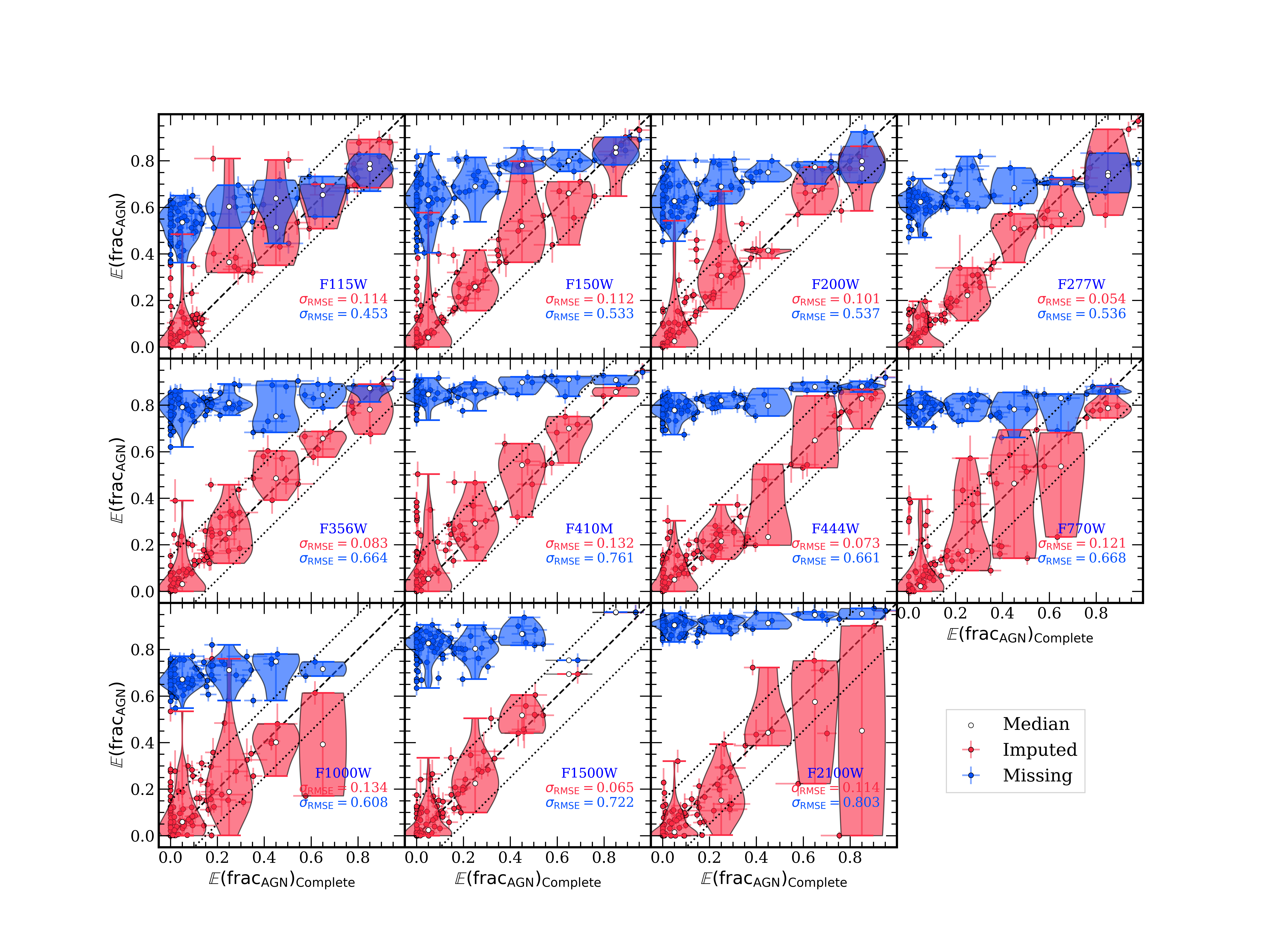}
\caption{ Comparison of {\tt AGNBoost} $\text{frac}_{\text{AGN}}$ predictions with missing photometry (blue) and SGAIN-imputed photometry (red), and for a sample of $100$ MEGA galaxies. Each panel shows performance when one specific NIRCam or MIRI band is absent. Violin plot elements are the same as in Figure \ref{fig:sim_perf}. Missing bands cause {\tt AGNBoost} to systematically overestimate $\text{frac}_{\text{AGN}}$ regardless of wavelength, while SGAIN imputation recovers accurate estimates across all bands, though with greater scatter at longer wavelengths\label{fig:fagn_imput_all}}
\end{figure*}

Figure \ref{fig:fagn_imput} in the main text demonstrates that {\tt AGNBoost} systematically overestimates $\text{frac}_{\text{AGN}}$ when MIRI bands are randomly missing and not replaced with statistical imputation. Here we investigate whether this bias depends on which specific photometric band is absent, or whether the effect is independent of wavelength.

Figure \ref{fig:fagn_imput_all} compares {\tt AGNBoost} $\text{frac}_{\text{AGN}}$ predictions when each individual NIRCam and MIRI band is missing (without imputation) to predictions using SGAIN-imputed values and to predictions from complete photometry, for a sample of $100$ MEGA galaxies. Each panel shows results for a single missing band, systematically testing all 11 bands (7 NIRCam + 4 MIRI). When any photometric band is missing, {\tt AGNBoost} systematically overestimates $\text{frac}_{\text{AGN}}$, defaulting toward AGN classifications regardless of which band is absent. SGAIN imputation successfully recovers accurate $\text{frac}_{\text{AGN}}$ estimates in all cases, though with greater scatter for longer wavelength bands. 

These results indicate that the bias introduced by missing photometry is not wavelength-dependent, rather the absence of any band causes {\tt AGNBoost} to produce systematically biased predictions through its sparsity-aware split finding algorithm. The direction and magnitude of this bias appear consistent across all bands, highlighting the importance of either ensuring complete photometry or using imputation methods when applying {\tt AGNBoost} to incomplete datasets.

\section{Outlier {\tt CIGALE} SED fit comparisons} \label{app:ambiguous_seds}
In Section \ref{subsec:mega_results}, we identified 28 MEGA sources with significant disagreements between {\tt AGNBoost} and {\tt CIGALE} $\text{frac}_{\text{AGN}}$ estimates. These include 7 sources (purple stars in Figure \ref{fig:mega_agn_performance}) where {\tt CIGALE} identified AGN candidates ($\text{frac}_{\text{AGN}}^{\text{bayes}} > 0.3$) but {\tt AGNBoost} classified as SFGs ($\text{frac}_{\text{AGN}} \leq 0.3$), and 21 sources (blue squares in Figure \ref{fig:mega_agn_performance}) where {\tt AGNBoost} identified AGN candidates but {\tt CIGALE} classified as SFGs. To assess whether the available photometry could uniquely determine the presence of AGN emission for these sources, we refit them with {\tt CIGALE} both without an AGN component (forced SFG fit) and with a forced AGN component.

\begin{figure}[!htb]
\gridline{\fig{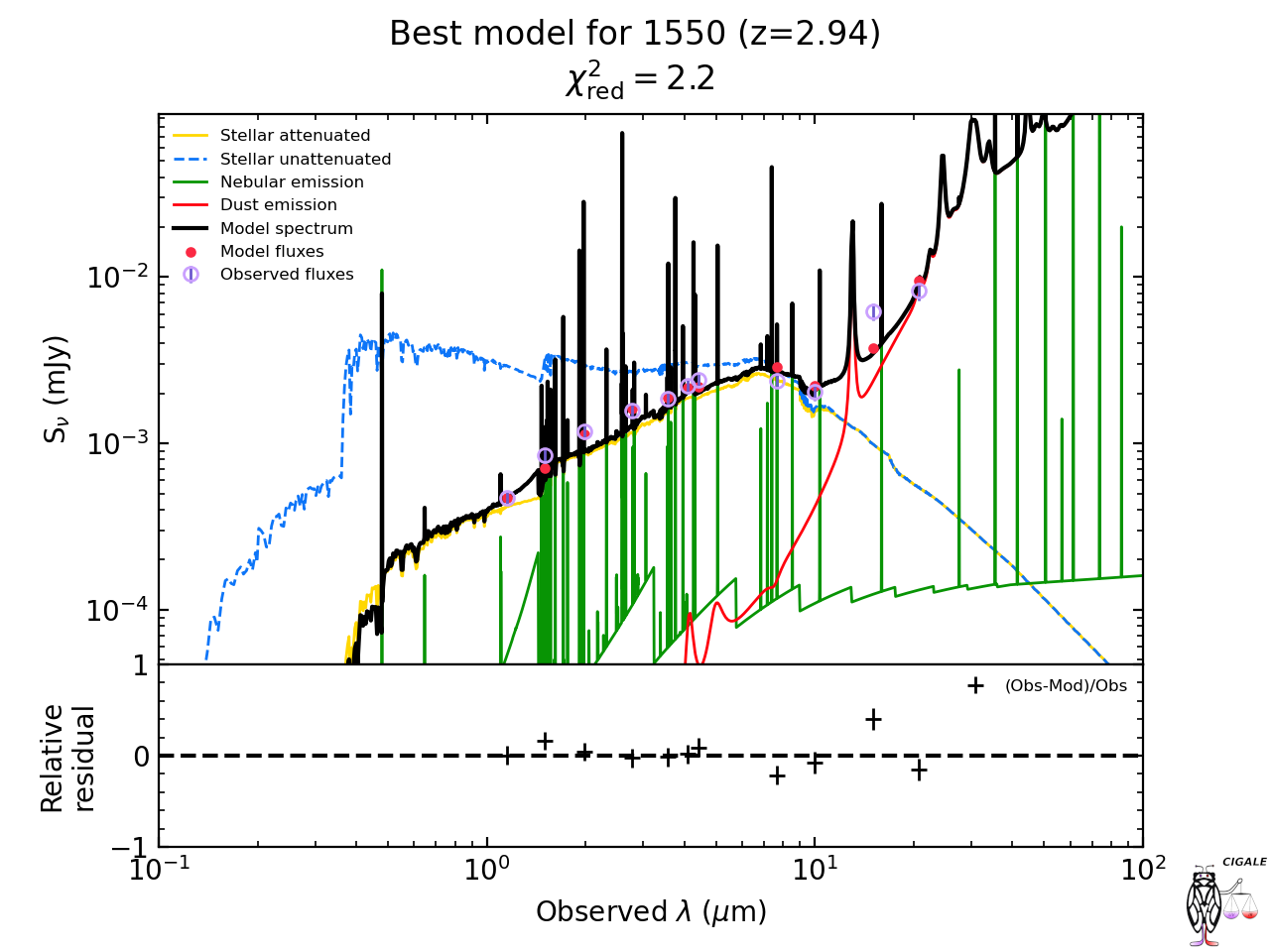}{0.45\textwidth}{(a)}
          \fig{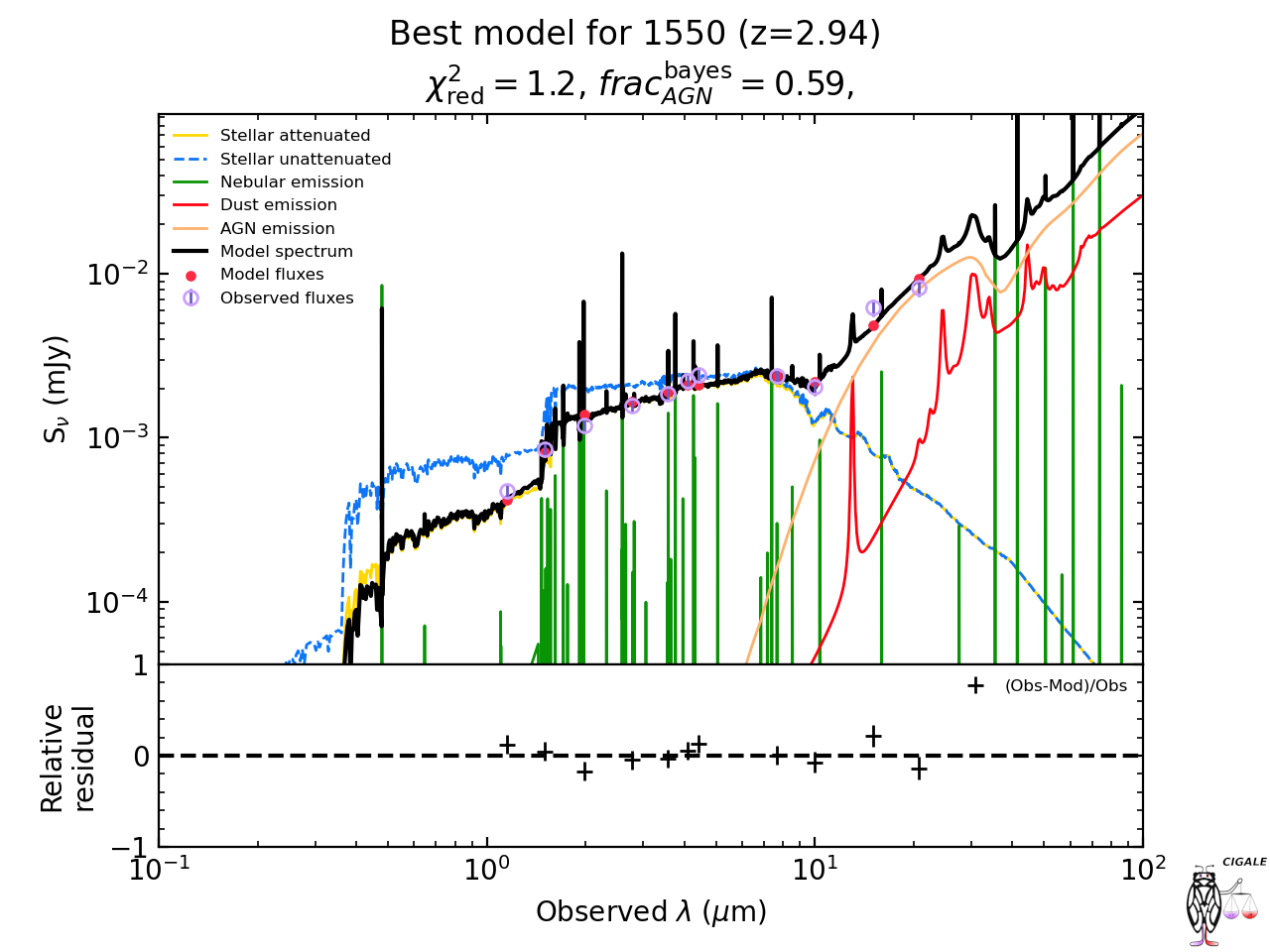}{0.45\textwidth}{(b)}}
\gridline{\fig{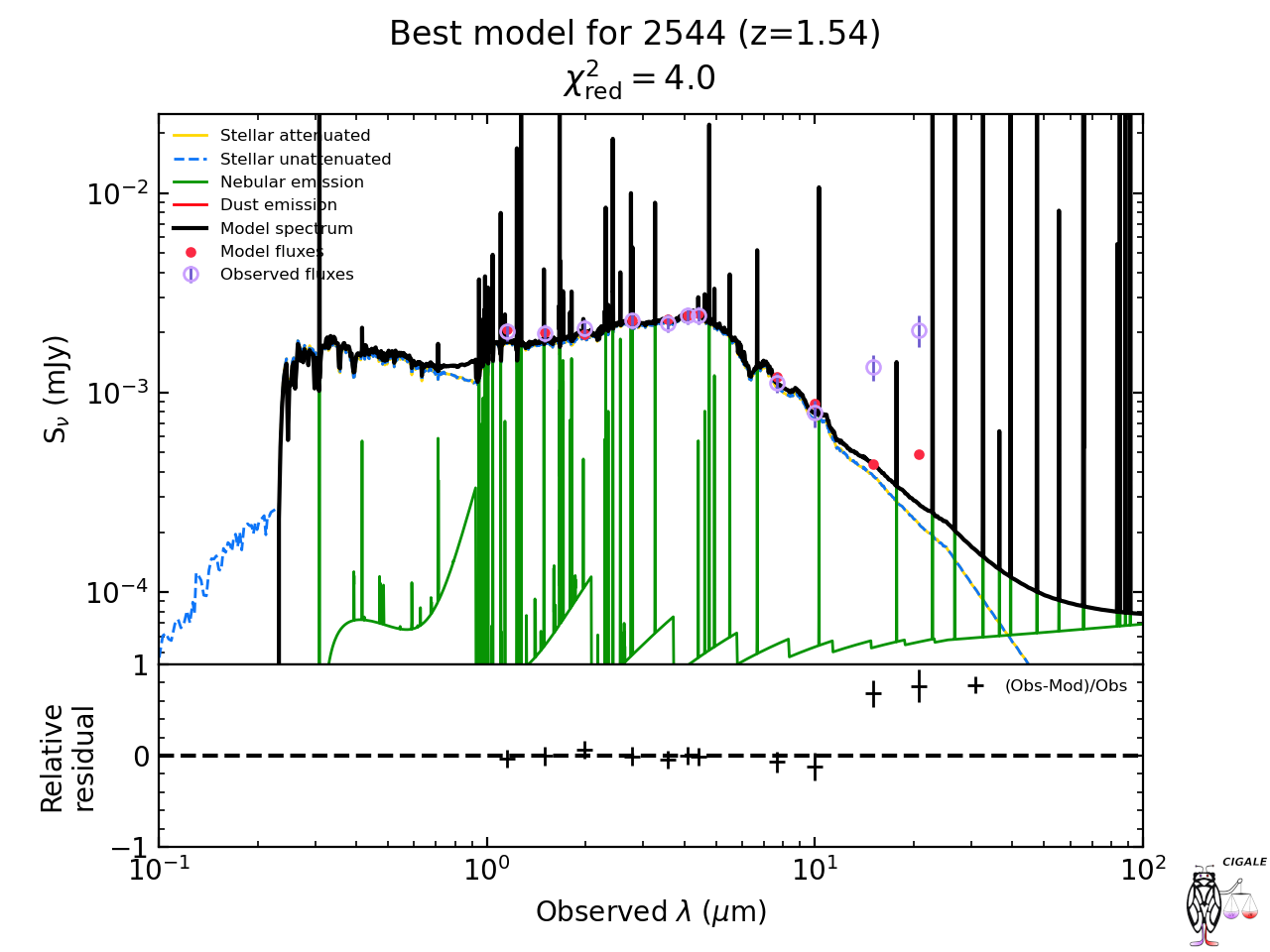}{0.45\textwidth}{(c)}
          \fig{2544_purplestar_AGN.png}{0.45\textwidth}{(d)}}
\caption{Example SED comparisons for two MEGA sources where {\tt CIGALE} identifies AGN candidates but {\tt AGNBoost} classifies as SFGs (purple stars in Figure \ref{fig:mega_agn_performance}). Left column: Forced SFG-only {\tt CIGALE} fits. Right column: Forced AGN {\tt CIGALE} fits. Top row: Source with nearly identical fit quality, demonstrating photometric ambiguity. Bottom row: Source with significant mid-IR residuals in the forced SFG fit, suggesting AGN emission is required. \label{fig:ambiguous_purplestars}}
\end{figure}

\begin{figure}[!htb]
\gridline{\fig{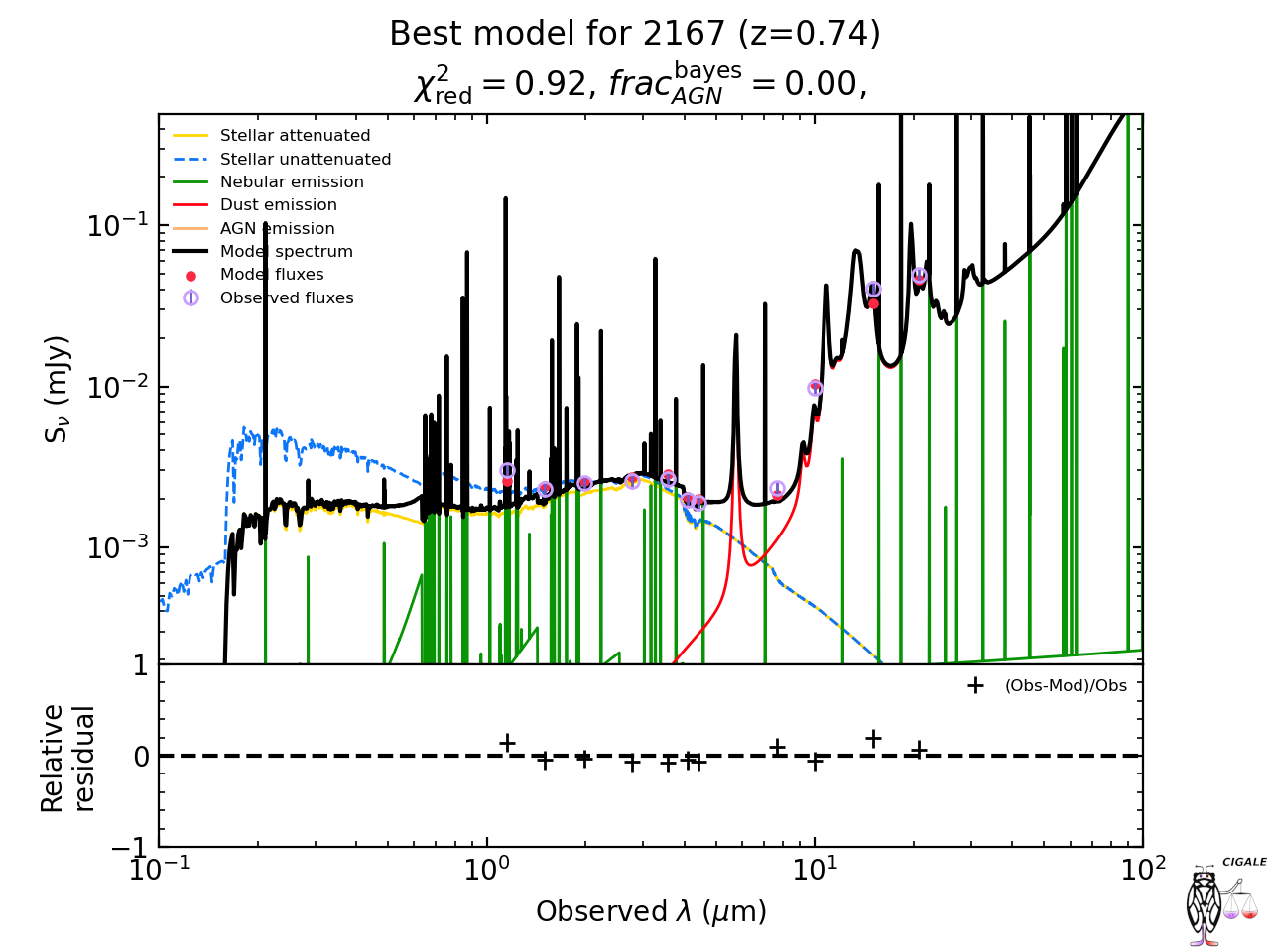}{0.45\textwidth}{(a)}
          \fig{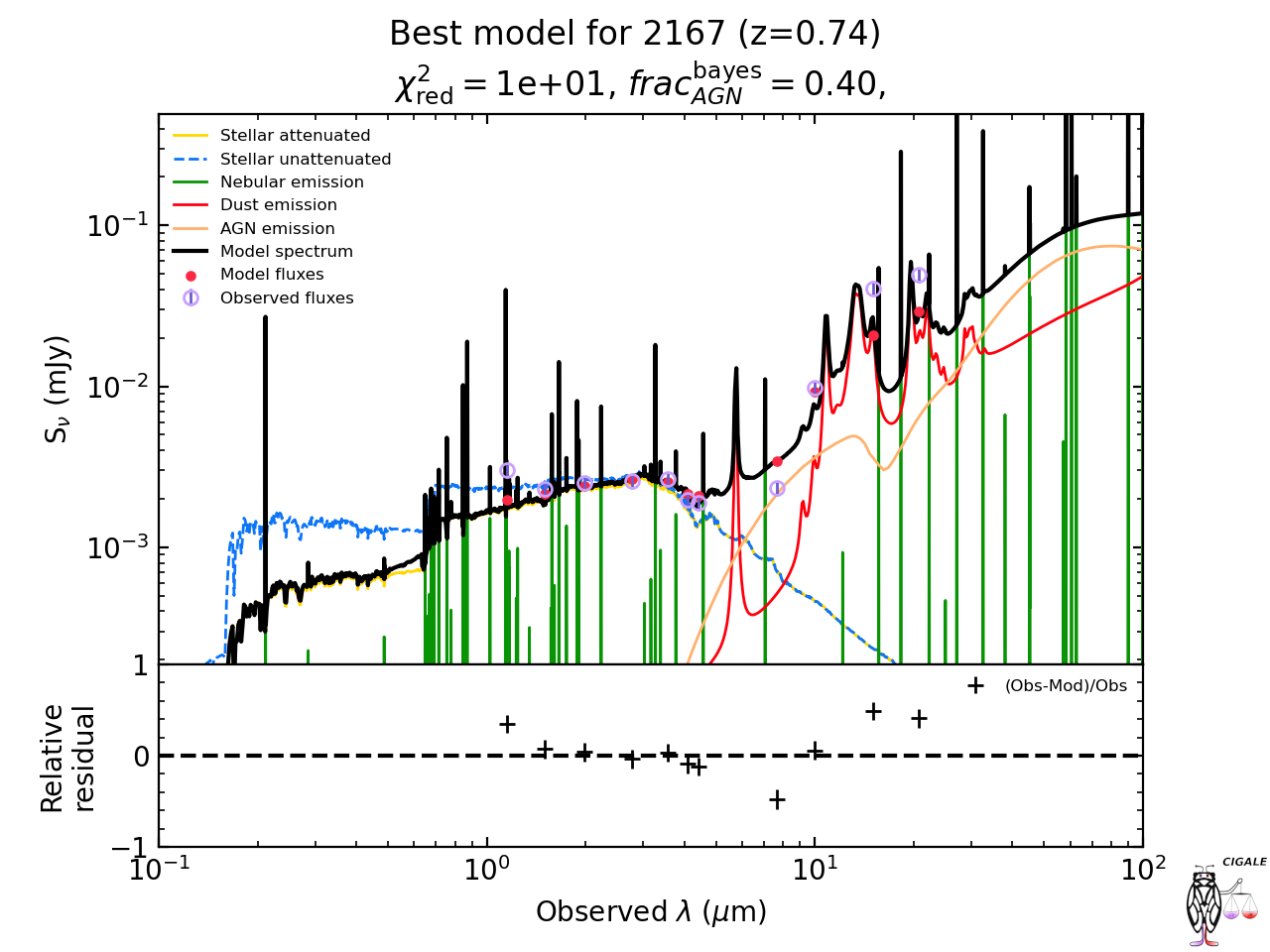}{0.45\textwidth}{(b)}
          }
\gridline{\fig{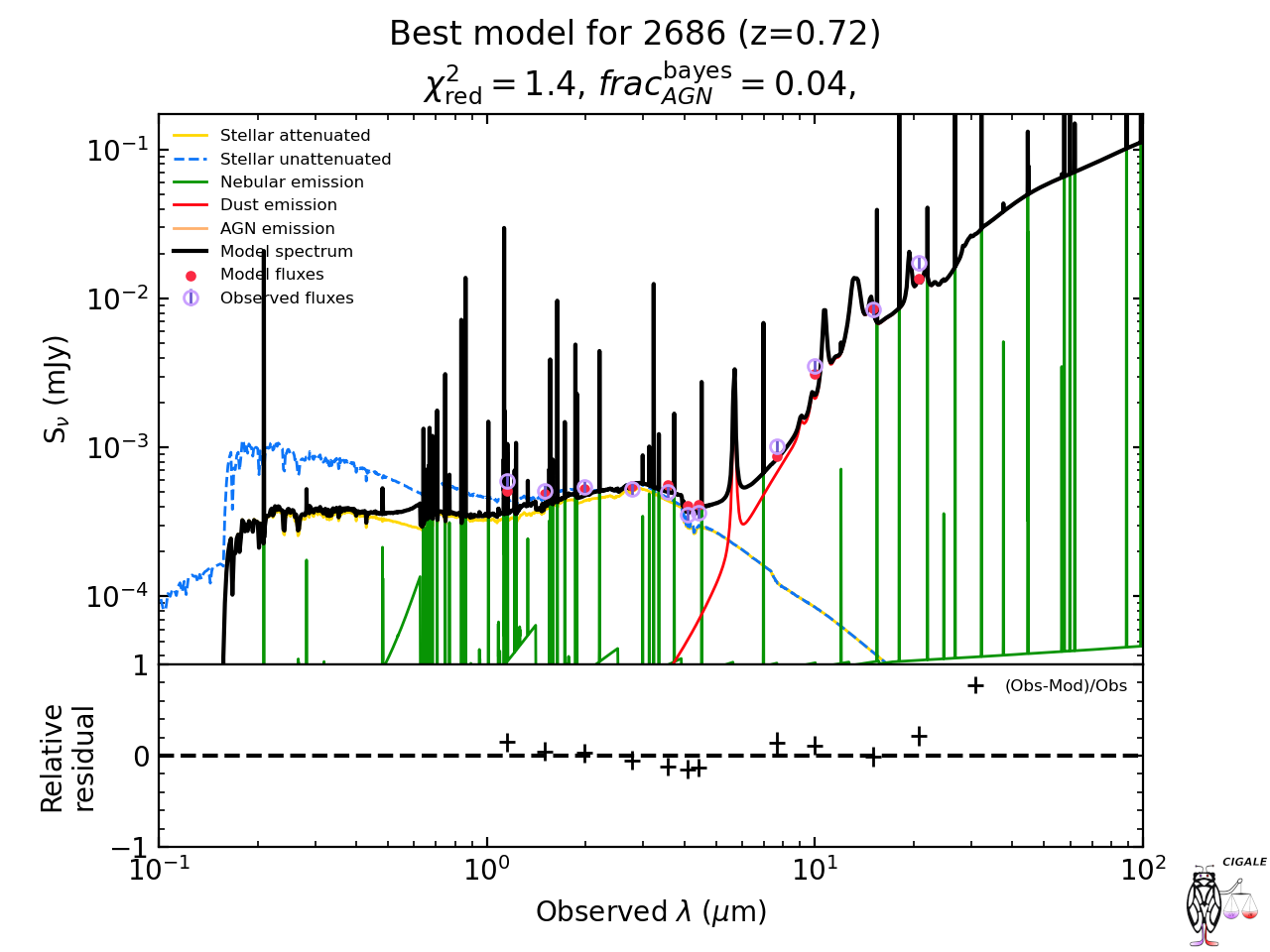}{0.45\textwidth}{(c)}
          \fig{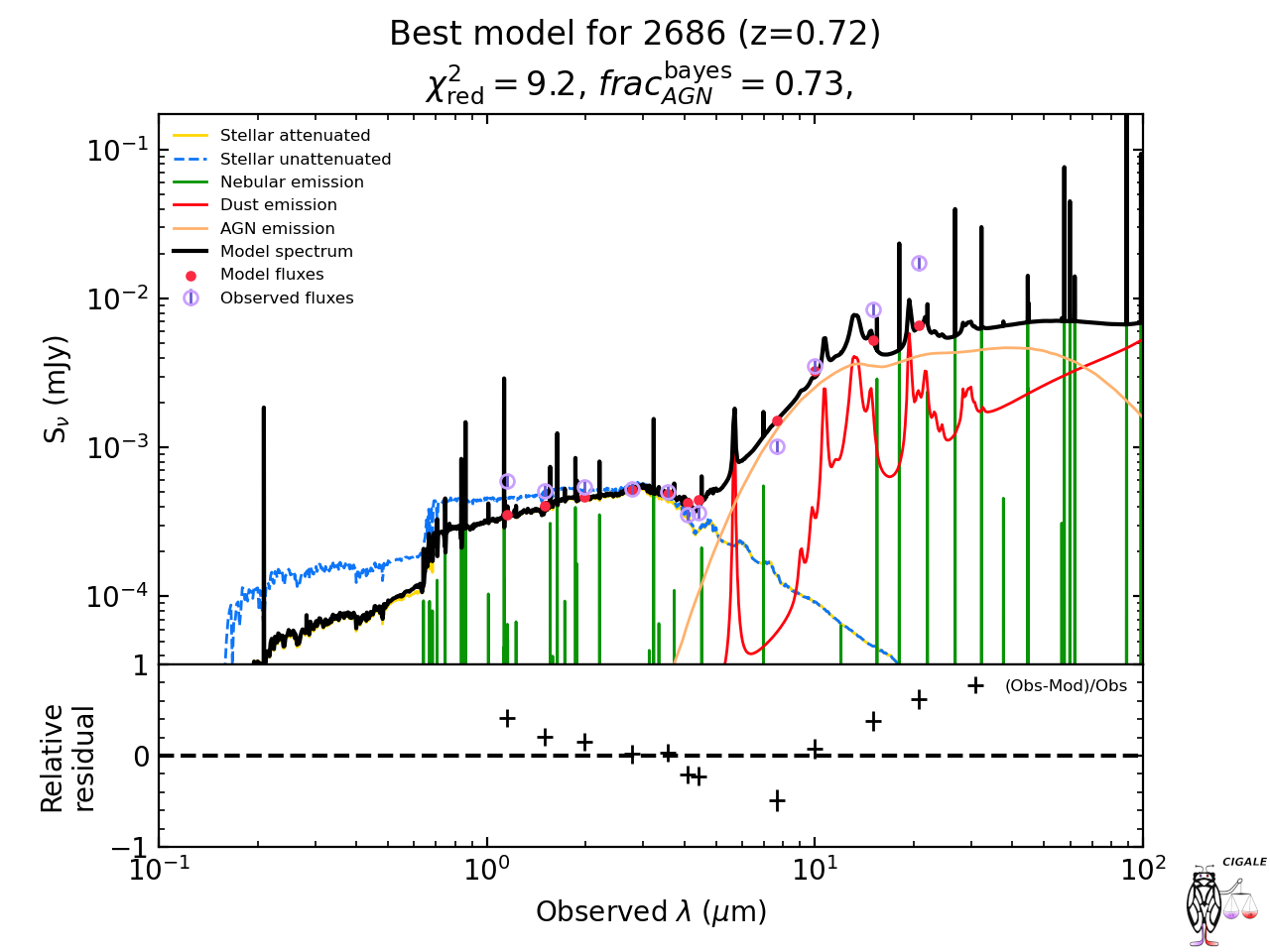}{0.45\textwidth}{(d)}}
\caption{Example SED comparisons for two MEGA sources where {\tt AGNBoost} identifies AGN candidates but {\tt CIGALE} classifies as SFGs (blue squares in Figure \ref{fig:mega_agn_performance}). Left column: Forced SFG-only {\tt CIGALE} fits. Right column: Forced AGN {\tt CIGALE} fits. Both sources show poor fits when an AGN component is forced, with strong mid-IR residuals indicating the forced AGN models do not accurately represent the observed photometry. \label{fig:bad_bluesquares}}
\end{figure}

\begin{figure}[!htb]
\gridline{\fig{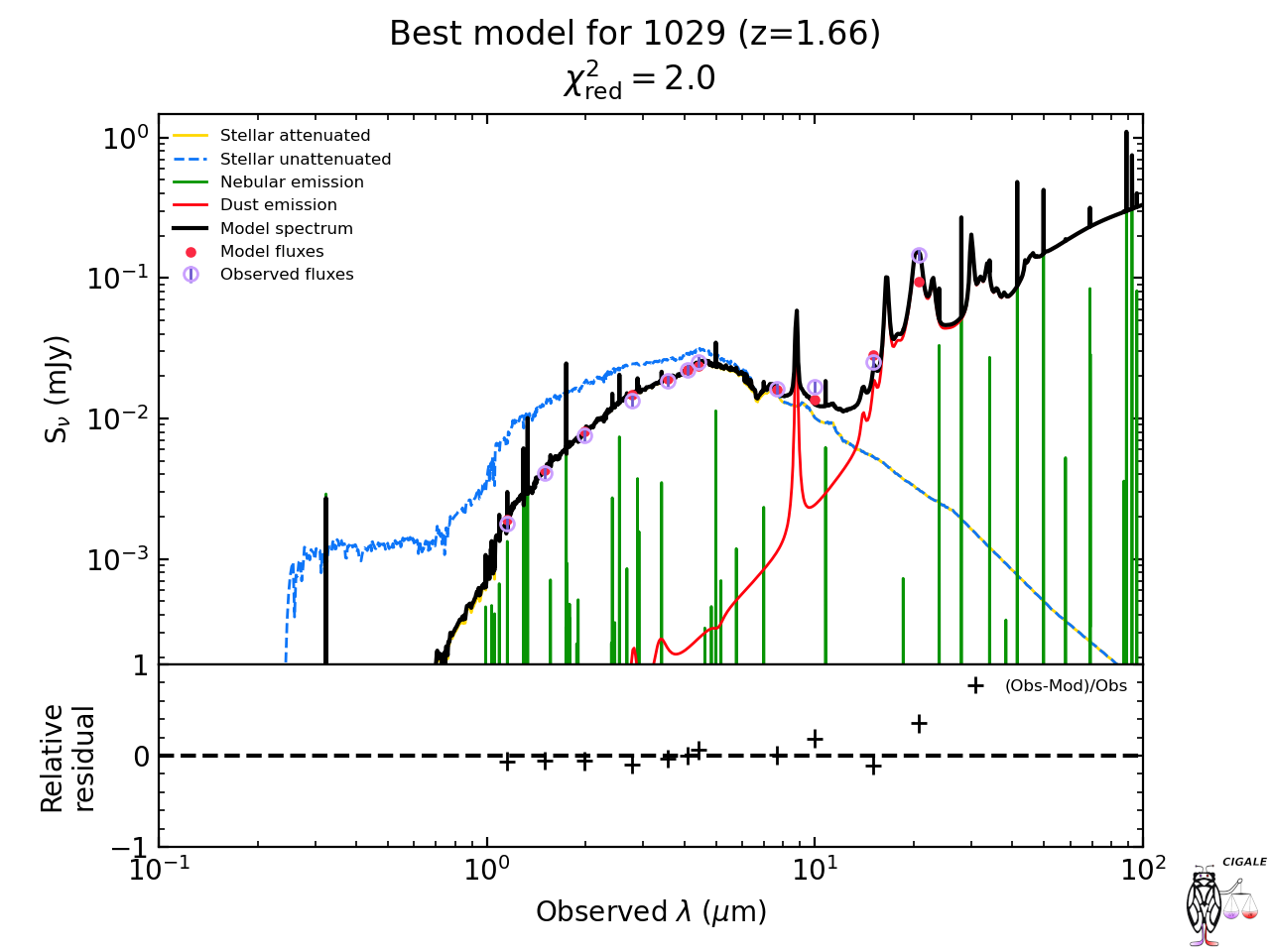}{0.45\textwidth}{(a)}
          \fig{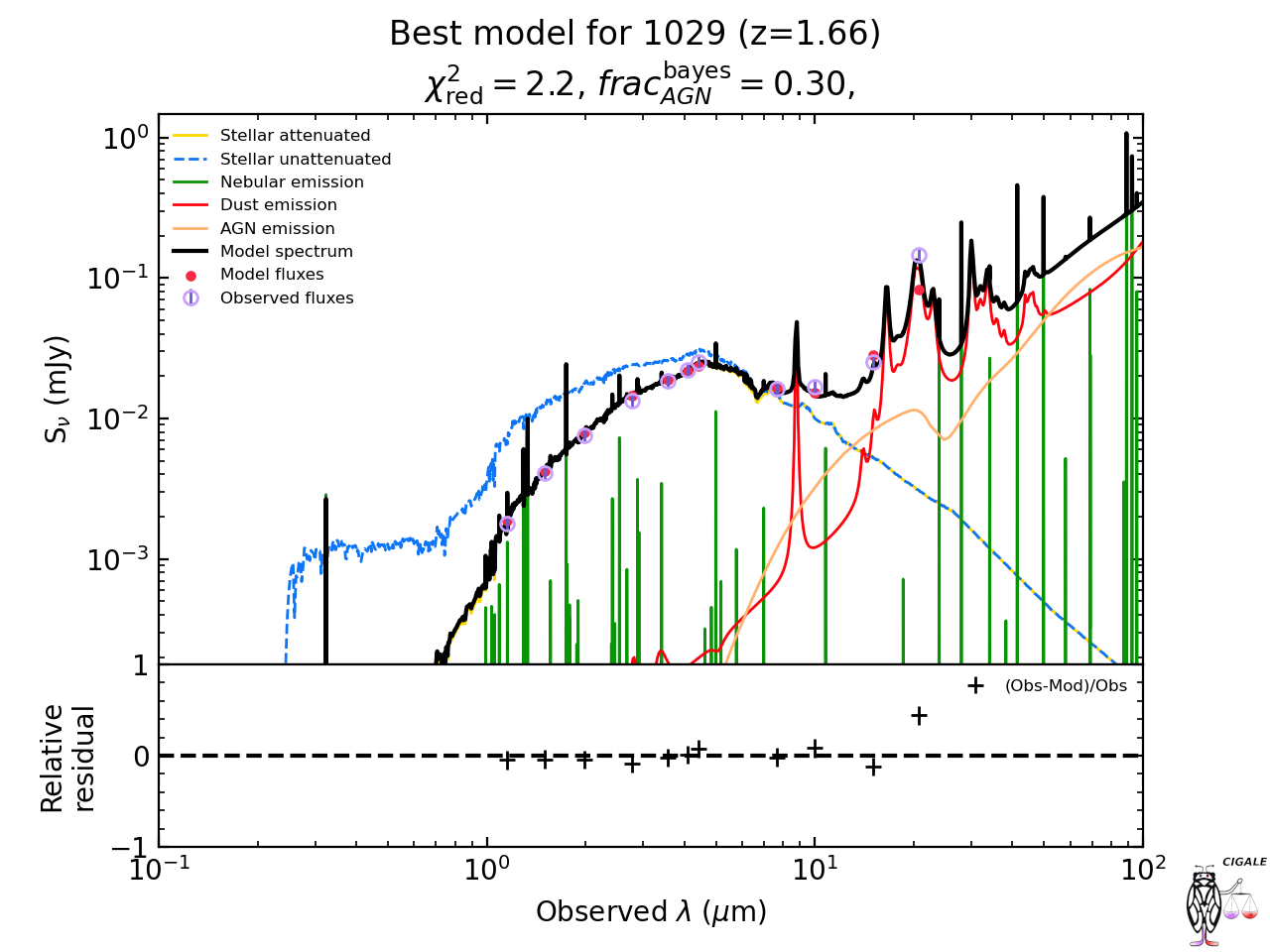}{0.45\textwidth}{(b)}
          }
\gridline{\fig{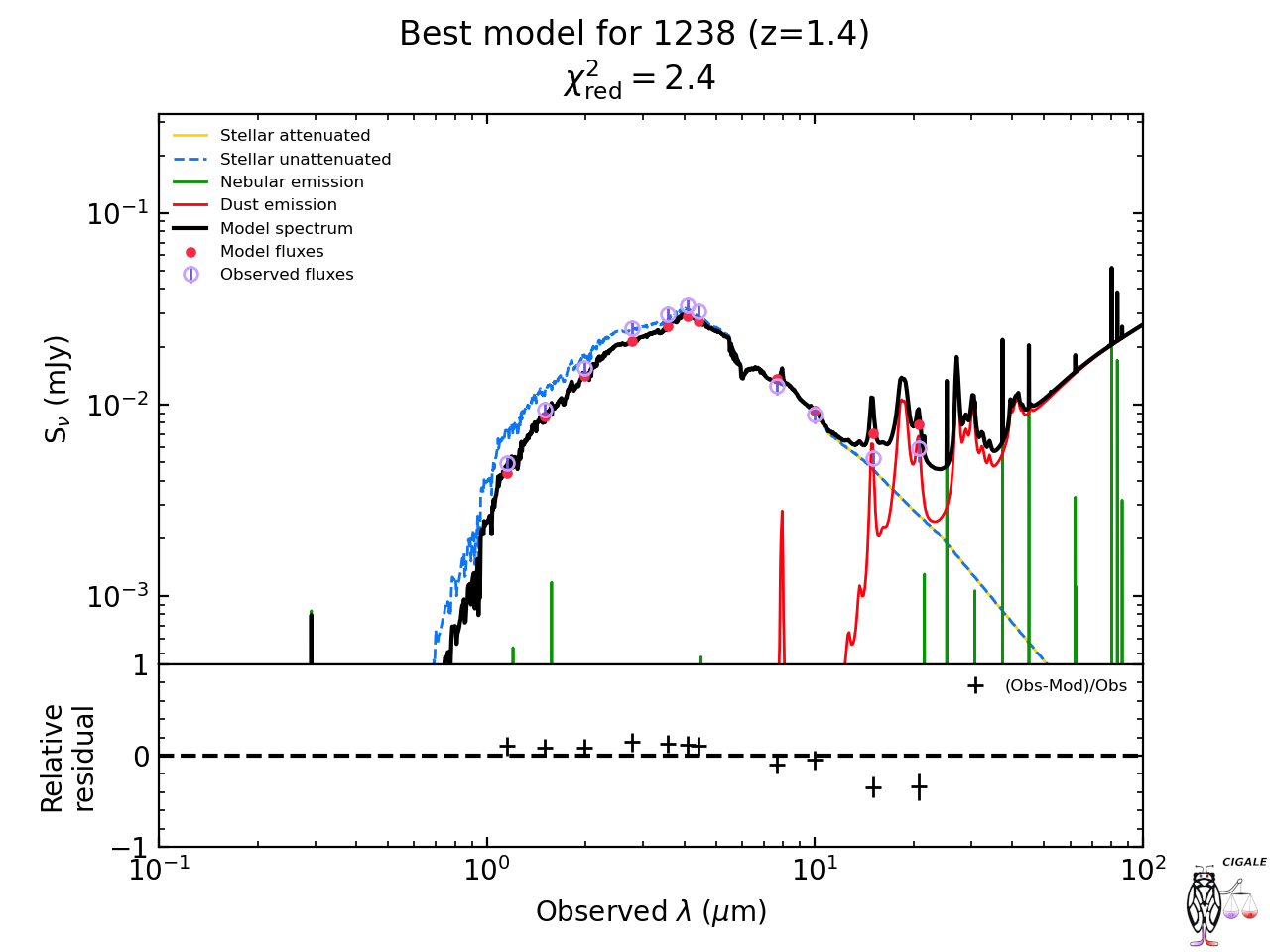}{0.45\textwidth}{(c)}
          \fig{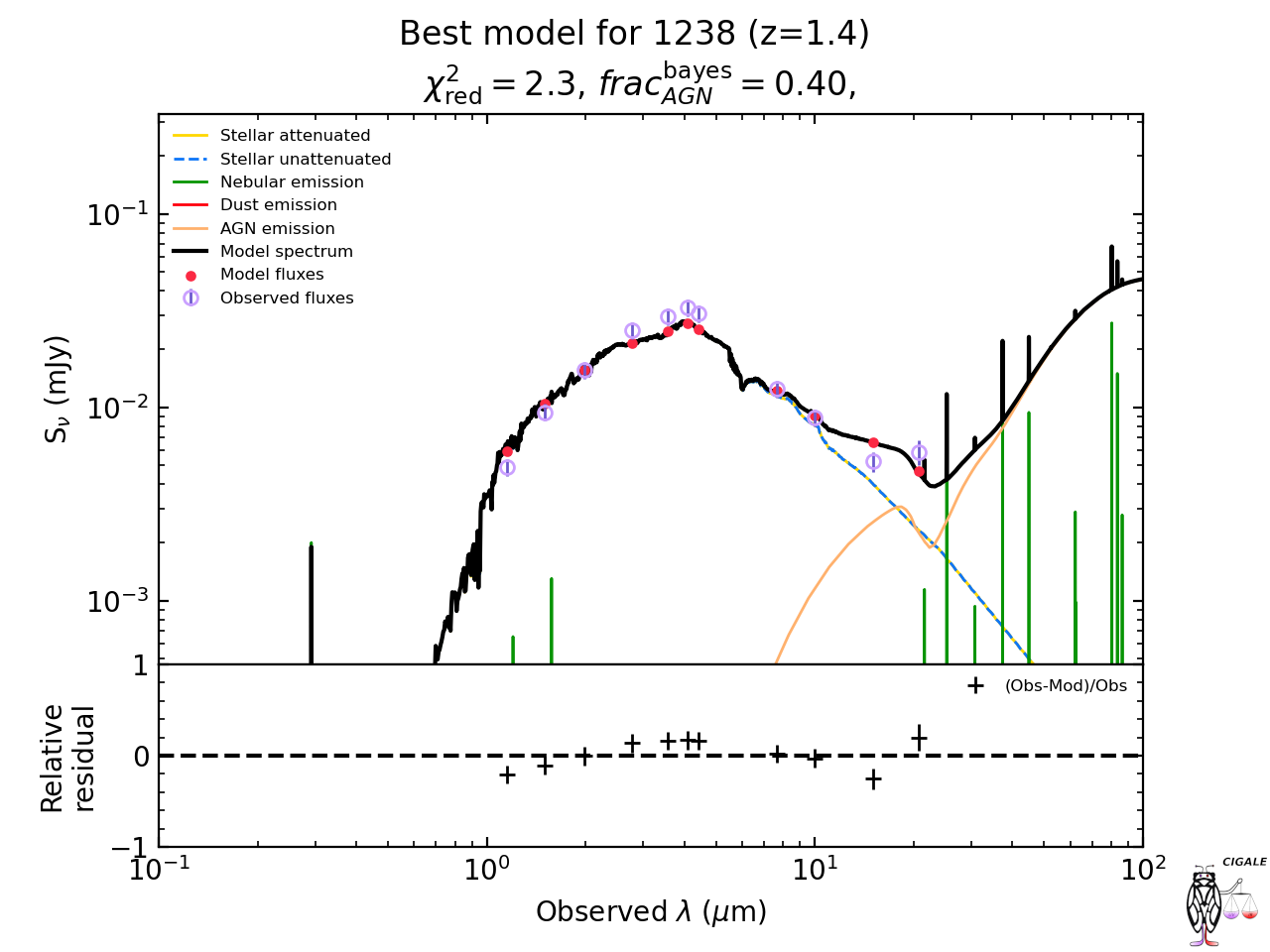}{0.45\textwidth}{(d)}}
\caption{Example SED comparisons for two MEGA sources where {\tt AGNBoost} identifies AGN candidates but {\tt CIGALE} classifies as SFGs (blue squares in Figure \ref{fig:mega_agn_performance}). Left column: Forced SFG-only {\tt CIGALE} fits. Right column: Forced AGN {\tt CIGALE} fits. Both sources show nearly identical fit quality with and without AGN components, demonstrating that the available photometry does not uniquely constrain the presence or strength of AGN emission. \label{fig:ambiguous_bluesquares}}
\end{figure}

Figures \ref{fig:ambiguous_purplestars}--\ref{fig:ambiguous_bluesquares} show example SEDs for six of these sources. Figure \ref{fig:ambiguous_purplestars} shows two sources where {\tt CIGALE} originally identified AGN candidates but {\tt AGNBoost} classified as SFGs. The top row demonstrates a case where both the forced SFG and forced AGN fits achieve nearly identical $\chi^2_{\text{red}}$ values, indicating that the available photometry does not uniquely constrain the presence of AGN emission. The bottom row shows a case where the forced SFG fit exhibits significant mid-IR residuals, suggesting that AGN emission is required to match the observed photometry.

Figure \ref{fig:bad_bluesquares} shows two sources where {\tt AGNBoost} identified AGN candidates but {\tt CIGALE} classified as SFGs. Both sources show poor fits when an AGN component is forced, with mid-IR residuals indicating that the forced AGN models do not accurately represent the observed SEDs.

Figure \ref{fig:ambiguous_bluesquares} shows two additional sources where {\tt AGNBoost} identified AGN candidates but {\tt CIGALE} classified as SFGs. Unlike the examples in Figure \ref{fig:bad_bluesquares}, both the forced SFG and forced AGN fits achieve comparable $\chi^2_{\text{red}}$ values, demonstrating that the available NIRCam+MIRI photometry does not provide sufficient information to uniquely determine whether AGN emission contributes to these sources' SEDs.

These examples illustrate the inherent ambiguity in photometric AGN identification for some sources, where limited wavelength coverage and photometric uncertainties prevent definitive classification.

\section{Effects of signal-to-noise on {\tt AGNBoost} outputs}
\label{appendix:snr_sensitivity}

Figure \ref{fig:sim_perf_witherr} in the main text shows {\tt AGNBoost} performance on {\tt CIGALE} mock data with realistic photometric uncertainties, applying an S/N $> 3$ cut in F770W, F1500W, and F2100W to match the selection criteria used for the MEGA sample. Here we assess how sensitive {\tt AGNBoost} performance is to the choice of S/N threshold.

\begin{figure*}[!htb]
\plottwo{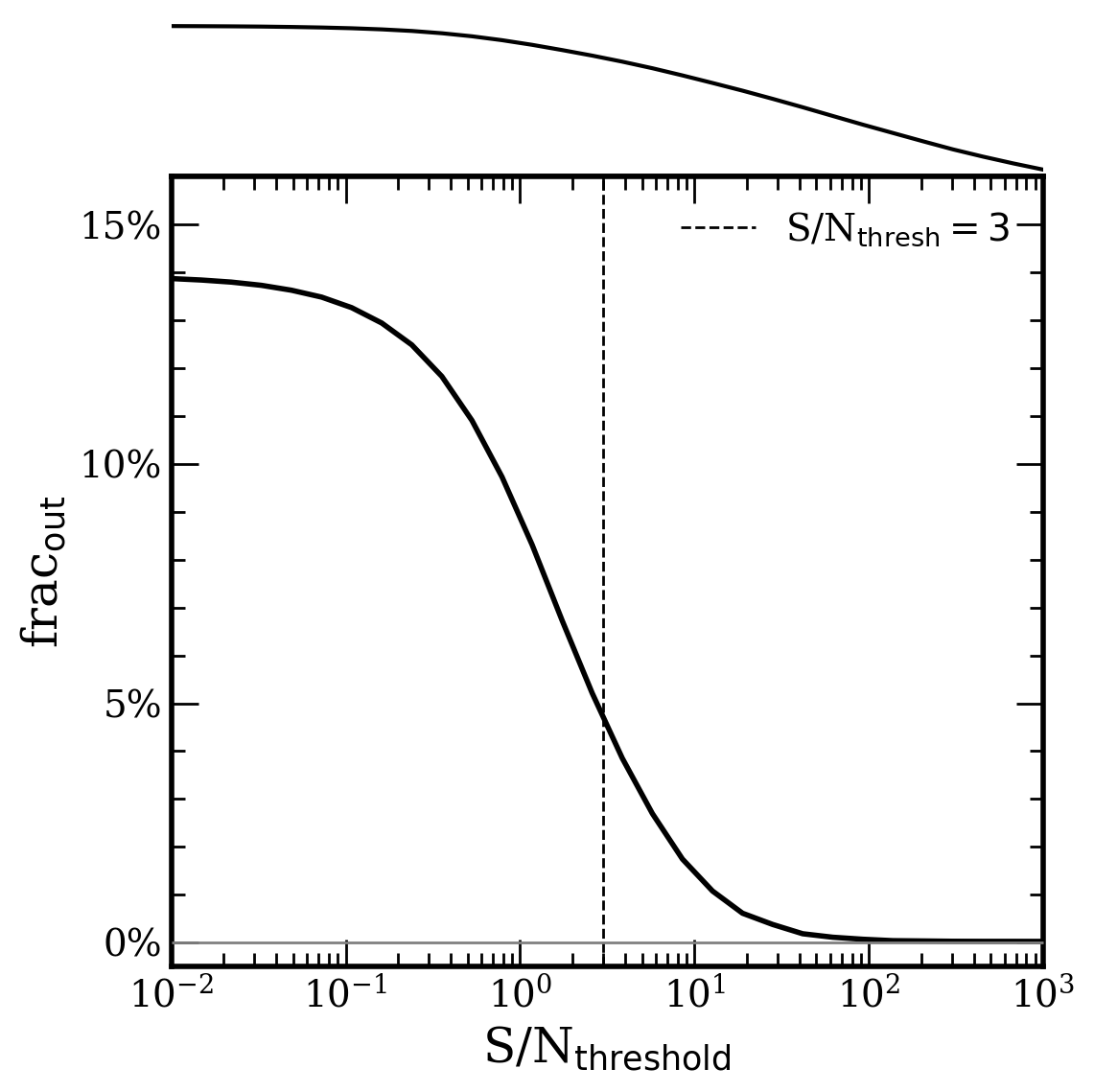}{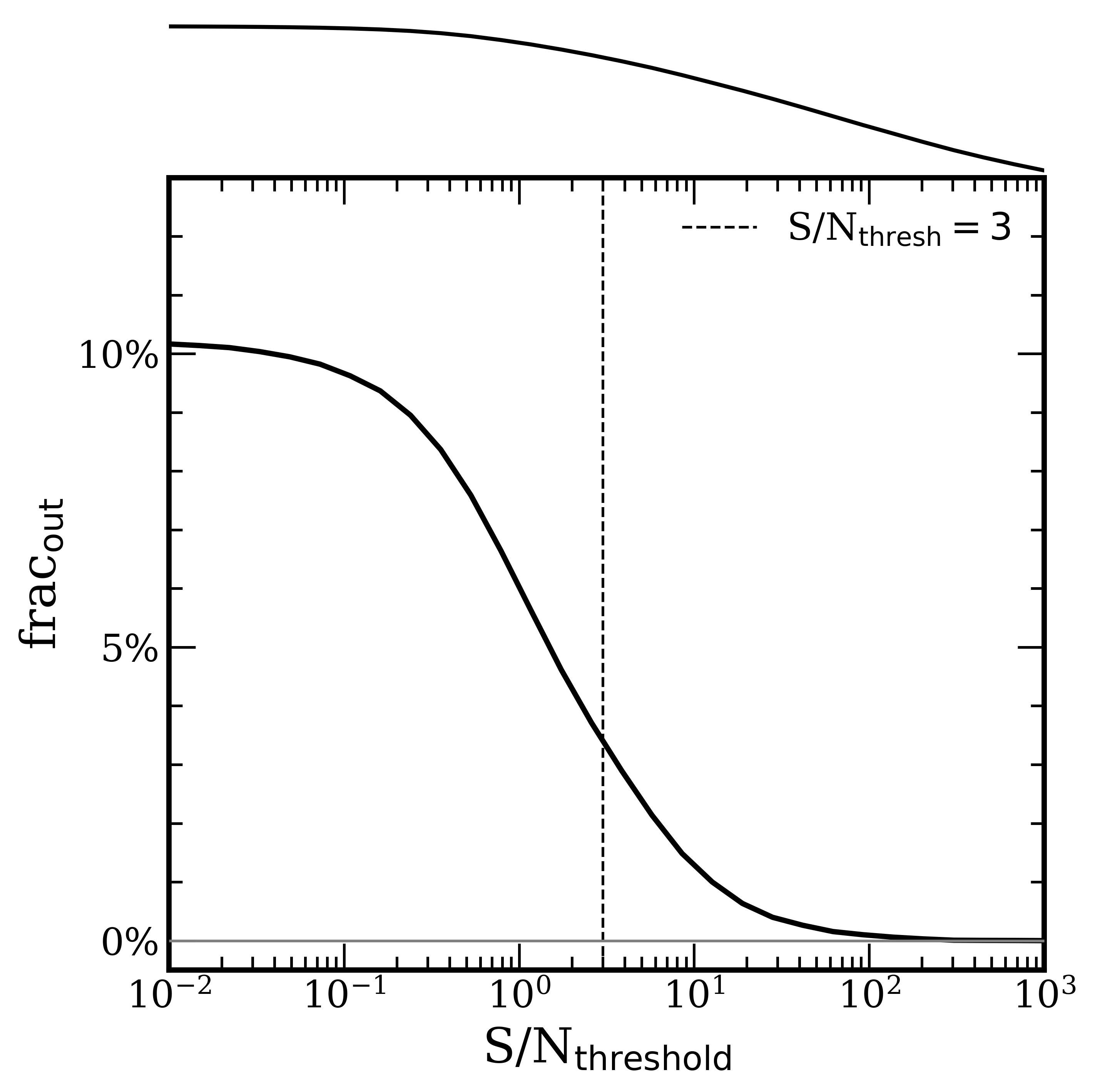}
\caption{Sensitivity of {\tt AGNBoost} performance to S/N threshold for $\text{frac}_{\text{AGN}}$ (left) and redshift (right) predictions. The marginal distributions along the top of each panel show sample sizes as a function of S/N threshold, demonstrating how stricter cuts reduce the number of sources. The main panels show the $15\%$ outlier fraction ($\text{frac}_{\text{out}}$, defined as predictions deviating by more than $15\%$ from the 1:1 relation) as a function of S/N threshold. Both models exhibit sigmoidal behavior, with outlier fractions decreasing from maxima of $\sim 14\%$ ($\text{frac}_{\text{AGN}}$) and $\sim 10\%$ (redshift) at low S/N to nearly zero by S/N $\sim 100$. The vertical dashed lines mark the S/N $> 3$ threshold adopted in our analysis, which lies near the inflection point of both curves.
\label{fig:snr_sensitivity}}
\end{figure*}

To quantify this sensitivity, we repeat the analysis of Figure \ref{fig:sim_perf_witherr} across a wide range of S/N thresholds from $10^{-2}$ to $10^3$. Figure \ref{fig:snr_sensitivity} shows the $15\%$ outlier fraction ($\text{frac}_{\text{out}}$) as a function of S/N threshold for both $\text{frac}_{\text{AGN}}$ (left) and redshift (right) predictions. The vertical dashed lines indicate the S/N $> 3$ threshold adopted in our analysis. The distributions along the top of each panel show how the sample size changes as a function of S/N threshold, with stricter cuts dramatically reducing the number of sources.

Both models exhibit sigmoidal behavior as a function of S/N threshold. At very low S/N thresholds (effectively no cut), the outlier fractions reach their maxima of $\sim 14\%$ for $\text{frac}_{\text{AGN}}$ and $\sim 10\%$ for redshift. The outlier fractions decrease to approximately half their maximum values by S/N $\sim 2$ and approach zero by S/N $\sim 100$. Notably, the S/N $> 3$ threshold we adopted lies close to the inflection point of both curves, representing a practical balance between sample size and prediction accuracy. Stricter cuts (S/N $> 5$--$10$) would marginally improve outlier fractions but at the cost of significantly reducing sample sizes, while more lenient cuts (S/N $> 1$--$2$) would retain more sources but with noticeably degraded performance.

\newpage

\bibliography{paper25}

\begin{thebibliography}{}
\expandafter\ifx\csname natexlab\endcsname\relax\def\natexlab#1{#1}\fi
\providecommand{\url}[1]{\href{#1}{#1}}
\providecommand{\dodoi}[1]{doi:~\href{http://doi.org/#1}{\nolinkurl{#1}}}
\providecommand{\doeprint}[1]{\href{http://ascl.net/#1}{\nolinkurl{http://ascl.net/#1}}}
\providecommand{\doarXiv}[1]{\href{https://arxiv.org/abs/#1}{\nolinkurl{https://arxiv.org/abs/#1}}}

\bibitem[{Aird {et~al.}(2010)Aird, Nandra, Laird, Georgakakis, Ashby, Barmby, Coil, Huang, Koekemoer, Steidel, \& Willmer}]{aird_evolution_2010}
Aird, J., Nandra, K., Laird, E.~S., {et~al.} 2010, Monthly Notices of the Royal Astronomical Society, 401, 2531, \dodoi{10.1111/j.1365-2966.2009.15829.x}

\bibitem[{Akiba {et~al.}(2019)Akiba, Sano, Yanase, Ohta, \& Koyama}]{optuna}
Akiba, T., Sano, S., Yanase, T., Ohta, T., \& Koyama, M. 2019, in Proceedings of the 25th ACM SIGKDD International Conference on Knowledge Discovery \& Data Mining, KDD '19 (New York, NY, USA: Association for Computing Machinery), 2623–2631, \dodoi{10.1145/3292500.3330701}

\bibitem[{Akins(2024)}]{lazy_jl}
Akins, H.~B. 2024, Lazy.jl: Fast photometric redshift fitting in Julia.
\newblock \url{https://github.com/hollisakins/Lazy.jl}

\bibitem[{Alonso-Herrero {et~al.}(2016)Alonso-Herrero, Esquej, Roche, Ramos~Almeida, González-Martín, Packham, Levenson, Mason, Hernán-Caballero, Pereira-Santaella, Alvarez, Aretxaga, López-Rodríguez, Colina, Díaz-Santos, Imanishi, Rodríguez~Espinosa, \& Perlman}]{alonso-herrero_mid-infrared_2016}
Alonso-Herrero, A., Esquej, P., Roche, P.~F., {et~al.} 2016, Monthly Notices of the Royal Astronomical Society, 455, 563, \dodoi{10.1093/mnras/stv2342}

\bibitem[{Assef {et~al.}(2013)Assef, Stern, Kochanek, Blain, Brodwin, Brown, Donoso, Eisenhardt, Jannuzi, Jarrett, Stanford, Tsai, Wu, \& Yan}]{assef_mid-infrared_2013}
Assef, R.~J., Stern, D., Kochanek, C.~S., {et~al.} 2013, The Astrophysical Journal, 772, 26, \dodoi{10.1088/0004-637X/772/1/26}

\bibitem[{{Assef} {et~al.}(2018){Assef}, {Prieto}, {Stern}, {Cutri}, {Eisenhardt}, {Graham}, {Jun}, {Rest}, {Flewelling}, {Kaiser}, {Kudritzki}, \& {Waters}}]{assef_2018}
{Assef}, R.~J., {Prieto}, J.~L., {Stern}, D., {et~al.} 2018, \apj, 866, 26, \dodoi{10.3847/1538-4357/aaddf7}

\bibitem[{{Astropy Collaboration} {et~al.}(2013){Astropy Collaboration}, {Robitaille}, {Tollerud}, {Greenfield}, {Droettboom}, {Bray}, {Aldcroft}, {Davis}, {Ginsburg}, {Price-Whelan}, {Kerzendorf}, {Conley}, {Crighton}, {Barbary}, {Muna}, {Ferguson}, {Grollier}, {Parikh}, {Nair}, {Unther}, {Deil}, {Woillez}, {Conseil}, {Kramer}, {Turner}, {Singer}, {Fox}, {Weaver}, {Zabalza}, {Edwards}, {Azalee Bostroem}, {Burke}, {Casey}, {Crawford}, {Dencheva}, {Ely}, {Jenness}, {Labrie}, {Lim}, {Pierfederici}, {Pontzen}, {Ptak}, {Refsdal}, {Servillat}, \& {Streicher}}]{astropy:2013}
{Astropy Collaboration}, {Robitaille}, T.~P., {Tollerud}, E.~J., {et~al.} 2013, \aap, 558, A33, \dodoi{10.1051/0004-6361/201322068}

\bibitem[{{Astropy Collaboration} {et~al.}(2018){Astropy Collaboration}, {Price-Whelan}, {Sip{\H{o}}cz}, {G{\"u}nther}, {Lim}, {Crawford}, {Conseil}, {Shupe}, {Craig}, {Dencheva}, {Ginsburg}, {Vand erPlas}, {Bradley}, {P{\'e}rez-Su{\'a}rez}, {de Val-Borro}, {Aldcroft}, {Cruz}, {Robitaille}, {Tollerud}, {Ardelean}, {Babej}, {Bach}, {Bachetti}, {Bakanov}, {Bamford}, {Barentsen}, {Barmby}, {Baumbach}, {Berry}, {Biscani}, {Boquien}, {Bostroem}, {Bouma}, {Brammer}, {Bray}, {Breytenbach}, {Buddelmeijer}, {Burke}, {Calderone}, {Cano Rodr{\'\i}guez}, {Cara}, {Cardoso}, {Cheedella}, {Copin}, {Corrales}, {Crichton}, {D'Avella}, {Deil}, {Depagne}, {Dietrich}, {Donath}, {Droettboom}, {Earl}, {Erben}, {Fabbro}, {Ferreira}, {Finethy}, {Fox}, {Garrison}, {Gibbons}, {Goldstein}, {Gommers}, {Greco}, {Greenfield}, {Groener}, {Grollier}, {Hagen}, {Hirst}, {Homeier}, {Horton}, {Hosseinzadeh}, {Hu}, {Hunkeler}, {Ivezi{\'c}}, {Jain}, {Jenness}, {Kanarek}, {Kendrew}, {Kern}, {Kerzendorf}, {Khvalko}, {King}, {Kirkby}, {Kulkarni},
  {Kumar}, {Lee}, {Lenz}, {Littlefair}, {Ma}, {Macleod}, {Mastropietro}, {McCully}, {Montagnac}, {Morris}, {Mueller}, {Mumford}, {Muna}, {Murphy}, {Nelson}, {Nguyen}, {Ninan}, {N{\"o}the}, {Ogaz}, {Oh}, {Parejko}, {Parley}, {Pascual}, {Patil}, {Patil}, {Plunkett}, {Prochaska}, {Rastogi}, {Reddy Janga}, {Sabater}, {Sakurikar}, {Seifert}, {Sherbert}, {Sherwood-Taylor}, {Shih}, {Sick}, {Silbiger}, {Singanamalla}, {Singer}, {Sladen}, {Sooley}, {Sornarajah}, {Streicher}, {Teuben}, {Thomas}, {Tremblay}, {Turner}, {Terr{\'o}n}, {van Kerkwijk}, {de la Vega}, {Watkins}, {Weaver}, {Whitmore}, {Woillez}, {Zabalza}, \& {Astropy Contributors}}]{astropy:2018}
{Astropy Collaboration}, {Price-Whelan}, A.~M., {Sip{\H{o}}cz}, B.~M., {et~al.} 2018, \aj, 156, 123, \dodoi{10.3847/1538-3881/aabc4f}

\bibitem[{{Astropy Collaboration} {et~al.}(2022){Astropy Collaboration}, {Price-Whelan}, {Lim}, {Earl}, {Starkman}, {Bradley}, {Shupe}, {Patil}, {Corrales}, {Brasseur}, {N{"o}the}, {Donath}, {Tollerud}, {Morris}, {Ginsburg}, {Vaher}, {Weaver}, {Tocknell}, {Jamieson}, {van Kerkwijk}, {Robitaille}, {Merry}, {Bachetti}, {G{"u}nther}, {Aldcroft}, {Alvarado-Montes}, {Archibald}, {B{'o}di}, {Bapat}, {Barentsen}, {Baz{'a}n}, {Biswas}, {Boquien}, {Burke}, {Cara}, {Cara}, {Conroy}, {Conseil}, {Craig}, {Cross}, {Cruz}, {D'Eugenio}, {Dencheva}, {Devillepoix}, {Dietrich}, {Eigenbrot}, {Erben}, {Ferreira}, {Foreman-Mackey}, {Fox}, {Freij}, {Garg}, {Geda}, {Glattly}, {Gondhalekar}, {Gordon}, {Grant}, {Greenfield}, {Groener}, {Guest}, {Gurovich}, {Handberg}, {Hart}, {Hatfield-Dodds}, {Homeier}, {Hosseinzadeh}, {Jenness}, {Jones}, {Joseph}, {Kalmbach}, {Karamehmetoglu}, {Ka{l}uszy{'n}ski}, {Kelley}, {Kern}, {Kerzendorf}, {Koch}, {Kulumani}, {Lee}, {Ly}, {Ma}, {MacBride}, {Maljaars}, {Muna}, {Murphy}, {Norman}, {O'Steen},
  {Oman}, {Pacifici}, {Pascual}, {Pascual-Granado}, {Patil}, {Perren}, {Pickering}, {Rastogi}, {Roulston}, {Ryan}, {Rykoff}, {Sabater}, {Sakurikar}, {Salgado}, {Sanghi}, {Saunders}, {Savchenko}, {Schwardt}, {Seifert-Eckert}, {Shih}, {Jain}, {Shukla}, {Sick}, {Simpson}, {Singanamalla}, {Singer}, {Singhal}, {Sinha}, {Sip{H{o}}cz}, {Spitler}, {Stansby}, {Streicher}, {{{S}}umak}, {Swinbank}, {Taranu}, {Tewary}, {Tremblay}, {Val-Borro}, {Van Kooten}, {Vasovi{'c}}, {Verma}, {de Miranda Cardoso}, {Williams}, {Wilson}, {Winkel}, {Wood-Vasey}, {Xue}, {Yoachim}, {Zhang}, {Zonca}, \& {Astropy Project Contributors}}]{astropy:2022}
{Astropy Collaboration}, {Price-Whelan}, A.~M., {Lim}, P.~L., {et~al.} 2022, \apj, 935, 167, \dodoi{10.3847/1538-4357/ac7c74}

\bibitem[{{Backhaus} {et~al.}(2025){Backhaus}, {Kirkpatrick}, {Yang}, {Troiani}, {Hamblin}, {Kartaltepe}, {Kocevski}, {Koekemoer}, {Lambrides}, {Papovich}, \& {Ronayne}}]{backhaus_2025}
{Backhaus}, B.~E., {Kirkpatrick}, A., {Yang}, G., {et~al.} 2025, \aj, 170, 300, \dodoi{10.3847/1538-3881/ae0cc4}

\bibitem[{Bai {et~al.}(2018)Bai, Liu, Wang, \& Yang}]{bai_machine_2018}
Bai, Y., Liu, J., Wang, S., \& Yang, F. 2018, AJ, 157, 9, \dodoi{10.3847/1538-3881/aaf009}

\bibitem[{{Bertin} \& {Arnouts}(1996)}]{SourceExtractor}
{Bertin}, E., \& {Arnouts}, S. 1996, \aaps, 117, 393, \dodoi{10.1051/aas:1996164}

\bibitem[{{Boquien} {et~al.}(2019){Boquien}, {Burgarella}, {Roehlly}, {Buat}, {Ciesla}, {Corre}, {Inoue}, \& {Salas}}]{boquien_cigale}
{Boquien}, M., {Burgarella}, D., {Roehlly}, Y., {et~al.} 2019, \aap, 622, A103, \dodoi{10.1051/0004-6361/201834156}

\bibitem[{{Bovy} {et~al.}(2011){Bovy}, {Hennawi}, {Hogg}, {Myers}, {Kirkpatrick}, {Schlegel}, {Ross}, {Sheldon}, {McGreer}, {Schneider}, \& {Weaver}}]{Bovy_2011}
{Bovy}, J., {Hennawi}, J.~F., {Hogg}, D.~W., {et~al.} 2011, \apj, 729, 141, \dodoi{10.1088/0004-637X/729/2/141}

\bibitem[{{Brammer} {et~al.}(2008){Brammer}, {van Dokkum}, \& {Coppi}}]{brammer_eazy}
{Brammer}, G.~B., {van Dokkum}, P.~G., \& {Coppi}, P. 2008, \apj, 686, 1503, \dodoi{10.1086/591786}

\bibitem[{{Brescia} {et~al.}(2019){Brescia}, {Salvato}, {Cavuoti}, {Ananna}, {Riccio}, {LaMassa}, {Urry}, \& {Longo}}]{Brescia_2019}
{Brescia}, M., {Salvato}, M., {Cavuoti}, S., {et~al.} 2019, \mnras, 489, 663, \dodoi{10.1093/mnras/stz2159}

\bibitem[{Bruzual \& Charlot(2003)}]{bruzual_stellar_2003}
Bruzual, G., \& Charlot, S. 2003, Monthly Notices of the Royal Astronomical Society, 344, 1000, \dodoi{10.1046/j.1365-8711.2003.06897.x}

\bibitem[{Calzetti {et~al.}(2000)Calzetti, Armus, Bohlin, Kinney, Koornneef, \& Storchi-Bergmann}]{calzetti_dust_2000}
Calzetti, D., Armus, L., Bohlin, R.~C., {et~al.} 2000, The Astrophysical Journal, 533, 682, \dodoi{10.1086/308692}

\bibitem[{{Chabrier}(2003)}]{chabrier_2003}
{Chabrier}, G. 2003, \pasp, 115, 763, \dodoi{10.1086/376392}

\bibitem[{Chen \& Guestrin(2016)}]{chen_xgboost}
Chen, T., \& Guestrin, C. 2016, in Proceedings of the 22nd ACM SIGKDD International Conference on Knowledge Discovery and Data Mining, KDD '16 (New York, NY, USA: Association for Computing Machinery), 785–794, \dodoi{10.1145/2939672.2939785}

\bibitem[{Conroy(2013)}]{conroy_sed_review}
Conroy, C. 2013, Annual Review of Astronomy and Astrophysics, 51, 393, \dodoi{https://doi.org/10.1146/annurev-astro-082812-141017}

\bibitem[{{Davis} {et~al.}(2007){Davis}, {Guhathakurta}, {Konidaris}, {Newman}, {Ashby}, {Biggs}, {Barmby}, {Bundy}, {Chapman}, {Coil}, {Conselice}, {Cooper}, {Croton}, {Eisenhardt}, {Ellis}, {Faber}, {Fang}, {Fazio}, {Georgakakis}, {Gerke}, {Goss}, {Gwyn}, {Harker}, {Hopkins}, {Huang}, {Ivison}, {Kassin}, {Kirby}, {Koekemoer}, {Koo}, {Laird}, {Le Floc'h}, {Lin}, {Lotz}, {Marshall}, {Martin}, {Metevier}, {Moustakas}, {Nandra}, {Noeske}, {Papovich}, {Phillips}, {Rich}, {Rieke}, {Rigopoulou}, {Salim}, {Schiminovich}, {Simard}, {Smail}, {Small}, {Weiner}, {Willmer}, {Willner}, {Wilson}, {Wright}, \& {Yan}}]{EGS_2007}
{Davis}, M., {Guhathakurta}, P., {Konidaris}, N.~P., {et~al.} 2007, \apjl, 660, L1, \dodoi{10.1086/517931}

\bibitem[{Dietterich(2000)}]{Dietterrich_2000}
Dietterich, T.~G. 2000, in Proceedings of the First International Workshop on Multiple Classifier Systems, MCS '00 (Berlin, Heidelberg: Springer-Verlag), 1–15

\bibitem[{Donley {et~al.}(2012)Donley, Koekemoer, Brusa, Capak, Cardamone, Civano, Ilbert, Impey, Kartaltepe, Miyaji, Salvato, Sanders, Trump, \& Zamorani}]{donley_identifying_2012}
Donley, J.~L., Koekemoer, A.~M., Brusa, M., {et~al.} 2012, The Astrophysical Journal, 748, 142, \dodoi{10.1088/0004-637X/748/2/142}

\bibitem[{{Draine} \& {Li}(2001)}]{Draine_2001}
{Draine}, B.~T., \& {Li}, A. 2001, \apj, 551, 807, \dodoi{10.1086/320227}

\bibitem[{Draine {et~al.}(2014)Draine, Aniano, Krause, Groves, Sandstrom, Braun, Leroy, Klaas, Linz, Rix, Schinnerer, Schmiedeke, \& Walter}]{draine_andromedas_2014}
Draine, B.~T., Aniano, G., Krause, O., {et~al.} 2014, The Astrophysical Journal, 780, 172, \dodoi{10.1088/0004-637X/780/2/172}

\bibitem[{{Duncan} {et~al.}(2018){Duncan}, {Jarvis}, {Brown}, \& {R{\"o}ttgering}}]{Duncan_2018}
{Duncan}, K.~J., {Jarvis}, M.~J., {Brown}, M. J.~I., \& {R{\"o}ttgering}, H. J.~A. 2018, \mnras, 477, 5177, \dodoi{10.1093/mnras/sty940}

\bibitem[{{Durodola} {et~al.}(2024){Durodola}, {Pacucci}, \& {Hickox}}]{Durodola_2024}
{Durodola}, E., {Pacucci}, F., \& {Hickox}, R.~C. 2024, arXiv e-prints, arXiv:2406.10329, \dodoi{10.48550/arXiv.2406.10329}

\bibitem[{{Finkelstein} {et~al.}(2024){Finkelstein}, {Leung}, {Bagley}, {Dickinson}, {Ferguson}, {Papovich}, {Akins}, {Arrabal Haro}, {Dav{\'e}}, {Dekel}, {Kartaltepe}, {Kocevski}, {Koekemoer}, {Pirzkal}, {Somerville}, {Yung}, {Amor{\'\i}n}, {Backhaus}, {Behroozi}, {Bisigello}, {Bromm}, {Casey}, {Ch{\'a}vez Ortiz}, {Cheng}, {Chworowsky}, {Cleri}, {Cooper}, {Davis}, {de la Vega}, {Elbaz}, {Franco}, {Fontana}, {Fujimoto}, {Giavalisco}, {Grogin}, {Holwerda}, {Huertas-Company}, {Hirschmann}, {Iyer}, {Jogee}, {Jung}, {Larson}, {Lucas}, {Mobasher}, {Morales}, {Morley}, {Mukherjee}, {P{\'e}rez-Gonz{\'a}lez}, {Ravindranath}, {Rodighiero}, {Rowland}, {Tacchella}, {Taylor}, {Trump}, \& {Wilkins}}]{Finkelstein_2024}
{Finkelstein}, S.~L., {Leung}, G. C.~K., {Bagley}, M.~B., {et~al.} 2024, \apjl, 969, L2, \dodoi{10.3847/2041-8213/ad4495}

\bibitem[{Finkelstein {et~al.}(2025)Finkelstein, Bagley, Haro, Dickinson, Ferguson, Kartaltepe, Kocevski, Koekemoer, Lotz, Papovich, Perez-Gonzalez, Pirzkal, Somerville, Trump, Yang, Yung, Fontana, Grazian, Grogin, Kewley, Kirkpatrick, Larson, Pentericci, Ravindranath, Wilkins, Almaini, Amorin, Barro, Bhatawdekar, Bisigello, Brooks, Buitrago, Calabro, Castellano, Cheng, Cleri, Cole, Cooper, Cooper, Costantin, Cox, Croton, Daddi, Davis, Dekel, Elbaz, Fernandez, Fujimoto, Gandolfi, Gardner, Gawiser, Giavalisco, Gomez-Guijarro, Guo, Gupta, Hathi, Harish, Henry, Hirschmann, Hu, Hutchison, Iyer, Jaskot, Jha, Jung, Kokorev, Kurczynski, Leung, Llerena, Long, Lucas, Lu, McGrath, McIntosh, Merlin, Morales, Napolitano, Pacucci, Pandya, Rafelski, Rodighiero, Rose, Santini, Seille, Simons, Shen, Straughn, Tacchella, Vanderhoof, Vega-Ferrero, Weiner, Willmer, Zhu, Bell, Wuyts, Holwerda, Wang, Wang, \& Zavala}]{finkelstein_cosmic_2025}
Finkelstein, S.~L., Bagley, M.~B., Haro, P.~A., {et~al.} 2025, The {Cosmic} {Evolution} {Early} {Release} {Science} {Survey} ({CEERS}),  arXiv, \dodoi{10.48550/arXiv.2501.04085}

\bibitem[{Fontana {et~al.}(2006)Fontana, Salimbeni, Grazian, Giallongo, Pentericci, Nonino, Fontanot, Menci, Monaco, Cristiani, Vanzella, de~Santis, \& Gallozzi}]{fontana_galaxy_2006}
Fontana, A., Salimbeni, S., Grazian, A., {et~al.} 2006, Astronomy and Astrophysics, 459, 745, \dodoi{10.1051/0004-6361:20065475}

\bibitem[{{F{\"o}rster Schreiber} {et~al.}(2003){F{\"o}rster Schreiber}, {Sauvage}, {Charmandaris}, {Laurent}, {Gallais}, {Mirabel}, \& {Vigroux}}]{m82_paper}
{F{\"o}rster Schreiber}, N.~M., {Sauvage}, M., {Charmandaris}, V., {et~al.} 2003, \aap, 399, 833, \dodoi{10.1051/0004-6361:20021719}

\bibitem[{{Fotopoulou} \& {Paltani}(2018)}]{Fotopoulou_2018}
{Fotopoulou}, S., \& {Paltani}, S. 2018, \aap, 619, A14, \dodoi{10.1051/0004-6361/201730763}

\bibitem[{{Franceschini} {et~al.}(1991){Franceschini}, {Toffolatti}, {Mazzei}, {Danese}, \& {de Zotti}}]{Franceschini_1991}
{Franceschini}, A., {Toffolatti}, L., {Mazzei}, P., {Danese}, L., \& {de Zotti}, G. 1991, \aaps, 89, 285

\bibitem[{Förster~Schreiber \& Wuyts(2020)}]{forster_schreiber_star-forming_2020}
Förster~Schreiber, N.~M., \& Wuyts, S. 2020, Annual Review of Astronomy and Astrophysics, 58, 661, \dodoi{10.1146/annurev-astro-032620-021910}

\bibitem[{Ganaie {et~al.}(2021)Ganaie, Hu, Tanveer, \& Suganthan}]{Ganaie_2021}
Ganaie, M.~A., Hu, M., Tanveer, M., \& Suganthan, P.~N. 2021, CoRR, abs/2104.02395

\bibitem[{{Gardner} {et~al.}(2006){Gardner}, {Mather}, {Clampin}, {Doyon}, {Greenhouse}, {Hammel}, {Hutchings}, {Jakobsen}, {Lilly}, {Long}, {Lunine}, {McCaughrean}, {Mountain}, {Nella}, {Rieke}, {Rieke}, {Rix}, {Smith}, {Sonneborn}, {Stiavelli}, {Stockman}, {Windhorst}, \& {Wright}}]{Gardner_2006}
{Gardner}, J.~P., {Mather}, J.~C., {Clampin}, M., {et~al.} 2006, \ssr, 123, 485, \dodoi{10.1007/s11214-006-8315-7}

\bibitem[{{Gardner} {et~al.}(2023){Gardner}, {Mather}, {Abbott}, {Abell}, {Abernathy}, {Abney}, {Abraham}, {Abraham}, {Abul-Huda}, {Acton}, {Adams}, {Adams}, {Adler}, {Adriaensen}, {Aguilar}, {Ahmed}, {Ahmed}, {Ahmed}, {Albat}, {Albert}, {Alberts}, {Aldridge}, {Allen}, {Allen}, {Altenburg}, {Altunc}, {Alvarez}, {{\'A}lvarez-M{\'a}rquez}, {Alves de Oliveira}, {Ambrose}, {Anandakrishnan}, {Andersen}, {Anderson}, {Anderson}, {Anderson}, {Anderson}, {Aprea}, {Archer}, {Arenberg}, {Argyriou}, {Arribas}, {Artigau}, {Arvai}, {Atcheson}, {Atkinson}, {Averbukh}, {Aymergen}, {Bacinski}, {Baggett}, {Bagnasco}, {Baker}, {Balzano}, {Banks}, {Baran}, {Barker}, {Barrett}, {Barringer}, {Barto}, {Bast}, {Baudoz}, {Baum}, {Beatty}, {Beaulieu}, {Bechtold}, {Beck}, {Beddard}, {Beichman}, {Bellagama}, {Bely}, {Berger}, {Bergeron}, {Bernier}, {Bertch}, {Beskow}, {Betz}, {Biagetti}, {Birkmann}, {Bjorklund}, {Blackwood}, {Blazek}, {Blossfeld}, {Bluth}, {Boccaletti}, {Boegner}, {Bohlin}, {Boia}, {B{\"o}ker}, {Bonaventura}, {Bond},
  {Bosley}, {Boucarut}, {Bouchet}, {Bouwman}, {Bower}, {Bowers}, {Bowers}, {Boyce}, {Boyer}, {Boyer}, {Boyer}, {Boyer}, {Bradley}, {Brady}, {Brandl}, {Brannen}, {Breda}, {Bremmer}, {Brennan}, {Bresnahan}, {Bright}, {Broiles}, {Bromenschenkel}, {Brooks}, {Brooks}, {Brown}, {Brown}, {Brown}, {Bruce}, {Bryson}, {Bujanda}, {Bullock}, {Bunker}, {Bureo}, {Burt}, {Bush}, {Bushouse}, {Bussman}, {Cabaud}, {Cale}, {Calhoon}, {Calvani}, {Canipe}, {Caputo}, {Cara}, {Carey}, {Case}, {Cesari}, {Cetorelli}, {Chance}, {Chandler}, {Chaney}, {Chapman}, {Charlot}, {Chayer}, {Cheezum}, {Chen}, {Chen}, {Cherinka}, {Chichester}, {Chilton}, {Chittiraibalan}, {Clampin}, {Clark}, {Clark}, {Clark}, {Claybrooks}, {Cleveland}, {Cohen}, {Cohen}, {Col{\'o}n}, {Coleman}, {Colina}, {Comber}, {Comeau}, {Comer}, {Conde Reis}, {Connolly}, {Conroy}, {Contos}, {Contreras}, {Cook}, {Cooper}, {Cooper}, {Correia}, {Correnti}, {Cossou}, {Costanza}, {Coulais}, {Cox}, {Coyle}, {Cracraft}, {Crew}, {Curtis}, {Cusveller}, {Da Costa Maciel}, {Dailey},
  {Daugeron}, {Davidson}, {Davies}, {Davis}, {Davis}, {Day}, {de Chambure}, {de Jong}, {De Marchi}, {Dean}, {Decker}, {Delisa}, {Dell}, \& {Dellagatta}}]{jwst_2023}
{Gardner}, J.~P., {Mather}, J.~C., {Abbott}, R., {et~al.} 2023, \pasp, 135, 068001, \dodoi{10.1088/1538-3873/acd1b5}

\bibitem[{{Grogin} {et~al.}(2011){Grogin}, {Kocevski}, {Faber}, {Ferguson}, {Koekemoer}, {Riess}, {Acquaviva}, {Alexander}, {Almaini}, {Ashby}, {Barden}, {Bell}, {Bournaud}, {Brown}, {Caputi}, {Casertano}, {Cassata}, {Castellano}, {Challis}, {Chary}, {Cheung}, {Cirasuolo}, {Conselice}, {Roshan Cooray}, {Croton}, {Daddi}, {Dahlen}, {Dav{\'e}}, {de Mello}, {Dekel}, {Dickinson}, {Dolch}, {Donley}, {Dunlop}, {Dutton}, {Elbaz}, {Fazio}, {Filippenko}, {Finkelstein}, {Fontana}, {Gardner}, {Garnavich}, {Gawiser}, {Giavalisco}, {Grazian}, {Guo}, {Hathi}, {H{\"a}ussler}, {Hopkins}, {Huang}, {Huang}, {Jha}, {Kartaltepe}, {Kirshner}, {Koo}, {Lai}, {Lee}, {Li}, {Lotz}, {Lucas}, {Madau}, {McCarthy}, {McGrath}, {McIntosh}, {McLure}, {Mobasher}, {Moustakas}, {Mozena}, {Nandra}, {Newman}, {Niemi}, {Noeske}, {Papovich}, {Pentericci}, {Pope}, {Primack}, {Rajan}, {Ravindranath}, {Reddy}, {Renzini}, {Rix}, {Robaina}, {Rodney}, {Rosario}, {Rosati}, {Salimbeni}, {Scarlata}, {Siana}, {Simard}, {Smidt}, {Somerville}, {Spinrad},
  {Straughn}, {Strolger}, {Telford}, {Teplitz}, {Trump}, {van der Wel}, {Villforth}, {Wechsler}, {Weiner}, {Wiklind}, {Wild}, {Wilson}, {Wuyts}, {Yan}, \& {Yun}}]{candels_2011}
{Grogin}, N.~A., {Kocevski}, D.~D., {Faber}, S.~M., {et~al.} 2011, \apjs, 197, 35, \dodoi{10.1088/0067-0049/197/2/35}

\bibitem[{Hamblin(2025)}]{dataverse_mockcatalog}
Hamblin, K. 2025, {AGNBoost CIGALE Mock Dataset}, V1,  Harvard Dataverse, \dodoi{10.7910/DVN/YYGZ3P}

\bibitem[{Harris {et~al.}(2020)Harris, Millman, van~der Walt, Gommers, Virtanen, Cournapeau, Wieser, Taylor, Berg, Smith, Kern, Picus, Hoyer, van Kerkwijk, Brett, Haldane, del R{\'{i}}o, Wiebe, Peterson, G{\'{e}}rard-Marchant, Sheppard, Reddy, Weckesser, Abbasi, Gohlke, \& Oliphant}]{numpy}
Harris, C.~R., Millman, K.~J., van~der Walt, S.~J., {et~al.} 2020, Nature, 585, 357, \dodoi{10.1038/s41586-020-2649-2}

\bibitem[{Hastie {et~al.}(2001)Hastie, Tibshirani, \& Friedman}]{hastie01statisticallearning}
Hastie, T., Tibshirani, R., \& Friedman, J. 2001, The Elements of Statistical Learning, Springer Series in Statistics (New York, NY, USA: Springer New York Inc.)

\bibitem[{Heckman {et~al.}(2004)Heckman, Kauffmann, Brinchmann, Charlot, Tremonti, \& White}]{heckman_present-day_2004}
Heckman, T.~M., Kauffmann, G., Brinchmann, J., {et~al.} 2004, The Astrophysical Journal, 613, 109, \dodoi{10.1086/422872}

\bibitem[{Hickox \& Alexander(2018)}]{hickox_araa}
Hickox, R.~C., \& Alexander, D.~M. 2018, Annual Review of Astronomy and Astrophysics, 56, 625, \dodoi{https://doi.org/10.1146/annurev-astro-081817-051803}

\bibitem[{{Holwerda} {et~al.}(2021){Holwerda}, {Wu}, {Keel}, {Young}, {Mullins}, {Hinz}, {Ford}, {Barmby}, {Chandar}, {Bailin}, {Peek}, {Pickering}, \& {B{\"o}ker}}]{holwerda_2021}
{Holwerda}, B.~W., {Wu}, J.~F., {Keel}, W.~C., {et~al.} 2021, \apj, 914, 142, \dodoi{10.3847/1538-4357/abffcc}

\bibitem[{Hopkins(2004)}]{hopkins_evolution_2004}
Hopkins, A.~M. 2004, The Astrophysical Journal, 615, 209, \dodoi{10.1086/424032}

\bibitem[{Hunter(2007)}]{matplotlib}
Hunter, J.~D. 2007, Computing in Science \& Engineering, 9, 90, \dodoi{10.1109/MCSE.2007.55}

\bibitem[{Hüllermeier \& Waegeman(2021)}]{hullermeier_aleatoric_2021}
Hüllermeier, E., \& Waegeman, W. 2021, Mach Learn, 110, 457, \dodoi{10.1007/s10994-021-05946-3}

\bibitem[{{Iyer} \& {Gawiser}(2017)}]{Iyer_2017}
{Iyer}, K., \& {Gawiser}, E. 2017, in Galaxy Evolution Across Time, 2, \dodoi{10.5281/zenodo.805863}

\bibitem[{{Iyer} {et~al.}(2025){Iyer}, {Pacifici}, {Calistro-Rivera}, \& {Lovell}}]{Iyer_2025}
{Iyer}, K.~G., {Pacifici}, C., {Calistro-Rivera}, G., \& {Lovell}, C.~C. 2025, arXiv e-prints, arXiv:2502.17680, \dodoi{10.48550/arXiv.2502.17680}

\bibitem[{{Kirkpatrick} {et~al.}(2015){Kirkpatrick}, {Pope}, {Sajina}, {Roebuck}, {Yan}, {Armus}, {D{\'\i}az-Santos}, \& {Stierwalt}}]{kirkpatrick_2015}
{Kirkpatrick}, A., {Pope}, A., {Sajina}, A., {et~al.} 2015, \apj, 814, 9, \dodoi{10.1088/0004-637X/814/1/9}

\bibitem[{Kirkpatrick {et~al.}(2012)Kirkpatrick, Pope, Alexander, Charmandaris, Daddi, Dickinson, Elbaz, Gabor, Hwang, Ivison, Mullaney, Pannella, Scott, Altieri, Aussel, Bournaud, Buat, Coia, Dannerbauer, Dasyra, Kartaltepe, Leiton, Lin, Magdis, Magnelli, Morrison, Popesso, \& Valtchanov}]{kirkpatrick_goods-herschel_2012}
Kirkpatrick, A., Pope, A., Alexander, D.~M., {et~al.} 2012, The Astrophysical Journal, 759, 139, \dodoi{10.1088/0004-637X/759/2/139}

\bibitem[{{Kirkpatrick} {et~al.}(2013){Kirkpatrick}, {Pope}, {Charmandaris}, {Daddi}, {Elbaz}, {Hwang}, {Pannella}, {Scott}, {Altieri}, {Aussel}, {Coia}, {Dannerbauer}, {Dasyra}, {Dickinson}, {Kartaltepe}, {Leiton}, {Magdis}, {Magnelli}, {Popesso}, \& {Valtchanov}}]{kirkpatrick_2013}
{Kirkpatrick}, A., {Pope}, A., {Charmandaris}, V., {et~al.} 2013, \apj, 763, 123, \dodoi{10.1088/0004-637X/763/2/123}

\bibitem[{{Kirkpatrick} {et~al.}(2017){Kirkpatrick}, {Alberts}, {Pope}, {Barro}, {Bonato}, {Kocevski}, {P{\'e}rez-Gonz{\'a}lez}, {Rieke}, {Rodr{\'\i}guez-Mu{\~n}oz}, {Sajina}, {Grogin}, {Mantha}, {Pandya}, {Pforr}, {Salvato}, \& {Santini}}]{kirkpatrick_2017}
{Kirkpatrick}, A., {Alberts}, S., {Pope}, A., {et~al.} 2017, \apj, 849, 111, \dodoi{10.3847/1538-4357/aa911d}

\bibitem[{Kirkpatrick {et~al.}(2023)Kirkpatrick, Yang, Le~Bail, Troiani, Bell, Cleri, Elbaz, Finkelstein, Hathi, Hirschmann, Holwerda, Kocevski, Lucas, McKinney, Papovich, Pérez-González, de~la Vega, Bagley, Daddi, Dickinson, Ferguson, Fontana, Grazian, Grogin, Arrabal~Haro, Kartaltepe, Kewley, Koekemoer, Lotz, Pentericci, Pirzkal, Ravindranath, Somerville, Trump, Wilkins, \& Yung}]{kirkpatrick_ceers_2023}
Kirkpatrick, A., Yang, G., Le~Bail, A., {et~al.} 2023, The Astrophysical Journal, 959, L7, \dodoi{10.3847/2041-8213/ad0b14}

\bibitem[{{Kocevski} {et~al.}(2025){Kocevski}, {Finkelstein}, {Barro}, {Taylor}, {Calabr{\`o}}, {Laloux}, {Buchner}, {Trump}, {Leung}, {Yang}, {Dickinson}, {P{\'e}rez-Gonz{\'a}lez}, {Pacucci}, {Inayoshi}, {Somerville}, {McGrath}, {Akins}, {Bagley}, {Bowler}, {Bisigello}, {Carnall}, {Casey}, {Cheng}, {Cleri}, {Costantin}, {Cullen}, {Davis}, {Donnan}, {Dunlop}, {Ellis}, {Ferguson}, {Fujimoto}, {Fontana}, {Giavalisco}, {Grazian}, {Grogin}, {Hathi}, {Hirschmann}, {Huertas-Company}, {Holwerda}, {Illingworth}, {Juneau}, {Kartaltepe}, {Koekemoer}, {Li}, {Lucas}, {Magee}, {Mason}, {McLeod}, {McLure}, {Napolitano}, {Papovich}, {Pirzkal}, {Rodighiero}, {Santini}, {Wilkins}, \& {Yung}}]{Kocevski_2025}
{Kocevski}, D.~D., {Finkelstein}, S.~L., {Barro}, G., {et~al.} 2025, \apj, 986, 126, \dodoi{10.3847/1538-4357/adbc7d}

\bibitem[{Kodra {et~al.}(2023)Kodra, Andrews, Newman, L.~Finkelstein, Fontana, Hathi, Salvato, Wiklind, Wuyts, Broussard, Chartab, Conselice, Cooper, Dekel, Dickinson, Ferguson, Gawiser, Grogin, Iyer, Kartaltepe, Kassin, Koekemoer, Koo, Lucas, Mantha, McIntosh, Mobasher, Pacifici, Pérez-González, \& Santini}]{Kodra_2023}
Kodra, D., Andrews, B.~H., Newman, J.~A., {et~al.} 2023, The Astrophysical Journal, 942, 36, \dodoi{10.3847/1538-4357/ac9f12}

\bibitem[{{Koekemoer} {et~al.}(2011){Koekemoer}, {Faber}, {Ferguson}, {Grogin}, {Kocevski}, {Koo}, {Lai}, {Lotz}, {Lucas}, {McGrath}, {Ogaz}, {Rajan}, {Riess}, {Rodney}, {Strolger}, {Casertano}, {Castellano}, {Dahlen}, {Dickinson}, {Dolch}, {Fontana}, {Giavalisco}, {Grazian}, {Guo}, {Hathi}, {Huang}, {van der Wel}, {Yan}, {Acquaviva}, {Alexander}, {Almaini}, {Ashby}, {Barden}, {Bell}, {Bournaud}, {Brown}, {Caputi}, {Cassata}, {Challis}, {Chary}, {Cheung}, {Cirasuolo}, {Conselice}, {Roshan Cooray}, {Croton}, {Daddi}, {Dav{\'e}}, {de Mello}, {de Ravel}, {Dekel}, {Donley}, {Dunlop}, {Dutton}, {Elbaz}, {Fazio}, {Filippenko}, {Finkelstein}, {Frazer}, {Gardner}, {Garnavich}, {Gawiser}, {Gruetzbauch}, {Hartley}, {H{\"a}ussler}, {Herrington}, {Hopkins}, {Huang}, {Jha}, {Johnson}, {Kartaltepe}, {Khostovan}, {Kirshner}, {Lani}, {Lee}, {Li}, {Madau}, {McCarthy}, {McIntosh}, {McLure}, {McPartland}, {Mobasher}, {Moreira}, {Mortlock}, {Moustakas}, {Mozena}, {Nandra}, {Newman}, {Nielsen}, {Niemi}, {Noeske}, {Papovich},
  {Pentericci}, {Pope}, {Primack}, {Ravindranath}, {Reddy}, {Renzini}, {Rix}, {Robaina}, {Rosario}, {Rosati}, {Salimbeni}, {Scarlata}, {Siana}, {Simard}, {Smidt}, {Snyder}, {Somerville}, {Spinrad}, {Straughn}, {Telford}, {Teplitz}, {Trump}, {Vargas}, {Villforth}, {Wagner}, {Wandro}, {Wechsler}, {Weiner}, {Wiklind}, {Wild}, {Wilson}, {Wuyts}, \& {Yun}}]{candels_2011_anton}
{Koekemoer}, A.~M., {Faber}, S.~M., {Ferguson}, H.~C., {et~al.} 2011, \apjs, 197, 36, \dodoi{10.1088/0067-0049/197/2/36}

\bibitem[{Kormendy \& Ho(2013)}]{kormendy_coevolution_2013}
Kormendy, J., \& Ho, L.~C. 2013, Annual Review of Astronomy and Astrophysics, 51, 511, \dodoi{10.1146/annurev-astro-082708-101811}

\bibitem[{Lacy {et~al.}(2007)Lacy, Petric, Sajina, Canalizo, Storrie-Lombardi, Armus, Fadda, \& Marleau}]{lacy_optical_2007}
Lacy, M., Petric, A.~O., Sajina, A., {et~al.} 2007, The Astronomical Journal, 133, 186, \dodoi{10.1086/509617}

\bibitem[{Lacy {et~al.}(2004)Lacy, Storrie-Lombardi, Sajina, Appleton, Armus, Chapman, Choi, Fadda, Fang, Frayer, Heinrichsen, Helou, Im, Marleau, Masci, Shupe, Soifer, Surace, Teplitz, Wilson, \& Yan}]{lacy_obscured_2004}
Lacy, M., Storrie-Lombardi, L.~J., Sajina, A., {et~al.} 2004, The Astrophysical Journal Supplement Series, 154, 166, \dodoi{10.1086/422816}

\bibitem[{Laurent {et~al.}(2000)Laurent, Mirabel, Charmandaris, Gallais, Madden, Sauvage, Vigroux, \& Cesarsky}]{laurent_mid-infrared_2000}
Laurent, O., Mirabel, I.~F., Charmandaris, V., {et~al.} 2000, Astronomy and Astrophysics, 359, 887, \dodoi{10.48550/arXiv.astro-ph/0005376}

\bibitem[{Leitherer {et~al.}(2002)Leitherer, Li, Calzetti, \& Heckman}]{leitherer_global_2002}
Leitherer, C., Li, I.~H., Calzetti, D., \& Heckman, T.~M. 2002, The Astrophysical Journal Supplement Series, 140, 303, \dodoi{10.1086/342486}

\bibitem[{Leung {et~al.}(2025)Leung, Finkelstein, Pérez-González, Morales, Taylor, Barro, Kocevski, Akins, Carnall, Chávez~Ortiz, Cleri, Cullen, Donnan, Dunlop, Ellis, Grogin, Hirschmann, Koekemoer, Kokorev, Lucas, McLeod, Papovich, \& Yung}]{Leung_2025}
Leung, G. C.~K., Finkelstein, S.~L., Pérez-González, P.~G., {et~al.} 2025, The Astrophysical Journal, 992, 26, \dodoi{10.3847/1538-4357/adfcce}

\bibitem[{Lundberg \& Lee(2017)}]{lundberg_shap}
Lundberg, S.~M., \& Lee, S.-I. 2017, in Proceedings of the 31st International Conference on Neural Information Processing Systems, NIPS'17 (Red Hook, NY, USA: Curran Associates Inc.), 4768–4777

\bibitem[{Lundberg {et~al.}(2020)Lundberg, Erion, Chen, DeGrave, Prutkin, Nair, Katz, Himmelfarb, Bansal, \& Lee}]{shap_tree}
Lundberg, S.~M., Erion, G., Chen, H., {et~al.} 2020, Nature Machine Intelligence, 2, 2522

\bibitem[{Luo {et~al.}(2024)Luo, Tang, Chen, Fu, Du, Zhang, Gong, Shu, Lu, Li, Meng, Zhou, \& Fan}]{luo_imputation}
Luo, Z., Tang, Z., Chen, Z., {et~al.} 2024, Monthly Notices of the Royal Astronomical Society, 531, 3539, \dodoi{10.1093/mnras/stae1397}

\bibitem[{{Lyu} {et~al.}(2016){Lyu}, {Rieke}, \& {Alberts}}]{Lyu_2016}
{Lyu}, J., {Rieke}, G.~H., \& {Alberts}, S. 2016, \apj, 816, 85, \dodoi{10.3847/0004-637X/816/2/85}

\bibitem[{{Lyu} {et~al.}(2024){Lyu}, {Alberts}, {Rieke}, {Shivaei}, {P{\'e}rez-Gonz{\'a}lez}, {Sun}, {Hainline}, {Baum}, {Bonaventura}, {Bunker}, {Egami}, {Eisenstein}, {Florian}, {Ji}, {Johnson}, {Morrison}, {Rieke}, {Robertson}, {Rujopakarn}, {Tacchella}, {Scholtz}, \& {Willmer}}]{Lyu_2024}
{Lyu}, J., {Alberts}, S., {Rieke}, G.~H., {et~al.} 2024, \apj, 966, 229, \dodoi{10.3847/1538-4357/ad3643}

\bibitem[{Madau \& Dickinson(2014)}]{madau_cosmic_2014}
Madau, P., \& Dickinson, M. 2014, Annual Review of Astronomy and Astrophysics, 52, 415, \dodoi{10.1146/annurev-astro-081811-125615}

\bibitem[{Magdis {et~al.}(2012)Magdis, Daddi, Béthermin, Sargent, Elbaz, Pannella, Dickinson, Dannerbauer, da~Cunha, Walter, Rigopoulou, Charmandaris, Hwang, \& Kartaltepe}]{magdis_evolving_2012}
Magdis, G.~E., Daddi, E., Béthermin, M., {et~al.} 2012, The Astrophysical Journal, 760, 6, \dodoi{10.1088/0004-637X/760/1/6}

\bibitem[{{Maiolino} {et~al.}(2024){Maiolino}, {Scholtz}, {Curtis-Lake}, {Carniani}, {Baker}, {de Graaff}, {Tacchella}, {{\"U}bler}, {D'Eugenio}, {Witstok}, {Curti}, {Arribas}, {Bunker}, {Charlot}, {Chevallard}, {Eisenstein}, {Egami}, {Ji}, {Jones}, {Lyu}, {Rawle}, {Robertson}, {Rujopakarn}, {Perna}, {Sun}, {Venturi}, {Williams}, \& {Willott}}]{Maiolino_2024}
{Maiolino}, R., {Scholtz}, J., {Curtis-Lake}, E., {et~al.} 2024, \aap, 691, A145, \dodoi{10.1051/0004-6361/202347640}

\bibitem[{Mechbal {et~al.}(2024)Mechbal, Ackermann, \& Kowalski}]{mechbal_machine_2024}
Mechbal, S., Ackermann, M., \& Kowalski, M. 2024, A\&A, 685, A107, \dodoi{10.1051/0004-6361/202346557}

\bibitem[{{Merz} {et~al.}(2025){Merz}, {Liu}, {Schmidt}, {Malz}, {Zhang}, {Branton}, {Burke}, {Delucchi}, {Ejjagiri}, {Kubica}, {Liu}, {Lynn}, {Oldag}, \& {LSST Dark Energy Science Collaboration}}]{Merz_2025}
{Merz}, G., {Liu}, X., {Schmidt}, S., {et~al.} 2025, The Open Journal of Astrophysics, 8, 40, \dodoi{10.33232/001c.136809}

\bibitem[{Messias {et~al.}(2012)Messias, Afonso, Salvato, Mobasher, \& Hopkins}]{messias_new_2012}
Messias, H., Afonso, J., Salvato, M., Mobasher, B., \& Hopkins, A.~M. 2012, The Astrophysical Journal, 754, 120, \dodoi{10.1088/0004-637X/754/2/120}

\bibitem[{März(2019)}]{marz_xgboostlss}
März, A. 2019, XGBoostLSS -- An extension of XGBoost to probabilistic forecasting.
\newblock \doarXiv{1907.03178}

\bibitem[{Neves {et~al.}(2021)Neves, Naik, \& Proen{\c{c}}a}]{SGAIN_paper}
Neves, D.~T., Naik, M.~G., \& Proen{\c{c}}a, A. 2021, in Computational Science -- ICCS 2021, ed. M.~Paszynski, D.~Kranzlm{\"u}ller, V.~V. Krzhizhanovskaya, J.~J. Dongarra, \& P.~M.~A. Sloot (Cham: Springer International Publishing), 98--113

\bibitem[{{Ni} {et~al.}(2021){Ni}, {Brandt}, {Yang}, {Leja}, {Chen}, {Luo}, {Matharu}, {Sun}, {Vito}, {Xue}, \& {Zhang}}]{Rieke_miri}
{Ni}, Q., {Brandt}, W.~N., {Yang}, G., {et~al.} 2021, \mnras, 500, 4989, \dodoi{10.1093/mnras/staa3514}

\bibitem[{{Pacifici} {et~al.}(2023){Pacifici}, {Iyer}, {Mobasher}, {da Cunha}, {Acquaviva}, {Burgarella}, {Calistro Rivera}, {Carnall}, {Chang}, {Chartab}, {Cooke}, {Fairhurst}, {Kartaltepe}, {Leja}, {Ma{\l}ek}, {Salmon}, {Torelli}, {Vidal-Garc{\'\i}a}, {Boquien}, {Brammer}, {Brown}, {Capak}, {Chevallard}, {Circosta}, {Croton}, {Davidzon}, {Dickinson}, {Duncan}, {Faber}, {Ferguson}, {Fontana}, {Guo}, {Haeussler}, {Hemmati}, {Jafariyazani}, {Kassin}, {Larson}, {Lee}, {Mantha}, {Marchi}, {Nayyeri}, {Newman}, {Pandya}, {Pforr}, {Reddy}, {Sanders}, {Shah}, {Shahidi}, {Stevans}, {Triani}, {Tyler}, {Vanderhoof}, {de la Vega}, {Wang}, \& {Weston}}]{Pacifici_2023}
{Pacifici}, C., {Iyer}, K.~G., {Mobasher}, B., {et~al.} 2023, \apj, 944, 141, \dodoi{10.3847/1538-4357/acacff}

\bibitem[{{Padovani} {et~al.}(2017){Padovani}, {Alexander}, {Assef}, {De Marco}, {Giommi}, {Hickox}, {Richards}, {Smol{\v{c}}i{\'c}}, {Hatziminaoglou}, {Mainieri}, \& {Salvato}}]{padovani_2017}
{Padovani}, P., {Alexander}, D.~M., {Assef}, R.~J., {et~al.} 2017, \aapr, 25, 2, \dodoi{10.1007/s00159-017-0102-9}

\bibitem[{pandas~development team(2020)}]{reback2020pandas}
pandas~development team, T. 2020, pandas-dev/pandas: Pandas, latest,  Zenodo, \dodoi{10.5281/zenodo.3509134}

\bibitem[{Pedregosa {et~al.}(2011)Pedregosa, Varoquaux, Gramfort, Michel, Thirion, Grisel, Blondel, Prettenhofer, Weiss, Dubourg, Vanderplas, Passos, Cournapeau, Brucher, Perrot, \& Duchesnay}]{scikit-learn}
Pedregosa, F., Varoquaux, G., Gramfort, A., {et~al.} 2011, Journal of Machine Learning Research, 12, 2825

\bibitem[{Peeters {et~al.}(2004)Peeters, Spoon, \& Tielens}]{peeters_polycyclic_2004}
Peeters, E., Spoon, H. W.~W., \& Tielens, A. G. G.~M. 2004, The Astrophysical Journal, 613, 986, \dodoi{10.1086/423237}

\bibitem[{{P{\'e}rez-Gonz{\'a}lez} {et~al.}(2024{\natexlab{a}}){P{\'e}rez-Gonz{\'a}lez}, {Rinaldi}, {Caputi}, {{\'A}lvarez-M{\'a}rquez}, {Annunziatella}, {Langeroodi}, {Moutard}, {Boogaard}, {Iani}, {Melinder}, {Costantin}, {{\"O}stlin}, {Colina}, {Greve}, {Wright}, {Alonso-Herrero}, {Bik}, {Bosman}, {Crespo G{\'o}mez}, {Dicken}, {Eckart}, {Garc{\'\i}a-Mar{\'\i}n}, {Gillman}, {G{\"u}del}, {Henning}, {Hjorth}, {Jermann}, {Labiano}, {Meyer}, {Pei{\ensuremath{\beta}}ker}, {Pye}, {Ray}, {Tikkanen}, {Walter}, \& {van der Werf}}]{Perez_2024_dark}
{P{\'e}rez-Gonz{\'a}lez}, P.~G., {Rinaldi}, P., {Caputi}, K.~I., {et~al.} 2024{\natexlab{a}}, \apjl, 969, L10, \dodoi{10.3847/2041-8213/ad517b}

\bibitem[{{P{\'e}rez-Gonz{\'a}lez} {et~al.}(2024{\natexlab{b}}){P{\'e}rez-Gonz{\'a}lez}, {Barro}, {Rieke}, {Lyu}, {Rieke}, {Alberts}, {Williams}, {Hainline}, {Sun}, {Pusk{\'a}s}, {Annunziatella}, {Baker}, {Bunker}, {Egami}, {Ji}, {Johnson}, {Robertson}, {Rodr{\'\i}guez Del Pino}, {Rujopakarn}, {Shivaei}, {Tacchella}, {Willmer}, \& {Willott}}]{perez_2024_highz}
{P{\'e}rez-Gonz{\'a}lez}, P.~G., {Barro}, G., {Rieke}, G.~H., {et~al.} 2024{\natexlab{b}}, \apj, 968, 4, \dodoi{10.3847/1538-4357/ad38bb}

\bibitem[{{Polletta} {et~al.}(2007){Polletta}, {Tajer}, {Maraschi}, {Trinchieri}, {Lonsdale}, {Chiappetti}, {Andreon}, {Pierre}, {Le F{\`e}vre}, {Zamorani}, {Maccagni}, {Garcet}, {Surdej}, {Franceschini}, {Alloin}, {Shupe}, {Surace}, {Fang}, {Rowan-Robinson}, {Smith}, \& {Tresse}}]{SWIRE}
{Polletta}, M., {Tajer}, M., {Maraschi}, L., {et~al.} 2007, \apj, 663, 81, \dodoi{10.1086/518113}

\bibitem[{Pope {et~al.}(2008)Pope, Chary, Alexander, Armus, Dickinson, Elbaz, Frayer, Scott, \& Teplitz}]{pope_mid-infrared_2008}
Pope, A., Chary, R.-R., Alexander, D.~M., {et~al.} 2008, The Astrophysical Journal, 675, 1171, \dodoi{10.1086/527030}

\bibitem[{Pérez-González {et~al.}(2008)Pérez-González, Rieke, Villar, Barro, Blaylock, Egami, Gallego, Gil~de Paz, Pascual, Zamorano, \& Donley}]{perez-gonzalez_stellar_2008}
Pérez-González, P.~G., Rieke, G.~H., Villar, V., {et~al.} 2008, The Astrophysical Journal, 675, 234, \dodoi{10.1086/523690}

\bibitem[{{Rieke} {et~al.}(2017){Rieke}, {Alberts}, {Lyu}, {Morrison}, \& {Shivaei}}]{SMILES}
{Rieke}, G., {Alberts}, S., {Lyu}, J., {Morrison}, J., \& {Shivaei}, I. 2017, {MIRI in the Hubble Ultra-Deep Field}, JWST Proposal. Cycle 1, ID. \#1207

\bibitem[{Rigby {et~al.}(2023)Rigby, Perrin, McElwain, Kimble, Friedman, Lallo, Doyon, Feinberg, Ferruit, Glasse, Rieke, Rieke, Wright, Willott, Colon, Milam, Neff, Stark, Valenti, Abell, Abney, Abul-Huda, Acton, Adams, Adler, Aguilar, Ahmed, Albert, Alberts, Aldridge, Allen, Altenburg, Álvarez Márquez, Alves~de Oliveira, Andersen, Anderson, Anderson, Argyriou, Armstrong, Arribas, Artigau, Arvai, Atkinson, Bacon, Bair, Banks, Barrientes, Barringer, Bartosik, Bast, Baudoz, Beatty, Bechtold, Beck, Bergeron, Bergkoetter, Bhatawdekar, Birkmann, Blazek, Blome, Boccaletti, Böker, Boia, Bonaventura, Bond, Bosley, Boucarut, Bourque, Bouwman, Bower, Bowers, Boyer, Bradley, Brady, Braun, Breda, Bresnahan, Bright, Britt, Bromenschenkel, Brooks, Brooks, Brown, Brown, Brown, Bunker, Burger, Bushouse, Cale, Cameron, Cameron, Canipe, Caplinger, Caputo, Cara, Carey, Carniani, Carrasquilla, Carruthers, Case, Catherine, Chance, Chapman, Charlot, Charlow, Chayer, Chen, Cherinka, Chichester, Chilton, Chonis, Clampin, Clark,
  Clark, Coe, Coleman, Comber, Comeau, Connolly, Cooper, Cooper, Coppock, Correnti, Cossou, Coulais, Coyle, Cracraft, Curti, Cuturic, Davis, Davis, Dean, DeLisa, deMeester, Dencheva, Dencheva, DePasquale, Deschenes, Hunor~Detre, Diaz, Dicken, DiFelice, Dillman, Dixon, Doggett, Donaldson, Douglas, DuPrie, Dupuis, Durning, Easmin, Eck, Edeani, Egami, Ehrenwinkler, Eisenhamer, Eisenhower, Elie, Elliott, Elliott, Ellis, Engesser, Espinoza, Etienne, Etxaluze, Falini, Feeney, Ferry, Filippazzo, Fincham, Fix, Flagey, Florian, Flynn, Fontanella, Ford, Forshay, Fox, Franz, Fu, Fullerton, Galkin, Galyer, García~Marín, Gardner, Gardner, Garland, Garrett, Gasman, Gaspar, Gaudreau, Gauthier, Geers, Geithner, Gennaro, Giardino, Girard, Giuliano, Glassmire, \& Glauser}]{rigby_science_2023}
Rigby, J., Perrin, M., McElwain, M., {et~al.} 2023, Publications of the Astronomical Society of the Pacific, 135, 048001, \dodoi{10.1088/1538-3873/acb293}

\bibitem[{{Roster} {et~al.}(2024){Roster}, {Salvato}, {Krippendorf}, {Saxena}, {Shirley}, {Buchner}, {Wolf}, {Dwelly}, {Bauer}, {Aird}, {Ricci}, {Assef}, {Anderson}, {Liu}, {Merloni}, {Weller}, \& {Nandra}}]{Roster_2024}
{Roster}, W., {Salvato}, M., {Krippendorf}, S., {et~al.} 2024, \aap, 692, A260, \dodoi{10.1051/0004-6361/202452361}

\bibitem[{Rousseeuw {et~al.}(1999)Rousseeuw, Ruts, \& and}]{Roussee_bagplot}
Rousseeuw, P.~J., Ruts, I., \& and, J. W.~T. 1999, The American Statistician, 53, 382, \dodoi{10.1080/00031305.1999.10474494}

\bibitem[{Sajina {et~al.}(2005)Sajina, Lacy, \& Scott}]{sajina_simulating_2005}
Sajina, A., Lacy, M., \& Scott, D. 2005, The Astrophysical Journal, 621, 256, \dodoi{10.1086/426536}

\bibitem[{Sajina {et~al.}(2012)Sajina, Yan, Fadda, Dasyra, \& Huynh}]{sajina_spitzer-_2012}
Sajina, A., Yan, L., Fadda, D., Dasyra, K., \& Huynh, M. 2012, The Astrophysical Journal, 757, 13, \dodoi{10.1088/0004-637X/757/1/13}

\bibitem[{{Saxena} {et~al.}(2024){Saxena}, {Salvato}, {Roster}, {Shirley}, {Buchner}, {Wolf}, {Kohl}, {Starck}, {Dwelly}, {Comparat}, {Malyali}, {Krippendorf}, {Zenteno}, {Lang}, {Schlegel}, {Zhou}, {Dey}, {Valdes}, {Myers}, {Assef}, {Ricci}, {Temple}, {Merloni}, {Koekemoer}, {Anderson}, {Morrison}, {Liu}, \& {Nandra}}]{Saxena_2024}
{Saxena}, A., {Salvato}, M., {Roster}, W., {et~al.} 2024, \aap, 690, A365, \dodoi{10.1051/0004-6361/202450886}

\bibitem[{{Scargle}(1998)}]{Bayesian_blocks}
{Scargle}, J.~D. 1998, \apj, 504, 405, \dodoi{10.1086/306064}

\bibitem[{Shankar {et~al.}(2009)Shankar, Weinberg, \& Miralda-Escudé}]{shankar_self-consistent_2009}
Shankar, F., Weinberg, D.~H., \& Miralda-Escudé, J. 2009, The Astrophysical Journal, 690, 20, \dodoi{10.1088/0004-637X/690/1/20}

\bibitem[{Stalevski {et~al.}(2012)Stalevski, Fritz, Baes, Nakos, \& Popović}]{stalevski_3d_2012}
Stalevski, M., Fritz, J., Baes, M., Nakos, T., \& Popović, L.~C. 2012, Monthly Notices of the Royal Astronomical Society, 420, 2756, \dodoi{10.1111/j.1365-2966.2011.19775.x}

\bibitem[{Stalevski {et~al.}(2016)Stalevski, Ricci, Ueda, Lira, Fritz, \& Baes}]{stalevski_dust_2016}
Stalevski, M., Ricci, C., Ueda, Y., {et~al.} 2016, Monthly Notices of the Royal Astronomical Society, 458, 2288, \dodoi{10.1093/mnras/stw444}

\bibitem[{Stasinopoulos \& Rigby(2007)}]{stasinopoulos_generalized_2007}
Stasinopoulos, D.~M., \& Rigby, R.~A. 2007, Journal of Statistical Software, 23, 1, \dodoi{10.18637/jss.v023.i07}

\bibitem[{Stern {et~al.}(2005)Stern, Eisenhardt, Gorjian, Kochanek, Caldwell, Eisenstein, Brodwin, Brown, Cool, Dey, Green, Jannuzi, Murray, Pahre, \& Willner}]{stern_mid-infrared_2005}
Stern, D., Eisenhardt, P., Gorjian, V., {et~al.} 2005, ApJ, 631, 163, \dodoi{10.1086/432523}

\bibitem[{{Taylor}(2005)}]{topcat}
{Taylor}, M.~B. 2005, in Astronomical Society of the Pacific Conference Series, Vol. 347, Astronomical Data Analysis Software and Systems XIV, ed. P.~{Shopbell}, M.~{Britton}, \& R.~{Ebert}, 29

\bibitem[{{Taylor}(2006)}]{STILTS}
{Taylor}, M.~B. 2006, in Astronomical Society of the Pacific Conference Series, Vol. 351, Astronomical Data Analysis Software and Systems XV, ed. C.~{Gabriel}, C.~{Arviset}, D.~{Ponz}, \& S.~{Enrique}, 666

\bibitem[{Toba {et~al.}(2014)Toba, Oyabu, Matsuhara, Malkan, Gandhi, Nakagawa, Isobe, Shirahata, Oi, Ohyama, Takita, Yamauchi, \& Yano}]{toba_luminosity_2014}
Toba, Y., Oyabu, S., Matsuhara, H., {et~al.} 2014, ApJ, 788, 45, \dodoi{10.1088/0004-637X/788/1/45}

\bibitem[{Ustimenko {et~al.}(2020)Ustimenko, Prokhorenkova, \& Malinin}]{malinin_uncertainty}
Ustimenko, A., Prokhorenkova, L., \& Malinin, A. 2020, CoRR, abs/2006.10562

\bibitem[{Vidal {et~al.}(2025)Vidal, Sajina, Banks, Béthermin, Ferkinhoff, Petric, Pope, Lyu, U, Yung, \& Patil}]{vidal_2025}
Vidal, E.~P., Sajina, A., Banks, A.~R., {et~al.} 2025.
\newblock \doarXiv{2509.15331}

\bibitem[{Virtanen {et~al.}(2020)Virtanen, Gommers, Oliphant, Haberland, Reddy, Cournapeau, Burovski, Peterson, Weckesser, Bright, {van der Walt}, Brett, Wilson, Millman, Mayorov, Nelson, Jones, Kern, Larson, Carey, Polat, Feng, Moore, {VanderPlas}, Laxalde, Perktold, Cimrman, Henriksen, Quintero, Harris, Archibald, Ribeiro, Pedregosa, {van Mulbregt}, \& {SciPy 1.0 Contributors}}]{2020SciPy-NMeth}
Virtanen, P., Gommers, R., Oliphant, T.~E., {et~al.} 2020, Nature Methods, 17, 261, \dodoi{10.1038/s41592-019-0686-2}

\bibitem[{Wu {et~al.}(2012)Wu, Hao, Jia, Zhang, \& Peng}]{wu_sdss_2012}
Wu, X.-B., Hao, G., Jia, Z., Zhang, Y., \& Peng, N. 2012, The Astronomical Journal, 144, 49, \dodoi{10.1088/0004-6256/144/2/49}

\bibitem[{Yang {et~al.}(2019)Yang, Boquien, Buat, Burgarella, Ciesla, Duras, Stalevski, Brandt, \& Papovich}]{yang_2020}
Yang, G., Boquien, M., Buat, V., {et~al.} 2019, Monthly Notices of the Royal Astronomical Society, 491, 740, \dodoi{10.1093/mnras/stz3001}

\bibitem[{Yang {et~al.}(2023)Yang, Caputi, Papovich, Arrabal~Haro, Bagley, Behroozi, Bell, Bisigello, Buat, Burgarella, Cheng, Cleri, Davé, Dickinson, Elbaz, Ferguson, Finkelstein, Grogin, Hathi, Hirschmann, Holwerda, Huertas-Company, Hutchison, Iani, Kartaltepe, Kirkpatrick, Kocevski, Koekemoer, Kokorev, Larson, Lucas, Pérez-González, Rinaldi, Shen, Trump, de~la Vega, Yung, \& Zavala}]{yang_ceers_2023}
Yang, G., Caputi, K.~I., Papovich, C., {et~al.} 2023, The Astrophysical Journal Letters, 950, L5, \dodoi{10.3847/2041-8213/acd639}

\bibitem[{Zhou(2012)}]{Zhou_2012}
Zhou, Z.-H. 2012, Ensemble Methods: Foundations and Algorithms, 1st edn. (Chapman \& Hall/CRC)

\end{thebibliography}
\bibliographystyle{aasjournal}



\end{document}